\documentclass[usedcolumn,useAMS,usenatbib,a4paper]{mn2e}

\usepackage{graphicx,times}
\usepackage{epsfig}
\usepackage{multirow}
\usepackage{booktabs}
\usepackage{mathptmx}
\usepackage{amssymb,amsmath}
\usepackage{xcolor}
\usepackage{rotating}

\newcommand{\Msun}{{\rm M}_{\odot}}

\newcommand{\kms}{{\rm ~km~s^{-1}}}

\def \fof{\textsc{fof}}
\def \gadget{\textsc{gadget-2}}
\def \sag{\textsc{sag}}
\def \subfind{\textsc{subfind}}

\def \SAG0{\textsc{sag0}}
\def \SAG1S{\textsc{sag1s}}
\def \SAG1TH5{\textsc{sag1th5}}
\def \SAG1TH100{\textsc{sag1th100}}
\def \SAG1TH1000{\textsc{sag1th1000}}

\title[Chemo-Archaeological Downsizing: Impact of a Top Heavy IGIMF]{Chemo-Archaeological Downsizing in a Hierarchical Universe: Impact of a Top Heavy IGIMF}
\author[I. D. Gargiulo et al.]{I. D. Gargiulo$^{1}$\thanks{E-mail: gargiulo@fcaglp.unlp.edu.ar}, S. A. Cora$^{1,2,3}$, N. D. Padilla$^{4,5}$, 
A. M. Mu\~noz Arancibia$^{4}$,
A. N. Ruiz$^{3,6,7}$, \newauthor A. A. Orsi$^{4,5}$, T. E. Tecce$^{4,5}$,
C. Weidner$^{8,9}$  
and G. Bruzual$^{10}$\\
$^{1}$Instituto de Astrof\'isica de La Plata (CCT La Plata, CONICET, UNLP), Paseo del Bosque s/n, B1900FWA, La Plata, Argentina.\\
$^{2}$Facultad de Ciencias Astron\'omicas y Geof\'{\i}sicas, Universidad Nacional de La Plata, Paseo del Bosque s/n, B1900FWA, La Plata, Argentina.\\
$^{3}$Consejo Nacional de Investigaciones Cient\'{\i}ficas y T\'ecnicas, Rivadavia 1917, C1033AAJ Buenos Aires, Argentina.\\
$^{4}$Instituto de Astrof\'{\i}sica, Pontificia Universidad Cat\'olica de Chile, Av. Vicu\~na Mackenna 4860, Santiago, Chile.\\
$^{5}$Centro de Astro-Ingenier\'{\i}a, Pontificia Universidad Cat\'olica de Chile, Av. Vicu\~na Mackenna 4860, Santiago, Chile.\\
$^{6}$Instituto de Astronom\'{\i}a Te\'orica y Experimental (CCT C\'ordoba, CONICET, UNC), Laprida 854, X5000BGR, C\'ordoba, Argentina.\\
$^{7}$Observatorio Astron\'omico de C\'ordoba, Universidad Nacional de C\'ordoba, Laprida 854, X5000BGR, C\'ordoba, Argentina.\\
$^{8}$ Instituto de Astrof\'{\i}sica de Canarias, Calle V\'{\i}a L\'actea, E38205 - La Laguna (Tenerife), Espa\~na.\\ 
$^{9}$ Departamento de Astrof\'{\i}sica, Universidad de La Laguna (ULL), E-38206 La Laguna, Tenerife, Espa\~na.\\
$^{10}$Centro de Radioastronom\'{\i}a y Astrof\'{\i}sica, UNAM, Campus Morelia, M\'exico.\\ 
}

\begin{document}

\date{Accepted ... . Received ... ; in original form ... }

\pagerange{\pageref{firstpage}--\pageref{lastpage}} \pubyear{2014}

\maketitle

\label{firstpage}

\begin{abstract}

We make use of a semi-analytical model of galaxy formation to investigate 
the origin of the observed correlation 
between [$\alpha /{\rm Fe}$] abundance ratios 
and stellar mass in elliptical galaxies.
We implement a new galaxy-wide stellar initial mass function 
(Top Heavy Integrated Galaxy Initial Mass Function, TH-IGIMF) 
in the semi-analytic model \sag~and evaluate its impact on the 
chemical evolution
of galaxies.
The SFR-dependence of the slope of the TH-IGIMF is found to be key 
to reproducing the correct [$\alpha /{\rm Fe}$]-stellar mass relation.
Massive galaxies reach higher [$\alpha /{\rm Fe}$]
abundance ratios because they are characterized by more top-heavy
IMFs as a result of their higher SFR.
As a consequence of our analysis, the value of the minimum embedded 
star cluster mass and of the slope of the
embedded cluster mass function, which are free parameters involved in the TH-IGIMF theory,
are found to be as low as $5\,\Msun$ and $2$, respectively. 
A mild downsizing trend is present for galaxies generated assuming either
a universal IMF or a variable TH-IGIMF.
We find that, regardless of galaxy mass, 
older galaxies (with formation redshifts $\gtrsim 2$) 
are formed in shorter time-scales ($\lesssim 2\,{\rm Gyr}$), thus
achieving larger [$\alpha /{\rm Fe}$] values. 
Hence, the time-scale of galaxy formation alone cannot explain the slope of 
the [$\alpha /{\rm Fe}$]-galaxy mass relation, but is responsible for
the big dispersion  of [$\alpha /{\rm Fe}$] abundance ratios
at fixed stellar mass. We further test the hyphothesis of a 
TH-IGIMF in elliptical galaxies by looking into mass-to-light ratios, 
and luminosity functions.
Models with a TH-IGIMF are also favoured by these constraints. In particular,
mass-to-light ratios agree with observed values for massive galaxies
while being overpredicted for less massive ones; this overprediction
is present regardless of the IMF considered. \\

\end{abstract}

\begin{keywords}
Galaxies: abundances - Galaxies: formation - Methods: numerical
\end{keywords}

\section{Introduction}
\label{sec:intro}

The chemical signatures imprinted in stellar populations  
are 
key features in the understanding of
the formation and evolution 
of galaxies.  
The alpha-element abundance relative to iron has been recognized 
for a long time as an indicator of the 
formation time-scale
of stellar 
populations.  It is well known that the alpha-elements are produced
by alpha capture exclusively in 
core-collapse supernovae (SNe CC),
whose 
progenitors have lifetimes of tens of Megayears, while Fe 
and Fe-peak elements are mainly the yields of type Ia supernovae (SNe Ia),
whose progenitors are long lived 
\citep[$0.5 - 3 \,{\rm Gyr}$,][]{Greggio05}. A composite stellar 
population with an excess of alpha-elements relative to iron is therefore 
assumed to form in a short period of time ($\lesssim 1 \,{\rm Gyr}$) 
\citep{Matteucci94},
given that the bulk of stars in the system form before the interstellar
medium (ISM) is contamined with the SNe Ia yields. 
The early works of \citet{Tinsley76, Tinsley79} already considered
this abundance ratio in order to explain the formation
time-scales of the solar neighbourhood and the Milky Way stellar halo. 
This argument has been also considered to explain the
trend observed among elliptical galaxies 
in which more massive ellipticals (with higher velocity dispersions) 
have systematically higher alpha-elements to iron ratios
\citep[e.g.][]{Worthey1992}.
\citet{Faber1973} 
first observed that some absorption features related to 
the abundance of Mg increase 
monotonically with increasing luminosity. Since then, a large amount of work 
has been done measuring alpha-element abundances to study this trend in all 
sort of environments 
\citep[e.g.][]{Worthey1992, Trager2000, Kuntschner2000, Bernardi2006, SanchezBlazquez2006, Spolaor2010, Thomas2010}. 

In the context of the $\Lambda$-CDM 
paradigm, the observed trend of $[\alpha/{\rm Fe}]$ as a function of mass could
be considered troublesome. In this framework, the growth of structure is 
hierarchical. Dark matter (DM) haloes form via dissipationless gravitational
collapse \citep{White-Rees1978} and grow in a bottom-up fashion, with more
massive haloes taking longer periods of time to build-up their masses.
Considering that galaxies form in the centre of DM haloes, a priori 
it is not expected that more massive galaxies have higher 
abundances of $\alpha$-elements relative to iron.  Nevertheless, the existence
of baryonic processes like gas cooling, star formation (SF) 
and feedback from SNe 
and active galactic nuclei (AGN) can detach the actual formation of the galaxy 
from the assembly of its parent DM halo. This ``anti-hierarchical" 
behavior, usually called {\it downsizing}, manifests in different ways in 
galaxies: the characteristic mass of star forming galaxies increases with 
redshift \citep[e.g.][]{Cowie-Barger2008}, the evolution of the stellar mass 
functions seems to indicate that less massive galaxies assembly their mass 
at later times \citep[e.g.][]{PerezGonzalez2008}, more massive galaxies 
have older stellar populations 
than less massive ones \citep[e.g.][]{Gallazzi2005}. 
\citet{Fontanot2009} classify the different observational manifestations of
downsizing, suggesting the possibility that not all these phenomena are 
related to the same underlying physical process.
In particular, they coined the term 
{\it{chemo-archaeological downsizing}} to refer to the trend 
of abundance ratios with stellar mass shown by
elliptical galaxies. 

The pointed tension between observations and the 
hierarchical paradigm has been addressed by different groups using analytic
and semi-analytic models of galaxy formation (SAMs).     
\citet{Thomas1999} 
was the first in modelling
the chemical enrichment of elliptical galaxies
 within the framework of hierarchical galaxy formation, but 
considering a closed box
 model for each galaxy and ignoring the merger history of galaxies by using
pre-calculated star formation histories. 
He fails to reproduce the observed trend
of $[\alpha/{\rm Fe}]$ versus velocity dispersion.
\citet{Nagashima+2005b} study the metal enrichment of elliptical galaxies
using a semi-analytic model 
with merger trees.
They find that a top heavy initial mass
function (IMF) in starbursts
is needed in order to reproduce the observed  $[\alpha/{\rm Fe}]$ abundance 
ratio in $L_{\star}$ 
ellipticals, but none of their models reproduce the observed trend of 
this ratio 
with velocity dispersion, $\sigma$. 

\citet{Pipino+2009} 
 focussed on the $[\alpha/{\rm Fe}]$-stellar mass relation
in elliptical galaxies by including the enrichment of supernovae type Ia and II,
and of low- and intermediate- mass stars, as well as the 
corresponding delay times for SNe Ia in the semi-analytic model of galaxy 
formation GalICs. They find a 
better agreement with the observed $[\alpha/{\rm Fe}]$-$\sigma$ relation 
when taking into 
account AGN feedback in their models, although still marginal.
\citet{Calura2009} implement a chemical model on top of the star formation 
histories extracted from a semi-analytic model, 
and evaluate the impact on the $[\alpha/{\rm Fe}]$-$\sigma$ relation
of the AGN feedback and an ad-hoc variable IMF that becomes top heavy in 
starbursts, finding that they are both needed to obtain a slope in the relation
similar to the observed one. 
This was the first attempt to include a variable IMF in a semi-analytic model
to study the $[\alpha/{\rm Fe}]$ ratios in ellipticals.
A similar assumption about the IMF has been considered by Baugh et al. (2005)
to explain the number counts of submillimiter galaxies at high redshift.
Later on, \citet{Calura2011} mention the importance of the harassment of 
satellite galaxies at high redshift as a trigger of SF in the 
progenitors of massive galaxies that form stars enhanced in $\alpha$-elements,
 contributing to build up the relation.
\citet{Arrigoni2010} also analyse the $[\alpha/{\rm Fe}]$-mass relation
by using the semi-analytic
model of \citet{Somerville2008} that includes AGN feedback, in which
they implement a chemical enrichment model that involves SNe Ia and SNe II 
yields and the corresponding delay times for the SNe Ia explosions. 
They 
find a good agreement with observations by combining a shallower 
slope for the IMF and a lower SNe Ia fraction.
All the results obtained from semi-analytic models of galaxy 
formation and evolution suggest that the time-scale argument,
which compares the time-scales of SF and metal 
ejection from SNe,
might be 
insufficient to explain the build-up of the $[\alpha/{\rm Fe}]$-mass 
relation of early type galaxies,
supporting the importance of the role of the IMF.

Numerous works are being published that show 
 the need to explore alternative
and variable stellar initial mass functions.
The stellar IMF has direct consequences on the 
evolution of galaxies.  
It determines the ratio of massive stars to low and intermediate-mass stars, 
imprinting the chemical abundance ratios of stellar populations in galaxies
\citep{Lucatello2005, Koppen2007}, and establishing the amount 
of reheated mass deposited in the ISM by SNe explosions. 
It also determines the mass-to-light ratio (M/L) of galaxies,
given that stars with different initial stellar masses contribute with
very different amounts of light along their lives.
In the last decade, a formalism for a galaxy-wide stellar IMF has been 
developed \citep[see][ or Subsection \ref{sec:varIMF} 
of this paper]{Kroupa2012}, where
the instantaneous star formation rate (SFR) of galaxies determines the final 
shape of the integrated IMF.          
\citet{Recchi+2009} evaluate the influence 
on the [$\alpha$/Fe]-$\sigma$ relation of the 
galaxy-wide integrated galactic initial mass function
\citep[IGIMF,][]{Kroupa2003} 
implemented in an analytical model of chemical 
enrichment of galaxies. They find that the downsizing effect
(shorter duration of the SF in larger galaxies) must be 
present to reproduce the observed trend, although it must be milder than
the one infered by \citet{Thomas2005} on the basis of
the observational [$\alpha$/Fe]-$\sigma$ relation. 
Several simplifying assumptions have been adopted in their work.
Moreover, they do not include the effect 
of a top heavy version of the IGIMF (TH-IGIMF)
\citep{Weidner2011, Weidner2013a}.
This particular IMF has been tested by \citet{Fontanot2014} 
in the semi-analytic model MORGANA
evaluating its impact on the stellar mass functions and star formation rates. 
These results are compared with those obtained from other
IMF with different dependencies on galaxy properties and environment,
finding that the predicted galaxy properties are mainly affected by
an increasing Top-Heavy IMF at increasing star formation rates.

The aim of this work is to address the chemical evolution of galaxies
in the framework of the $\Lambda$-CDM paradigm, tackling the problem
of $\alpha$-element enrichment in elliptical galaxies.
We evaluate the impact of the new theory of 
TH-IGIMF, never considered so far 
for this particular issue.
We also analyse the influence of the TH-IGIMF
in the luminosity
function and the mass-to-light ratios of elliptical galaxies 
analyzing the tilt of the fundamental plane (M/L vs M).
For this purpose, we use 
the state of the art semi-analityc model 
\sag~(acronym for Semi-Analytic Galaxies)
that couples the most relevant 
ingredients of chemical evolution together with the physics involved
in the formation and evolution of galaxies
(\citealt{Cora06}; \citealt*{lcp08}; \citealt{Tecce2010}).

This work is organized as follows. In Section 
\ref{sec:HybridModel}, we present the model
of galaxy formation used that combines a cosmological dark-matter only
simulation and the semi-analytic model of galaxy formation \sag,
giving a brief description of the latter with emphasis on
the formation of elliptical galaxies, the chemical aspects
and the new implementation of extended bursts of star formation
(Subsection \ref{sec:SAG});
Subsection \ref{sec:varIMF} is devoted to the detailed description of
 the implementation of the TH-IGIMF.
In Section \ref{sec:CalibrationSAG}, we present a summary of the 
calibration process involved in this work and show the first 
results of the [$\alpha/{\rm Fe}$]-stellar mass relation with a 
universal Salpeter IMF (\ref{sec:AlphaSalp}).
In Section \ref{sec:AlphaFeTHIGIMF}, 
we analyse the impact of a variable IMF on the developement of the
[$\alpha/{\rm Fe}$]-stellar mass relation,
discussing
the influence of the various
time-scales involved in the modelling of formation of elliptical galaxies,
such as starburst timescales and SNe delay times  (Subsection \ref{sec:starbTimeScales}). In Subsection \ref{sec:masstolight} we test the implementation of the
variable IMF by means of the resultant luminosity functions and mass-to-light
ratios.  
In Section \ref{sec:timescales} we study in detail the global formation 
timescales of galaxies and analize their impact on the development of the 
[$\alpha/{\rm Fe}$]-stellar mass relation and the existence of a downsizing 
pattern.
In Section \ref{discussion}, we mention previous investigations in the
field putting in context our own results, with the perspective of 
future research work related to the main
topic of this paper.  
We present our conclusion in Section \ref{conclusions}.
Finally, we add complementary information in appendix \ref{App:AppendixA}, 
where we detail the physical recipies implemented in the 
model and the link with the free parameters involved in the calibration 
procedure that allows to obtain the best-fitting values. 

\section[]{Model of galaxy formation}
\label{sec:HybridModel}
In the present work, we use a 
model
of galaxy formation that combines a
dissipationless cosmological
{\em N}-body simulation 
with a semi-analytic model of galaxy
formation and evolution.
In the following, we describe the features of the simulation used
and those aspects of the semi-analytic code that are relevant for
the present study.

\subsection{{\em N}-body simulation}

The DM only {\em N}-body simulation used for this study 
considers the standard $\Lambda$-CDM scenario, 
characterized by 
cosmological parameters $\Omega_{\rm m}=0.28$,
$\Omega_{\rm b}=0.046$, $\Omega_{\Lambda}=0.72$, $h=0.7$, $n=0.96$, 
$\sigma_{8}=0.82$, according 
to the WMAP7 cosmology \citep{jarosik2011}. 
The simulation was
run with \gadget~\citep{springel_gadget2_2005} 
using $640^3$ particles in a cubic box of comoving sidelength
$L=150\,h^{-1}{\rm Mpc}$. 
The initial conditions were generated using
{\small GRAFIC2} \citep{bertschinger_grafic2_2001}. The simulation was evolved from $z_{\rm ini}=61.2$
to the present epoch, storing 100 outputs equally spaced in $\log_{10}(a)$ between $z=20$ and $z=0$
\citep{benson_trees_2012}.

DM haloes were identified using a friends-of-friends (\fof) algorithm.
Then, the application of \subfind~algorithm
\citep{springel_subfind_2001}
allows to select self-bound substructures
(subhaloes) having at least 10 particles, with a mass of 
$1\times 10^{9}\,{h^{-1}}\,\Msun$ each. DM haloes and
subhaloes have masses in the range $1\times 10^{10} - 7.8 \times 10^{14}\,{h^{-1}} \,\Msun$.

\subsection[]{Semi-analytic model of galaxy formation}
\label{sec:SAG}

We use the semi-analytic model {\small SAG} 
described in the works of \citet{Cora06},
\citet{lcp08} and \citet{Tecce2010}, 
which is based on the version of the Munich semi-analytic model 
presented by \citet{springel_subfind_2001}.
In this section,
we describe briefly the main features of the model,
giving more details of aspects closely related with the aim of this research
work, that is, the chemical implementation and the channels of bulge formation
that give rise to the population of elliptical galaxies.
The time-scales involved in the formation of this stellar component
are crucial in the understanding of the development of the 
[$\alpha$/Fe]-stellar mass relation. Several modifications have to be included
for a proper treatment of this aspect, which are also described in detail.

\subsubsection{Main features of \sag}
\label{sec:MainFeatures}

The semi-analytic model {\small SAG} uses the subhalo merger trees provided
by the underlying simulation to generate the galaxy population.
Cold gas is settled in the centres of DM subhaloes as a result
of the radiative cooling suffered by the hot gas contained within them.
It is assumed that the 
cold gas
disc has an exponential density profile 
\citep{Tecce2010}.
When the density of cold gas becomes high enough, SF
in the disc is triggered according to the conditions given
by \citet{Croton2006} which are considered in the current version
of \sag. We refer to this mode of SF as quiescent SF.
One aspect that changed with respect to previous
versions of \sag~is the estimation of the gas cooling rate;
we now consider the total radiated power
per chemical element
given by \citet{foster2012}\footnote{
These modifications were already taken into account in the version of 
\sag~considered
to evaluate the capability of the calibration method used to tune the free
parameters of the model \citep{Ruiz2014}.}. 
The only galaxies for which 
infall of cooling gas from the intergalactic medium takes place are
central galaxies. These galaxies, also named {\it type 0}, 
reside within the
most massive subhalo within a \fof~halo.
Satellite galaxies are classified in two types, 
those residing in smaller subhaloes 
of the same \fof~halo ({\it type 1}), and those which DM subhalo
are below the resolution limit 
as a result of mass loss caused by tidal forces ({\it type 2}). 
Galaxies of the latter type merge 
with the central galaxy of their host subhalo after a dynamical 
friction time-scale \citep{BinneyTremaine87}. 
When a galaxy becomes a satellite of
either type, it suffers strangulation, that is, all of its hot gas halo 
is removed and transferred to the hot gas component of
the corresponding type~0 galaxy.
From a given star formation event and an IMF adopted, we can estimate
the number of SNe CC that contribute to energy feedback.
Effects of feedback from active galactic nuclei (AGN) are also included
\citep{lcp08}. The energy injection from both sources
regulates the star formation through  
the transfer of gas and metals from the cold to the hot gas phase.

\subsubsection{Formation of elliptical galaxies}
\label{sec:ellip_form}

The morphological classification of galaxies generated by the model
is based on the ratio between bulge and total {\em r}-band luminosity, B/T. 
Elliptical galaxies are those characterized by a
ratio ${\rm B/T}>{{(\rm B/T)}_{\rm thresh}}$, where the parameter
${(\rm B/T)}_{\rm thresh}$
may adopt values in the range $\sim 0.7-0.85$.
Thus, the formation of elliptical galaxies is directly connected with
bulge formation, which in \sag~takes place 
through global disk instabilities and 
two different kinds of galactic 
mergers.

When a galaxy merger occurs, 
the stellar mass ratio between the satellite galaxy 
and the central galaxy, $M_{\rm sat}/M_{\rm cen}$, is evaluated.  
If $M_{\rm sat}/M_{\rm cen}>0.3$, 
then the merger is considered as a major one. In this case, 
all the gas in the remnant
galaxy is consumed in a starburst contributing to the bulge formation,
and the stellar disc is completely 
relaxed and transferred to the bulge.
The triggering of a starburst in a minor merger 
($M_{\rm sat}/M_{\rm cen}\leq 0.3$) will depend
on the fraction of cold gas present in the disc of the central galaxy,
as implemented by \citet{lcp08} following the work of
\citet{malbon2007}. If the ratio between the
cold gas and the disc mass of the central galaxy, 
$M_\text{cold,cen} / M_\text{disc,cen}$, is larger than a fixed parameter 
$f_\text{burst}=0.6$, 
then the perturbation introduced by the merging satellite 
drives all the cold gas from both galaxies into the
bulge component, where it is consumed in a starburst;
the disc mass is given by the sum of the cold gas and the stars formed
through quiescent SF. 
However, if the
satellite is much less massive than the central galaxy 
($M_{\rm sat}/M_{\rm cen}\leq$~0.05), no burst occurs.
In minor mergers, only the
stars of the merging satellite are transferred to the bulge component
of the central galaxy. 

The other channel that contributes to the bulge formation is the global
disk instability.
Some configurations of galactic 
discs do not remain stable with time.
When a galactic
disc is sufficiently massive that its self-gravity is dominant, it
becomes unstable. 
This condition is expressed in the
model through the Efstathiou-Lake-Negroponte \citep{Efstathiou+1982} criterium,
that is, the stability to bar formation is lost when 
\begin{equation}
\epsilon_{\rm d} \equiv \frac{V_{\rm disc} }{ (G M_{\rm disc} 
/ R_{\rm disc})^{1/2}} \le \epsilon_{\rm thresh},
\label{eq:diskinstab}
\end{equation}
where $M_{\rm disc}$ is the mass of the disc (cold gas plus stars),  
$R_{\rm disc}$ is the disc scale radius, 
and $V_{\rm disc}$ is the circular velocity of the disc.
For the latter, we use the velocity where the rotation curve flattens, 
which we approximate by the velocity calculated at $\sim 3\,R_{\rm disc}$
(see \citet{Tecce2010} for details concerning disc features).
The free parameter $\epsilon_{\rm thresh}$ has values close to unity;
in our model, we adopt $\epsilon_{\rm thresh}=1$.
We consider an additional free parameter that takes into account
the influence of a perturbing galaxy that effectively triggers the disc 
instability; this is modelled
by computing the mean separation between galaxies in a
\fof~group. 
When the mean separation is
smaller than $f_\text{pert}\,R_{\rm disc}$, 
being $f_\text{pert}$ a free parameter, we consider that  
a neighbouring galaxy perturbs the unstable disc. As a consequence
of this, all the stars
and cold gas in that disc are transferred to the bulge component, and
all the gas present is consumed in a starburst.

\subsubsection{Chemical enrichment model}
\label{sec:chem_model}


In our chemical model,
we follow the production of ten chemical elements
(H, $^4$He, $^{12}$C, $^{14}$N, $^{16}$O, $^{20}$Ne, $^{24}$Mg, $^{28}$Si, 
$^{32}$S, $^{40}$Ca) 
generated by stars
in different
mass ranges, from low- and intermediate-mass stars to quasi massive
and massive stars. 
This version of \sag~is characterized by a new set of stellar yields. 
We use the best 
combination of stellar yields reported by \citet{Romano2010}, selected to be in
accordance with the large number of constraints for the Milky Way.    
For low and intermediate-mass stars (LIMS), in the mass interval $1-8 \Msun$,
we use the yields of \citet{Karakas2010}. For the mass loss of pre-supernova 
stars (He and CNO elements), we use the yields computed by the Geneva group 
\citep{Hirschi2005}, and for the yields of the explosive nucleosynthesis due to
SNe CC, we use the results of \citet{Kobayashi2006}.
In all cases, we have adopted the total ejected mass of solar metallicity 
models, 
for which the solar abundances of \citet{AndersGrevesse89} are assumed,
with a solar composition $Z_{\odot} = 0.02$.

Ejecta from SNe Ia are also included, 
which are characterized
by high iron production ($\sim 0.6\, {\rm M}_{\odot}$); 
we consider the nucleosynthesis
prescriptions from the updated model W7 by \citet{Iwamoto99}.
The SNe Ia rates are estimated using the single degenerate model, in 
which a SN Ia occurs by carbon deflagration in C-O 
white dwarfs in binary systems
whose components
have masses between $0.8$ and $8\,\Msun$ 
 \citep{Greggio83}; 
we implement the formalism presented by \citet{Lia2002} (see their Section 4.2.2). 
The fraction of these binary systems,
$A_{\rm bin}$, is one of the free parameters of \sag.
We take into account 
the return time-scale of mass losses and ejecta from all sources considered.
For this purpose, we use the stellar lifetimes given by \citet{Padovani93} 
who use results from
\citet{Matteucci86} for masses over $6.6\,\Msun$, and from \citet{Renzini86}
for masses below this limit.  
This aspect becomes especially relevant for this work because of the different 
delay times that characterize different types of SNe, which
affect the abundance of $\alpha$-elements relative to iron.
For a detailed 
description of the implementation in \sag~of the chemical enrichment of
the different baryonic components, we refer the reader 
to \citet{Cora06}.

\subsubsection{Extended starbursts}
\label{sec:ExtendedBurstsInSAG}

In previous versions of the model, the formation of stars in the bulge occurs
through starbursts that consume the available cold gas in a single step 
\footnote{Differential equations in \sag~are integrated in
time-steps of equal size used to subdivide the intervals between 
simulation outputs.}.
We now consider that starbursts
are characterized by a given time-scale during which the cold gas
driven to the galactic center is consumed gradually in several steps.
Thus, this cold gas has
the possibility of being progressively contaminated by the stars 
formed in the bulge. This way, the $[\alpha/{\rm Fe}]$ abundance ratio
of successive generations of stars in the bulge 
is not only determined by the abundances
of the disc cold gas at the moment in which the starburst is triggered, but
is also modified by the relative contribution of different types of SNe,
which depends on the relation between 
the time-scale of the duration of
the starburst and the
lifetime of SNe progenitors.

The cold gas that will be eventually converted in bulge stars
is referred to as bulge cold gas, in order to differenciate it from the
disc cold gas.
The amount of bulge cold gas available allows 
us to make a first estimation of the mass of 
stars that will be formed in the bulge. Then, we compute the time-scale in
which they would form, which is chosen
to be the dynamical time considered in quiescent SF as a first 
approximation, that is, 
the one defined as the ratio
between the scale radius of the exponential profile that characterizes the 
galactic disc and its circular velocity.
This choice seems reasonable since a burst normally takes place via
a bar embedded in the galaxy disc.
Thus, from the ratio of the bulge cold gas mass 
and the time-scale estimated for the duration of the starburst, we
know the mass of stars that will be formed 
in each time step.
For simplicity, SF throughout the extended starburst process occurs
at a constant rate.
Other works in the literature that have taken into account the
starbursts time-scales consider that the
SFR in a burst decays exponentially with time after 
the burst is triggered \citep{Granato2000, Lacey2008}.
We justify our assumption considering that the bar formed during the 
instability, 
triggered either by a galaxy merger or
a disk instability, favours the gradual feeding of the bulge cold gas 
reservoir by the disc cold gas that is being driven 
to the galactic center 
\citep[e.g.][]{KormendyKennicutt2004}, 
keeping constant the conditions that give rise to SF. 
Note that the presence of a bar during the process of bulge formation
is only implicit in the assumption of constant SFR during starbursts.
Modelling bar formation is not an easy task
\citep[see][]{KormendyKennicutt2004, Athanassoula2013}, 
and is beyond the scope of this
paper. 

\subsection{Variable IMF}
\label{sec:varIMF}

The stellar IMF has a great influence on the chemical 
enrichment of galaxies since 
it defines the amount of stars formed in each stellar mass interval
for each star formation event
and, consequently, impacts on the amount
of metals returned to the ISM, where new stars will eventually
form. The IMF also determines the number of SNe CC involved in the
estimation of the
reheated mass that is transferred from the cold to the hot phase
during
SN feedback.
  Today, 
there is some consensus that in a simple stellar population 
the slope of the IMF above $1\,\Msun$ is not strictly 
different from that of Salpeter IMF ($\alpha \equiv 1+x = 2.35$); 
stars with 
masses over $1\,\Msun$ 
are the major contributors to chemical enrichment
\citep{Portinari98b}. Several attempts have been made in measuring this important 
distribution function in the solar neighbourhood 
\citep{Salpeter55, Chabrier2003} 
or directly in star clusters \citep{Kroupa2001, Kroupa2002}.
The latter is represented by a multi-component power-law IMF given by 
\begin{equation}
\xi(m) = k \left\{\begin{array}{ll}
k'\,\left(\,\frac{m}{m_{\rm H}} \right)^{-\alpha_{0}}&\hspace{-0.25cm},{m_{\rm low}} \le m/{M}_\odot < {m_{\rm H}}\\
\left(\frac{m}{m_{\rm H}} \right)^{-\alpha_{1}}&\hspace{-0.25cm},m_{\rm H} \le m/{M}_\odot < m_0,\\
\left(\frac{m_{0}}{m_{\rm H}} \right)^{-\alpha_{1}}
  \left(\frac{m}{m_{0}} \right)^{-\alpha_{2}}&\hspace{-0.25cm},m_0 \le m/{M}_\odot < m_1,\\ 
\left(\frac{m_{0}}{m_{\rm H}} \right)^{-\alpha_{1}}
    \left(\frac{m_{1}}{m_{0}} \right)^{-\alpha_{2}}
    \left(\frac{m}{m_{1}} \right)^{-\alpha_{3}}&\hspace{-0.25cm},m_1 \le m/{M}_\odot \le m_\mathrm{max},\\ 
\end{array} \right. 
\label{eq:4pow}
\end{equation}
\noindent with $m_{\rm low}=0.01$, $m_{\rm H}=0.08$, $m_{0}=0.5$ and
$m_{1}=1.0$, and exponents $\alpha_0=0.3$, $\alpha_1=1.3$, $\alpha_2=\alpha_3=2.35$ .
The normalization constant $k$ contains 
the desired scaling of the IMF (for example, to 
the total mass of the system). 
Note that  brown dwarfs are a separate population and contribute 
about 1.5 per
cent by mass only.
Therefore, 
they have a different normalisation factor, $k'$
\citep[$k' \sim 1/3$,][]{ThiesKroupa2007, ThiesKroupa2008}.
In the case of stellar 
systems with composite stellar populations, other aspects must  
be considered.

A galaxy-wide IMF may be represented by an 
integrated galactic initial mass function
(IGIMF) proposed by
\citet{Kroupa2003}, whose theory was 
developed in the last decade 
\citep[see][ for a review]{Kroupa2012}. 
This formalism states that star formation takes place exclusively in 
star clusters \citep{LadaLada2003}. 
This way, the stellar populations of each galaxy are 
composed by stars in surviving and
dissolved star clusters. 
Within each star cluster, stars form with the 
stellar IMF given by eq. \ref{eq:4pow}. 
The embedded star 
clusters in the galaxy also follow an initial distribution of masses of the 
form 
$\xi_{\rm ecl}(M_{\rm ecl})\, dM_{\rm ecl}\propto M_{\rm ecl}^{-\beta}\,dM_{\rm ecl}$, where $M_{\rm ecl}$ is the mass
of the embedded star cluster. Moreover, the most massive star in a cluster, 
$m_{\rm max}$, is related in a non-trivial way with the cluster mass 
\citep{Weidner2013b}; the mass of the most massive star is higher
in more massive clusters. Furthermore, observations 
indicate that higher SFRs lead to the
formation of brighter clusters \citep{Larsen2002}. \citet{Weidner2004} find
that this empirical correlation can be transformed into 
a relation between the SFR of 
the galaxy and the maximum embedded cluster mass of the form 
\begin{equation}
M_{\rm ecl}^{\rm max}(SFR)=8.5 \times 10^4 \, SFR^{0.75}\,\Msun.
\label{eq:MeclSFR}
\end{equation}

The IGIMF is the sum of all the new born stars in all of the
star clusters considering the above ingredients , that is,
\begin{equation}
\xi_{\rm IGIMF}(m,t)=\int_{M_{\rm ecl}^{\rm min}}^{M_{\rm ecl}^{\rm max(SFR(t))}}\xi(m\leq m_{\rm max(M_{\rm ecl})})\,\xi_{\rm ecl}(M_{\rm ecl})\,dM_{\rm ecl},
\label{eq:IGIMF}
\end{equation}
where $\xi(m\leq m_{\rm max(M_{\rm ecl})})$ is the 
stellar IMF given by eq. \ref{eq:4pow}, and 
$M_{\rm ecl}^{\rm min}$ 
defines the 
minimum mass of cluster that can be formed in a galaxy, which is
a free parameter of the IGIMF model.  
Now, for this formulation of the IGIMF, it is assumed that star formation
in all the star clusters at all epochs is produced with a canonical IMF. 
However, violent star formation conditions could drive crowding
in massive star clusters \citep{Elmegreen2004, Shadmehri2004}, 
and the formation 
of massive stars could be favoured under such conditions, given 
that the low-mass limit of star formation
would be higher.  
Following preliminary work of \citet{Marks2012},
\citet[][WKP11 hereafter]{Weidner2011} consider that, for 
embedded clusters   
$M_{\rm ecl}<2\times 10^5\,\Msun$, the slope of the 
canonical IMF (eq. \ref{eq:4pow}) 
for stars more massive than $1.3\,\Msun$ 
is $\alpha_3 = 2.35$. 
For larger cluster masses, 
the dependence of the stellar IMF slope
for stars in this mass range is given by
\begin{equation}
\alpha_{3}(M_{\rm ecl})=-1.67 \times {\rm log}_{\rm 10}\left(\frac{M_{\rm ecl}}{10^6\,\Msun}\right)+1.05,
\label{eq:alpha3Ecl}
\end{equation}
The effect of this assumption in the IGIMF is analysed in WKP11.
The high-mass end of the IGIMF is, in 
general, steeper than the stellar IMF since the formation of low mass stars 
is favoured. However, if one includes the effect of crowding in massive 
clusters, as we have seen, the slope of the canonical IMF for massive stars 
can become more top heavy than the Salpeter IMF for these massive clusters. 
When a starburst event with high SFR occurs in a galaxy, the 
formation of massive clusters is favoured and the formation of massive stars 
is highly enhanced, so the IGIMF can become also more top heavy
 than the Salpeter IMF for high mass stars .  

This formulation leads to the top heavy IGIMF (TH-IGIMF; WKP11).
The slope of this TH-IGIMF above $1.3\,\Msun$ can then be computed 
by a least-squares fit to the calculated IGIMF. Hence,
the TH-IGIMF can be translated into a power law of a form similar
to the canonical stellar IMF used to derive the IGIMF ,
but with an exponent $\alpha_{\rm TH}$ for high mass stars, that is,
\begin{equation}
\xi_{\rm TH}(m) = k \left\{\begin{array}{ll}
\left(\frac{m}{m_{\rm H}} \right)^{-\alpha_{1}}&\hspace{-0.25cm},m_{\rm H} \le \frac{m}{{M}_\odot} < m_0,\\
\left(\frac{m_{0}}{m_{\rm H}} \right)^{-\alpha_{1}}
  \left(\frac{m}{m_{0}} \right)^{-\alpha_{2}}&\hspace{-0.25cm},m_0 \le \frac{m}{{M}_\odot} < m_1,\\ 
\left(\frac{m_{0}}{m_{\rm H}} \right)^{-\alpha_{1}}
    \left(\frac{m_{1}}{m_{0}} \right)^{-\alpha_{2}}
    \left(\frac{m}{m_{1}} \right)^{-\alpha_{\rm TH}}&\hspace{-0.25cm},m_1 \le \frac{m}{{M}_\odot} \le m_\mathrm{max}.\\ 
\end{array} \right. 
\label{eq:THIGIMF-FIT}
\end{equation}
where $m_{\rm H} = 0.1$, $m_{\rm 0} = 0.5$ and $m_{\rm 1} = 1.3$. 

After having introduced the basic aspects of the TH-IGIMF theory,
we describe the way in which it is implemented in \sag.
Although the reference to 
``starbursts condition'' in the work of WKP11 implies a high level
of star formation rate in a starburst, 
we use the resulting dependence of the 
$\alpha_{\rm TH}$ with
SFR for any event of star formation in our model, that is, 
without making any distinction
between quiescent star formation mode and starbursts.
For each star formation event, 
 we assign to it the TH-IGIMF given by 
eq. \ref{eq:THIGIMF-FIT}, according to the corresponding SFR.  
The model considers small variations of log(SFR) so that
the slope of the TH-IGIMF is binned in steps of $0.05$.
We adopt $m_{\rm H}=0.1 M_{\odot}$ for the lower limit in the TH-IGIMF
according to our chemical implementation.

We consider 
embedded cluster mass functions characterized by
different 
values of the exponent $\beta$, and
of the minimum embedded 
cluster masses $M_{\rm ecl}^{\rm min}$.
We explore cases 
with  
(i) $\beta=2$ and $M_{\rm ecl}^{\rm min}=5 \, \Msun$,
(ii)  $\beta=2.1$ and $M_{\rm ecl}^{\rm min}=5 \, \Msun$, and
(iii) $\beta=2$ and $M_{\rm ecl}^{\rm min}=100 \, \Msun$.
It is worth noting that some stellar associations 
with masses
as low as
$5\,\Msun$ 
are found in the 
galaxy \citep[e.g. Taurus-Auriga star-forming regions,][]{Kirk2011} , 
thus favouring the lowest
value of $M_{\rm ecl}^{\rm min}$ considered here.  
Therefore, at least in quiescent star forming conditions, the 
$M_{\rm ecl}^{\rm min}$ is supported 
observationally. However, this constraint could not be valid in more 
extreme star-forming conditions. 

\begin{figure}
  \includegraphics[width=0.49\textwidth]{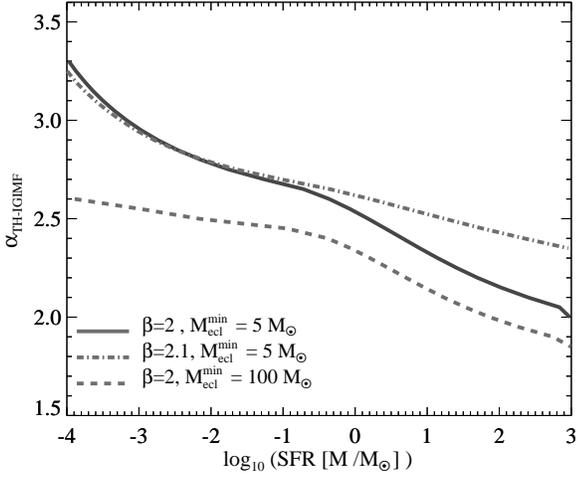}
 \vspace{1pt}
 \caption{Dependence of the
slope of the TH-IGIMF ($\alpha_{\rm TH}$) on the SFR
for different minimum embedded cluster masses and slopes of the
embedded star cluster mass function:
(i) $\beta=2$, $M_{\rm ecl}^{\rm min}=5 \Msun$ (solid line), 
(ii) $\beta=2.1$, $M_{\rm ecl}^{\rm min}=5 \Msun$ (dashed-dotted line),
 and
(iii) $\beta=2$, $M_{\rm ecl}^{\rm min}=100 \Msun$ (dashed line).
}
\label{fig:alphaTHIGIMF}
\end{figure}

Fig. \ref{fig:alphaTHIGIMF} shows the 
slope  
$\alpha_{\rm TH}$ 
as a function of the SFR 
for different 
values of $M_{\rm ecl}^{\rm min}$ and $\beta$.
It is clear from this plot that the slope $\alpha_{\rm TH}$
of the TH-IGIMF is larger than the slope of the Salpeter IMF
(characterized by a fixed value of $\alpha$ for all SFR) 
in lower star formation 
regimes.
The values of SFR at which the TH-IGIMF becomes more top heavy
than Salpeter depend on the value of $\beta$ and $M_{\rm ecl}^{\rm min}$,
being of the order of  
$1,\,10$ and $1000\ \Msun\,{\rm yr^{-1}}$ for 
$\beta=2$ and $M_{\rm ecl}^{\rm min}=100 \,\Msun$,  
$\beta=2$ and $M_{\rm ecl}^{\rm min}=5 \,\Msun$,  
and $\beta=2.1$ and $M_{\rm ecl}^{\rm min}=5 \,\Msun$,  
respectively. 

During periods of high star formation activity,
the formation of massive stars is favoured.
For higher SFRs, the
number of high mass stars in a TH-IGIMF becomes larger
than in the Salpeter IMF,
with the corresponding reduction of the number of low mass stars
required by the IMF normalization. 
This aspect is crucial to understand the build-up of the 
$[\alpha/{\rm Fe}]$-stellar mass relation, as we will see in Subsection 
\ref{sec:AlphaFeTHIGIMF}.     

\section{Calibration of the model and results for Salpeter IMF}
\label{sec:CalibrationSAG}

The semi-analytic modelling involves the tunning of the free
parameters through their 
calibration against a set of
observations.
For this purpose, we use the Particle Swarm
Optimization technique (Kennedy \& Eberhart 1995) applied to our
semi-analytic model \sag. The detailed description of the method
is given in \citet{Ruiz2014}. This procedure involves the variation
of a selected set of free parameters of the model within a given
range. 
With this novel method, we can explore
a large parameter space and find the best global maximum of the
likelihood surface.
The free parameters that are tuned are associated to the
modelling of the main physical processes included in the code,
summarized in Appendix \ref{App:AppendixA}. They are the
efficiency of SF ($\alpha$), the SNe feedback efficiency associated
to the SF taking place in the disc ($\epsilon$) 
and in the bulge ($\epsilon_{\rm bulge}$), 
the parameters related to the central supermassive black hole 
growth ($f_{\rm BH}$)
and AGN feedback efficiency ($\kappa_{\rm AGN}$),
the parameter involved in the triggering of disc instabilities 
($f_{\rm pert}$), and the fraction of binary stars ($A_{\rm bin}$).
 
\subsection{Observational constraints}
\label{sec:obs_const}

We evaluate the behaviour of the model considering two sets
of observational constraints to calibrate the free parameters. 
The first set 
involves 
the $z=0$ luminosity function in
the {\em r}-band ({\em r}-band LF), 
the relation between the mass of the central supermassive black hole
and the bulge mass (BHB relation), and 
the redshift evolution of SNe Ia and SN CC rates. The second set
also includes the 
$[\alpha / {\rm Fe}]$-stellar mass relation of elliptical galaxies.

The LF is a fundamental constraint for semi-analytic models of 
galaxy formation, since 
it reflects the influence of SF and feedback processes.
The low mass end of the LF is mainly associated to SNe feedback. 
This aspect is 
relevant to our work, since the different IMFs contribute with different 
numbers of SNe CC, being low (high) for IMFs with large (small) 
values of their slopes, and 
the feedback efficiency parameters,
$\epsilon$ and $\epsilon_{\rm bulge}$ must be properly tuned 
in order to produce results that do not underestimate
(overestimate) the effect of SN feedback.  
On the other hand, the high knee and the
drop in the high mass end of the LF of galaxies is mainly determined by 
AGN feedback.
This process is thought to be responsible for quenching the SF in
massive galaxies, contributing to the downsizing behaviour of 
galaxies. 
AGN feedback is also constrained through the BHB relation, since
the central supermassive black hole is 
responsible for this process.
Constraining 
the mass of bulges in galaxies through the BHB relation helps to
obtain an adequate mixture of galaxy morphologies. 

The evolution of the rates of SNe Ia and CC are particularly 
important in this study,
since the number of different types of SNe and their corresponding yields
have direct impact on the 
the $[\alpha / {\rm Fe}]$ abundance ratio of stars.
The evolution of SNe CC rate not only determines the 
amount of $\alpha$-elements produced along the galaxy lifetime,
but constitutes an alternative diagnostic of the evolution of
the SFR given the short lifetime of the progenitors of SNe CC.
Reproducing the observed SFR is a required condition for
any of our models, but is particularly relevant for those
involving a TH-IGIMF because of the
dependence of its slope on the SFR. In turn, the slope of the TH-IGIMF
determines the number of SNe CC and,
consequently, the SNe feedback that contributes to 
regulate the star formation process.  
One of the calibration parameters, $A_{{\rm bin}}$, 
determines the amount of 
binary systems that explode as SNe Ia, and therefore, 
the amount of iron recycled
into the ISM that will be available during the
formation of future generations of stars. 
Thus, it is crucial to constrain its value to avoid making wrong
conclusions with respect to the $[\alpha / {\rm Fe}]$ abundance ratio
in stars. We will resume the analysis of the SNe 
time-scales and 
the impact of the parameter $A_{{\rm bin}}$ on the delay time-scales of SNe Ia
when analysing the model with
TH-IGIMF in Subsection \ref{sec:SNeDelayTimes}.

It is important to
note that \sag~makes use of the luminosities of single stellar populations 
obtained 
with stellar population synthesis models,
for which we use models by \citet{Bruzual2003} and \citet{Bruzual2007}.
Since these models have an explicit dependence
on the adopted IMF, they were generated 
for all the possible slopes of the TH-IGIMF of interest in this work.
The unpublished 2007 models include the final prescription of \citet{Marigo2007}
and \citet{Marigo2008}
for the evolution of thermally
pulsating-asymptotic giant branch (TP-AGB) stars. 
This produces significantly redder 
near-IR colors, and hence younger ages and lower masses, 
for young and intermediate-age 
stellar populations, than their 2003 counterpart; see 
\citet{Bruzual2007}  
for details. 

\begin{table}
\caption{Values of the free parameters of models with Salpeter IMF: {\small SAGS-c1} and {\small SAGS-c2}. {\small SAGS-c1} is calibrated using the LFs in the
{\rm r}-band, the BHB-Mass relation and SNae rates as a function of redshift 
as observational constraints. For model {\small SAGS-c2} we add the 
$[\alpha/{\rm Fe}]$-mass relation to the set of constraints. }
     
\centering
\begin{tabular}{ c  c  c  } 
\hline
\hline
 & ${\rm Salpeter\,\, IMF}$ & ${\rm Salpeter\,\, IMF}$ \\ [.5mm]
\hline
Param & {\small SAGS-c1} & {\small SAGS-c2} \\  
\hline
$\alpha_{\rm SF}$ & 0.1271 & 0.2328  \\ 
$\epsilon$ & 0.2263 & 0.2541  \\
$\epsilon_{{\rm bulge}}$ & 0.0513 & 0.0778  \\
$frac_{{\rm BH}}$ & 0.0139 & 0.0141 \\
$k_{{\rm AGN}}$ & 2.32$\times 10^{-5}$ & 4.34$\times 10^{-4}$ \\
$A_{{\rm bin}}$ &  0.0341 & 0.0383 \\ 
$f_{{\rm pert}}$ & 44.62 & 47.63 \\
\hline
\hline
\end{tabular}
\label{table:table1} 
\end{table}

\begin{figure}
\centering
\includegraphics[scale=0.4]{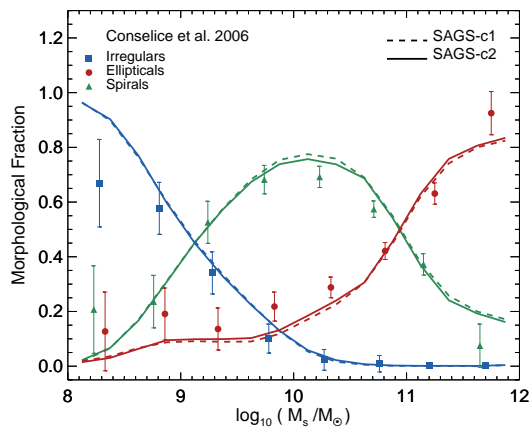}
\caption{Fraction of galaxies of different morphological types
as a function of the stellar mass for models {\small SAGS-c1} 
(dashed line)
and {\small SAGS-c2} (solid line) compared
with the observed morphological mixture in local
galaxies \citep{Conselice2006}.
Irregular, spiral and elliptical galaxies in the model are
represented by different colours, while the
corresponding fractions inferred from observations are represented by
triangles, rhombi and circles with the same colour code,
as indicated in the key.
Elliptical galaxies in the model are those
satisfying the condition ${\rm B/T}>({\rm B/T})_{\rm thresh} \equiv 0.8$.
}
\label{fig:morph}
\end{figure}

\begin{figure}
  \includegraphics[width=0.49\textwidth]{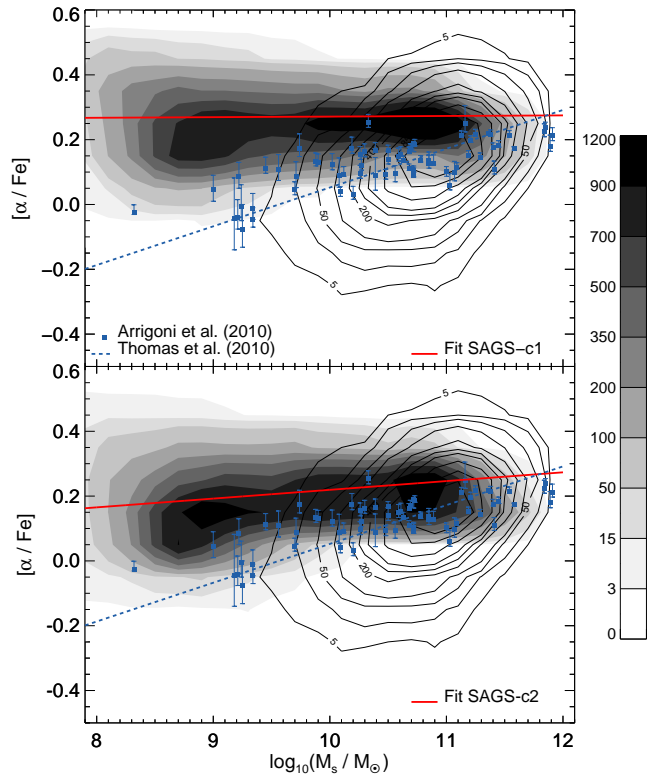}
 \vspace{1pt}
 \caption{[$\alpha/{\rm Fe}$]-stellar mass relation 
for elliptical galaxies in versions of \sag~with Salpeter IMF: 
{\small SAGS-c1}
(top panel)  and {\small SAGS-c2} (bottom panel).
The distributions of abundance ratios are
represented by grey density contours; the red solid
lines are the corresponding linear fits.
Blue line contours depict the relation traced by the T10
sample of elliptical galaxies; the fit to the mean values of their
[$\alpha/{\rm Fe}$] abundance ratios for bins of stellar mass is
represented by  
the blue dashed line.
Blue symbols correspond to data from \citet{Arrigoni2010}.
}
\label{fig:alphaSAGS-c1SAG1}
\end{figure}

The observational data used
for both sets of constraints are
the {\em r}-band LF of \citet{Blanton2005}, the BHB relation given by 
\citet{HaringRix2004} and \citet{Sani2010},
and the compilation of 
rates for both SNe Ia and SNe CC given by
\citet{Melinder2012}. 
In the second set of constraints, we also
consider the $[\alpha/{\rm Fe}]$ ratio of elliptical galaxies
presented in \citet[][ T10 hereafter]{Thomas2010}.
They analyse a subsample of 3360 galaxies out of 16502 early-type 
galaxies 
drawn from the Sloan Digital Sky Survey (SDSS) Data Release 4 
\citep[ DR4]{Adelman-McCarthy2006}, 
and morphologically selected 
by visual inspection.
The subsample is selected in the redshift range $0.05<z<0.06$ 
to avoid biases in 
velocity dispersion.
They obtain the $[\alpha/{\rm Fe}]$ 
abundance ratios for each galaxy in their sample 
by means of observed Lick spectral indices and stellar population synthesis
models that take into account element abundance ratio effects. 
They also compute the dynamical mass of each galaxy from the 
line-of-sight stellar velocity dispersion
using the scaling relation 
presented in \citet{Cappellari2006}. This dynamical mass
is considered a good proxy of the baryonic mass of elliptical galaxies, 
so we can compare
the $[\alpha/{\rm Fe}]$-stellar mass relation obtained from our
model with the $[\alpha/{\rm Fe}]$-dynamical mass relation presented by T10.
The average errors of the observed $[\alpha/{\rm Fe}]$
values are of the order of $0.06$~dex.
We also consider the observational data set from 
\citet{Arrigoni2010}, which was obtained by 
\citet{Trager2000} and re-analysed with the stellar population synthesis model 
described in \citet{Trager2008}.

\subsection{$[\alpha/{\rm Fe}]$-mass relation for a Salpeter IMF}
\label{sec:AlphaSalp}
The aim of this paper is to identify the main aspects involved
in the build-up of the [$\alpha/{\rm Fe}$]-stellar mass relation
of elliptical galaxies.
Our semi-analytic model provides the information needed to construct
this relation. We classify a galaxy as early-type if the galactic bulge 
represents more than 80 per cent of the total baryonic mass of the galaxy, 
that is, we adopt $({\rm B/T})_{\rm thresh}=0.8.$ 
(see Subsection \ref{sec:ellip_form}).
With this criterion, we estimate the fraction of early-type and
late type galaxies. Bulgeless galaxies are considered as irregulars.

Once we have selected the population of early-type galaxies in the
model, we estimate the abundance ratio [$\alpha/{\rm  Fe}$] of their stellar
component. 
Following \citet{Geisler2007},
we define the $[\alpha/{\rm Fe}]$ abundance ratio as
\begin{equation}
[\alpha/{\rm Fe]}=\frac{1}{6}\,\sum_{i=1}^6{[\alpha/{\rm Fe}]_i},
\label{eq:alphafe}
\end{equation}
where $[\alpha/{\rm Fe}]_i$ denotes 
the logarithm of the abundance ratio with respect to iron of each of the
six $\alpha$-elements available 
in our chemical implementation
($^{16}$O, $^{20}$Ne, $^{24}$Mg, $^{28}$Si, $^{32}$S, 
$^{40}$Ca).
For the sake of comparison with the observational data considered, 
we refer the abundance ratios of galaxies 
in the model to the solar 
abundances of \citet{Grevesse1996} as in T10 and \citet{Arrigoni2010}. 

We consider model \sag~with Salpeter IMF and calibrate it using the
two sets of observational constraints 
described in Section \ref{sec:obs_const}
giving place to two variants of the same model, referred to 
as {\small SAGS-c1}
and {\small SAGS-c2}. The inclusion of the
$[\alpha/{\rm Fe}]$-stellar mass relation as an additional
constraint in the second set
imposes more restrictions to the free parameters 
and their
values change with respect to those obtained from the first set
of constraints. The values of the parameters obtained for each calibration
are presented in Table \ref{table:table1}.

The fraction of galaxies
with different morphological types for model {\small SAGS-c1} 
and {\small SAGS-c2} are presented in Figure \ref{fig:morph}.
In both cases, they are 
in concordance with the observed fraction in the local Universe,
for which we use data from
\citet{Conselice2006}. The 
[$\alpha/{\rm Fe}$]-stellar mass relation 
for the population of elliptical galaxies 
is shown in Figure \ref{fig:alphaSAGS-c1SAG1}
for both {\small SAGS-c1} (top panel)
and {\small SAGS-c2} (bottom panel).
Model results are compared with
the $[\alpha/{\rm Fe}]$-dynamical mass relation 
of elliptical galaxies given by T10 which is
represented by blue line contours. The linear fit to this galaxy
population is shown by the blue dashed line.
The extrapolation of this fit to lower masses is in agreement with the
trend denoted by data from \citet{Arrigoni2010}, represented by blue squares.

We find that
using only the {\em r}-band LF, the BHB relation
and the evolution of the SN rates as constraints, the model 
is not able to reproduce the observed 
[$\alpha/{\rm Fe}$]-stellar mass relation.
In the galaxy population generated by {\small SAGS-c1},
less massive galaxies reach practically the same
[$\alpha/{\rm Fe}$] abundance ratios than more massive ones, 
giving rise to an [$\alpha/{\rm Fe}$]-stellar mass relation
with a flat trend, contrary to the observed one. 
This is reflected in the linear fit 
\footnote{We use the IDL routine robust\_linefit.pro\\ 
(http://idlastro.gsfc.nasa.gov/ftp/pro/robust/robust\_linefit.pro).}
to the 
distribution of the whole
early type galaxy population represented by the red line.
Grey coded contours depict the number of model
elliptical galaxies. 

Impossing 
the observed [$\alpha/{\rm Fe}$]-stellar mass relation 
as an additional constraint allows to
properly restrict the value of the SN feedback efficiency associated to
the stars in the bulge ($\epsilon_{\rm bulge}$), 
and a positive correlation between [$\alpha/{\rm Fe}$] abundances
and stellar mass is obtained
from the model {\small SAGS-c2}, 
in better agreement with the
observational trend. 
The larger value of $\epsilon_{\rm bulge}$ obtained from the second set
of constraints favours the ejection of bulge cold gas in less massive galaxies.
This gas reservoir may be later replenished for  
further star formation in the bulge with cold gas contaminated 
with higher iron abundances produced by the delayed contribution of SNe Ia. 
This change in  $\epsilon_{\rm bulge}$ is complemented with a 
higher SF efficiency, which in turn demands a higher 
AGN feedback efficiency ($k_{{\rm AGN}}$) in order to achieve the right
luminosities of high mass galaxies. 

Although the slope of the [$\alpha/{\rm Fe}$]-stellar mass relation for
model {\small SAGS-c2} is positive (a=0.025) it is still almost flat;
the slope of the fit to T10 data is $a=0.1184$.
This result highlights the difficulty of semi-analytic models 
to reproduce adequately this observed
relation when using an universal IMF.
Several groups have attempted to reproduce the 
slope of the [$\alpha/{\rm Fe}$]-stellar mass relation
using SAMs with a universal IMF during the last decade. Among the most recent
ones, \citet{Arrigoni2010}
find that with a top-heavier slope of the IMF, they can achieve a positive
slope in the [$\alpha/{\rm Fe}$]-stellar mass relation, close to the observed 
one. 
On the other hand, \citet{Calura2009} use another SAM and show that simply 
changing the slope of the IMF seems to be insufficient to obtain the right 
slope in the observed relation. They 
argue that the assumption of a constant IMF flatter than Salpeter only lift
the zero-point of the relation. We report here that when we change the slope 
of the IMF in \sag~, we also have as a result a variation in the zero point, with
no apreciable change in the slope, supporting the last conclusion. 
 \citet{Calura2011} include the effects of starbursts due to fly-by interactions
at high redshift 
following the model developed by \citet{CavaliereVittorini2000} and
studied by \citet{Menci2004} in their SAM.
They find that the slope of the [$\alpha/{\rm Fe}$]-stellar mass relation 
can be reproduced naturally when adding this effect, although the slope they
obtain is shallower than expected from observations (see their Figure 1).  
The effect of starbursts 
triggered by companions is present in our model 
\sag~through our particular implementation of disk instabilities 
(see Section \ref{sec:ellip_form}).
Forcing the free parameters of \sag~to
reproduce the [$\alpha/{\rm Fe}$]-stellar mass relation  
with a Salpeter IMF gives a similar slope than the best model of 
\citet{Calura2011}, although in both cases the slope is too shallow compared 
with observations. 
Even constraining the parameters with the relation itself,
it is not possible to 
achieve a satisfactory slope.

This procedure has shown the necessity of including the
[$\alpha/{\rm Fe}$]-stellar mass relation as a constraint.
From now on, we will calibrate the versions of \sag~with TH-IGIMF
using the second set of constraints, and compare their results with those
from model {\small SAGS-c2}, to which we will simply refer to as {\small SAGS}.

\section{[$\alpha/{\rm Fe}$]-mass relation: Impact of a TH-IGIMF}
\label{sec:AlphaFeTHIGIMF}

The main aim of this work is to evaluate the impact of the 
variable TH-IGIMF on the development of the 
[$\alpha/{\rm Fe}$]-mass relation in elliptical galaxies, 
motivated by the fact that a universal IMF is not able
to reproduce the right slope of this relation 
even when the relation itself is used as a constraint.
From the implementation of the TH-IGIMF in the updated version of
\sag, we have the possibility to evaluate the impact of
considering a variation of two different free parameters involved in the TH-IGIMF theory, 
that is, the minimun mass of the embedded star cluster and the slope of the
embedded star cluster mass function (see section \ref{sec:varIMF}). Thus, we
have three new different versions of \sag~corresponding to
i) $M_{\rm ecl}^{\rm min} = 5 \Msun$, $\beta = 2$, 
ii) $M_{\rm ecl}^{\rm min} = 5 \Msun$ ,$\beta = 2.1$  
and iii) $M_{\rm ecl}^{\rm min} = 100 \Msun$, $\beta = 2$, to which we refer to
as {\small SAGTH5B2}, {\small SAGTH5B21} and {\small SAGTH100B2}, 
respectively.

\begin{figure*}
\includegraphics[width=1.\textwidth]{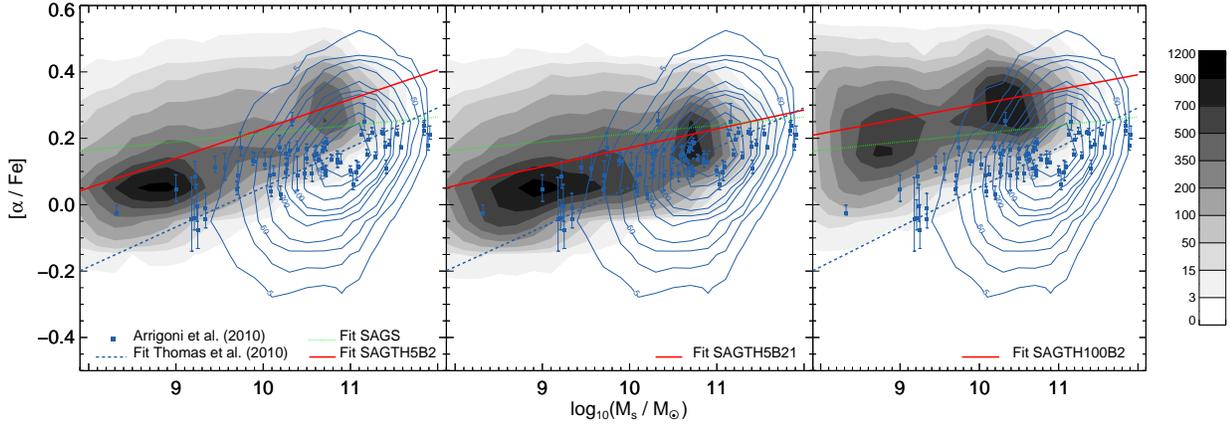}
\vspace{10mm}
\caption{ 
[$\alpha/{\rm Fe}$]-stellar mass relation in models {\small SAGSTH5B2}
(left panel), {\small SAGSTH5B21} (middle panel) and {\small SAGSTH100B21}
(right panel)
represented by grey density contours; the red solid
lines are the fit to the
distributions of abundance ratios.
Results from all models are compared with the fit 
to the [$\alpha/{\rm Fe}$]-stellar mass
distribution of model {\small SAGS}, represented by a green dotted line.
Blue line contours depict the relation traced by the T10
sample of elliptical galaxies; the fit to the observed distribution of 
[$\alpha/{\rm Fe}$] abundance ratios vs. mass is
represented by  
the blue dashed line.
Blue symbols correspond to data from \citet{Arrigoni2010}.
}
\label{fig:AlphaFeTHIGIMF}
\end{figure*}

\subsection[]{Impact of a TH-IGIMF}
\label{sec:impactVarIMF}

\begin{figure*}
\resizebox{0.33\hsize}{!}{\includegraphics{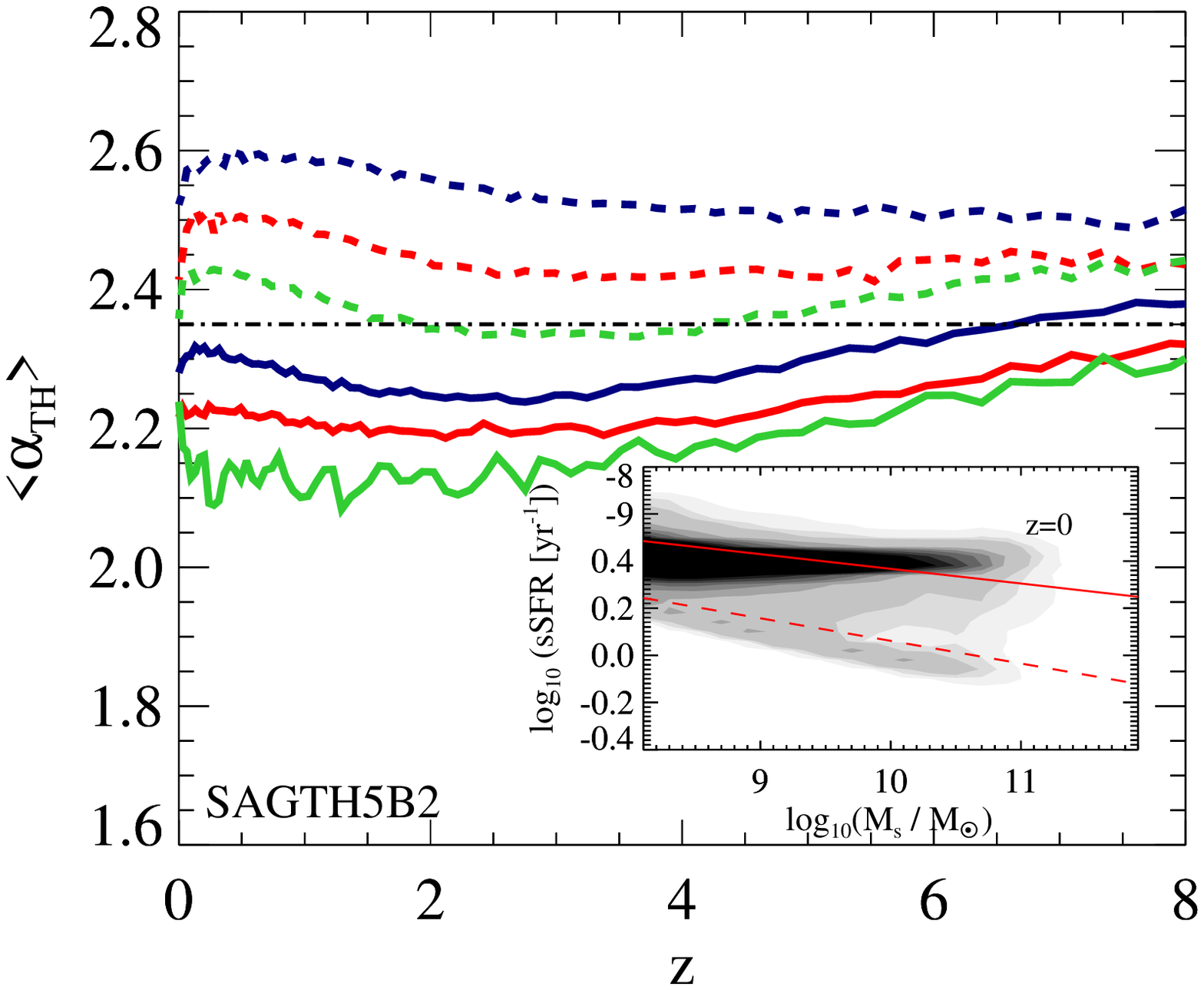}}
\hfill
\resizebox{0.33\hsize}{!}{\includegraphics{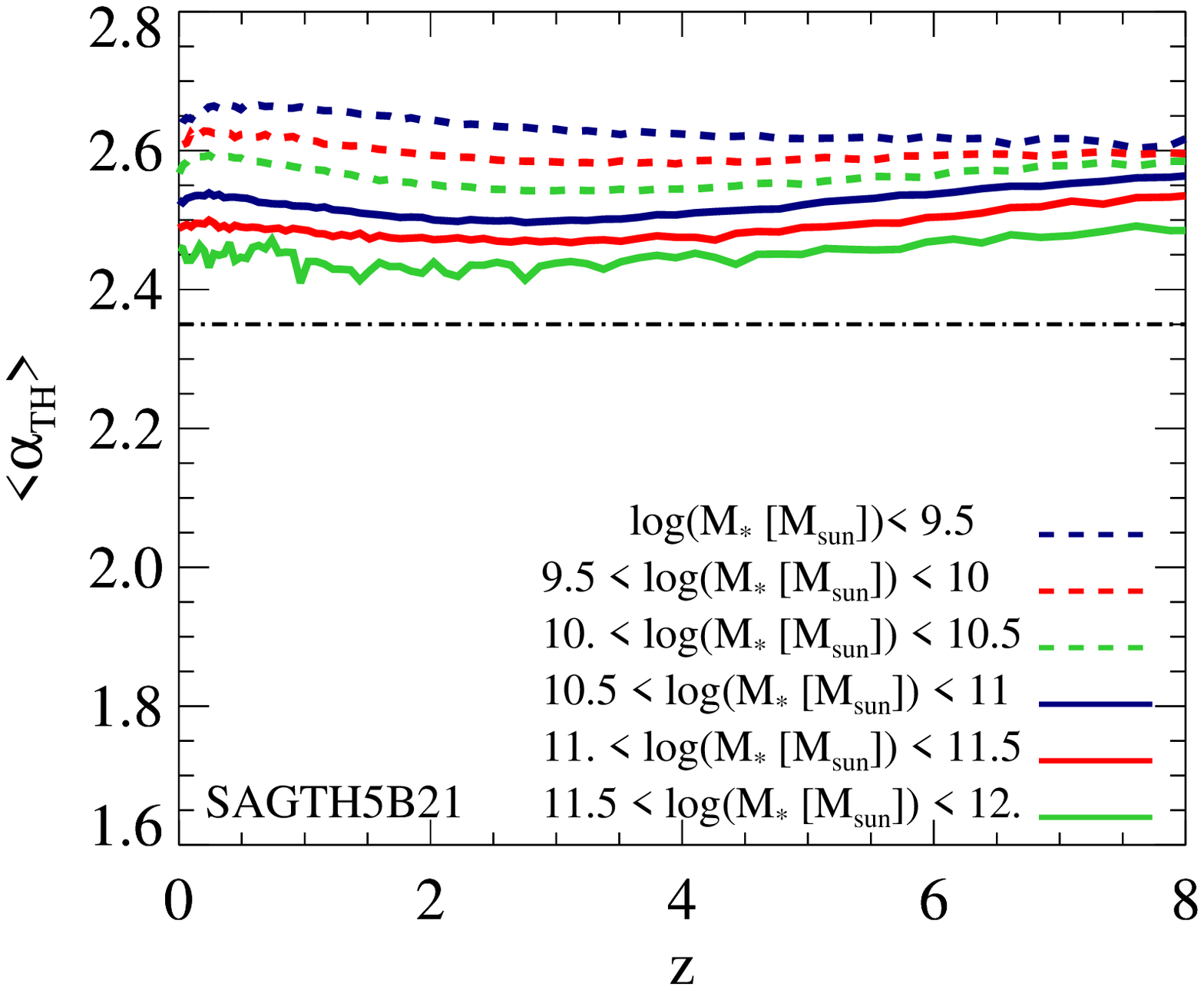}}
\hfill
\resizebox{0.33\hsize}{!}{\includegraphics{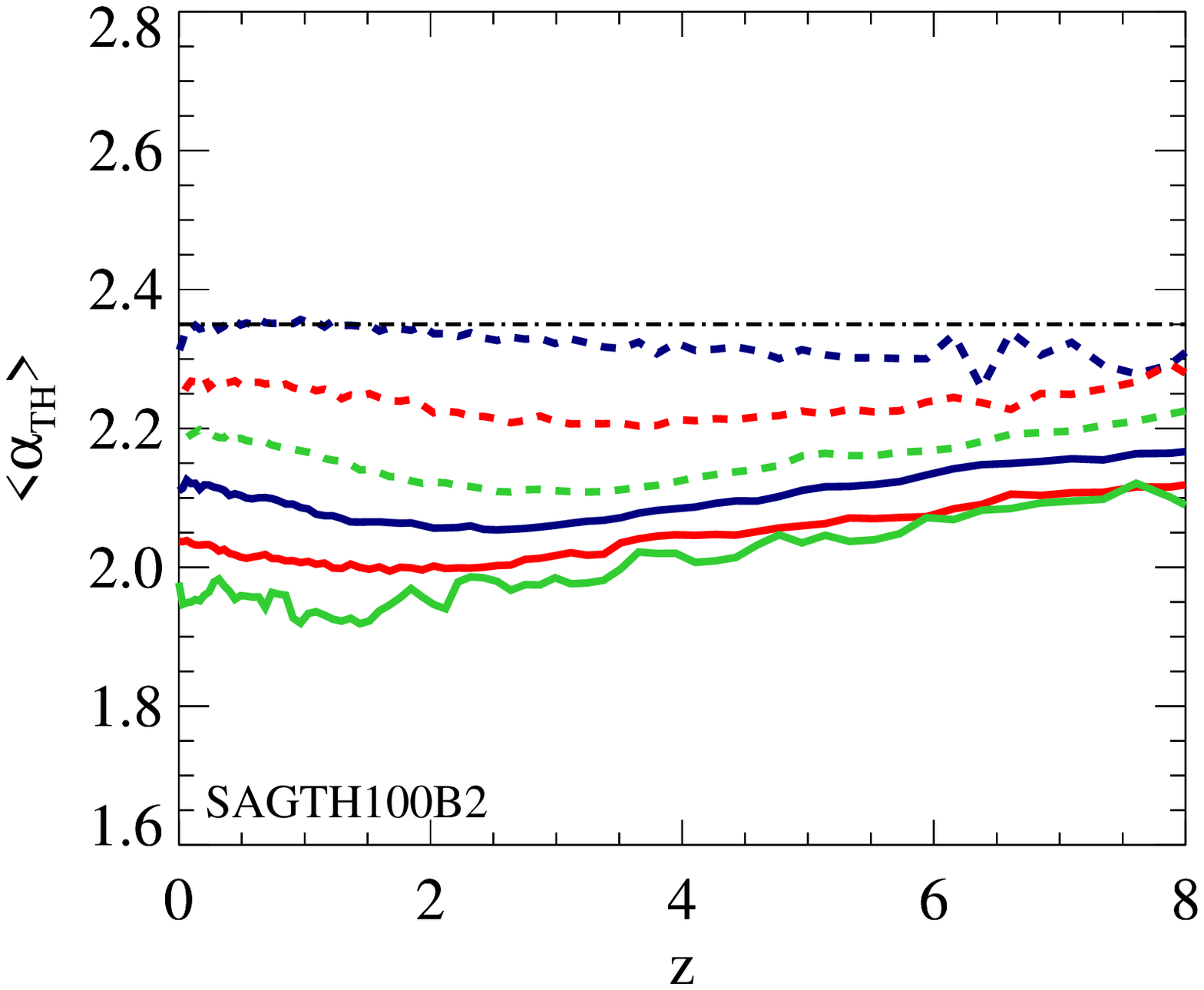}}
\hfill
\caption{Evolution of the mean slope of the TH-IGIMF ($\langle \alpha_{\rm TH} \rangle$,
see eq. \ref{eq:THIGIMF-FIT})
weighted with the stellar mass formed
in each event, which is characterized by a given SFR,
for galaxies within different
mass ranges, as indicated in the key; 
galaxies in each mass range are selected according to
their stellar mass at $z=0$.  
Results are shown for models {\small SAGTH5B2}
(left-hand panel), {\small SAGTH5B21} (middle panel) and {\small SAGTH100B2}
(right-hand panel), characterized by the parameters
$M_{\rm ecl}^{\rm min}=5 \Msun$ and $\beta=2$, 
$M_{\rm ecl}^{\rm min}=5 \Msun$ and 
$\beta=2.1$, and $M_{\rm ecl}^{\rm min}=100 \Msun$ and $\beta =2$, respectively.
The horizontal dashed-dotted line in each panel 
represents the value of the slope of the
Salpeter IMF.
The inset in the left-hand panel contains the distribution of
the specific star formation rate (sSFR) at $z=0$ as a funcion of stellar mass
(grey shaded contours).
The solid and dashed lines represent the linear fit to the
sSFR distribution for active and passive 
galaxies, respectively; the limit between the two populations satisfy the 
condition defined by \citet{Kimm11} (see their eq. 8).  
}
\label{fig:MeanSlope}
\end{figure*}

Testing new recipes to represent physical processes using a
fixed set of free parameters is a common practice in semi-analytic
modelling of galaxy 
formation. 
Generally,
the preferred values of these parameters are kept for a modified
version of the model in which some physical aspects are changed
in order to evaluate the impact of the modifications introduced on
the resulting galaxy properties. 
This is acceptable as long as the
new version gives results in good agreement with the selected 
observational constraints.
However, depending on the modifications introduced, this
procedure might not be maintained.

Since the IMF has a strong impact on the results, each of the models
with TH-IGIMF must be calibrated.
Table \ref{table:table2} shows the values of the parameters adopted in the
new calibrations, 
which include the [$\alpha/{\rm Fe}$]-stellar mass relation as a constraint.

\begin{table}
\caption{Calibration parameters for models {\small SAGTH5B2}, 
{\small SAGTH5B21} and {\small SAGTH100B2} which involve a TH-IGIMF
characterized by a particular combination of
the minimum embedded star cluster mass
and the slope of the
embedded star cluster mass function, that is,  
i) $M_{\rm ecl}^{\rm min}=5 \Msun$, $\beta=2$, 
ii) $M_{\rm ecl}^{\rm min}=5 \Msun$, $\beta=2.1$ 
and iii) $M_{\rm ecl}^{\rm min}=100 \Msun$, $\beta =2$, respectively.
} 
\centering
\begin{tabular}{ c c c c } 

\hline
\hline
& &TH-IGIMF &\\
\hline
Param & {\small SAGTH5B2}  & {\small SAGTH5B21} & {\small SAGTH100B2} \\ [.5mm] 
\hline
$\alpha_{\rm SF}$  & 0.0312 & 0.0338 & 0.0619 \\ 
$\epsilon$  & 0.3613 & 0.3532 & 0.2351   \\
$\epsilon_{{\rm bulge}}$  & 0.3163  & 0.413 & 0.0479   \\
$frac_{{\rm BH}}$  & 0.0516 & 0.0343 & 0.0765   \\
$k_{{\rm AGN}}$  & 7.92$\times 10^{-5}$ & 1.11$\times 10^{-4}$ & 9.09$\times 10^{-5}$   \\
$A_{{\rm bin}}$ & 0.0386 & 0.04 & 0.0415  \\ 
$f_{{\rm pert}}$  & 39.96 & 48.36 & 34.08 \\
\hline
\hline
\end{tabular}
\label{table:table2} 
\end{table}

Fig. \ref{fig:AlphaFeTHIGIMF} shows the results
of the abundance ratios [$\alpha/{\rm Fe}$] of elliptical galaxies
as a function of their stellar mass for models 
{\small SAGTH5B2}, {\small SAGTH5B21} and {\small SAGTH100B2}
(left, middle and right panels, respectively); the corresponding 
galaxy distributions 
are shown in grey density contours.
As mentioned in Subsection \ref{sec:AlphaSalp},  
elliptical galaxies in the model are those 
satisfying the condition ${\rm B/T}>({\rm B/T})_{\rm thresh} \equiv 0.8$. 
With this criterion, the fraction of elliptical
galaxies agree very well with the observed distribution in the
three versions of the model after being calibrated.
The [$\alpha/{\rm Fe}$]-stellar mass distributions from our models
are compared with   
the $[\alpha/{\rm Fe}]$-dynamical mass relation presented by T10
(blue line contours) and data from
\citet{Arrigoni2010} (blue squares).
The blue dashed line denotes the linear fit to the 
 $[\alpha/{\rm Fe}]$-dynamical mass relation of T10,
while the red solid line represents the linear fit to the 
distribution of model galaxies.

It is important to point out that the comparison of
abundances ratios obtained from models that include a TH-IGIMF
with the observed data considered here is not strictly fair.
The final observed abundance patterns are obtained 
by comparing 
the observed spectral features of galaxies 
with synthetic spectra from stellar population synthesis models
(SPSMs); 
these models assume
a fixed stellar IMF that is not SFR dependent.
The SPSMs models used by T10 are those of \citet{Thomas2003},
who assume a Salpeter IMF to compute the parameters of the 
stellar populations. \citet{Arrigoni2010} use recomputed abundances from 
\citet{Trager2008}, who also use the SPSMs corresponding to the Salpeter IMF.
Moreover,
the observed abundance ratios are computed using integrated spectra 
of galaxies in a specific wavelenght range. A fair comparison with model results  
should include luminosity weighted abundances in the corresponding bandwidth
from our model. However, \citet{Thomas1999} investigate the difference between
 luminosity and mass weighted abundances.  They found that the 
difference is negligible in the case of a fixed IMF. Later on, 
\citet{Recchi+2009} 
found that using an IGIMF, the difference between mass weighted and luminosity 
weighted abundances remain of the order of ~0.1 dex. In light of these 
results, we compare the observed data directly with the abundances of \sag,
that take into account the total mass of chemical elements in stars
. 

Keeping these inconsistencies in mind, we now analyse the features 
of the [$\alpha/{\rm Fe}$]-stellar mass relation that emerge
from our semi-analytic model for different values of $M_{\rm ecl}^{\rm min}$ and
$\beta$. Considering the plots in Fig. \ref{fig:AlphaFeTHIGIMF},
we see that the TH-IGIMF produces a positive change in the trend
of [$\alpha/{\rm Fe}$] abundance ratios with stellar mass (grey shaded contours
and fits represented by red solid lines) with
respect to the one obtained for {\small SAGS} which considers a Salpeter
IMF (green dotted line).
The SFR dependent IMF helps to recover the slope denoted by 
observational data. 
Although this is the case for all
the values considered for the minimum embeded star cluster mass and
slope of the embedded cluster mass function,
the agreement with the observational trend is particularly good for 
$M_{\rm ecl}^{\rm min} = 5\,\Msun$ and $\beta=2$ (right panel).
The slope of the fit to the [$\alpha/{\rm Fe}$]
abundance ratios given by {\small SAGTH5B2} ($a=0.088$)
is steeper than those obtained from models {\small SAGTH5B21} and  
{\small SAGTH100B2}
($a=0.0568$ and $a=0.044$, respectively). The former 
is closer to the slope of the fit
to T10 data ($a=0.1184$)
which was obtained taking into account all the galaxies in the sample;
the value of $0.1$ given by T10 corresponds to their subsample of red 
galaxies. All the combinations of parameters of the TH-IGIMF yield 
slopes of the fit to the [$\alpha/{\rm Fe}$] abundance ratios closer to the
 observed one, than {\small SAGS}, that yields a slope of $a=0.025$.  
However, if we consider median 
values, corresponding to the darker shaded contours, the model shows
an excess of [$\alpha/{\rm Fe}$] for all galaxy masses. This is 
related with a lack of {\rm Fe} in the galaxies in all of our models.
The modification of the SNIa delay times 
in the chemical implementation,
as done by \citet{Yates2013},
could improve this particular issue.  
We defer this analysis to a future work.

Considering that $M_{\rm ecl}^{\rm min}$ and $\beta$ are free parameters
of the TH-IGIMF theory,
the results that emerge from the analysis of Fig. \ref{fig:AlphaFeTHIGIMF}
support a value as small as
$5\,\Msun$ and 2 for these quantities.
As we mentioned in Subsection \ref{sec:varIMF}, there is evidence of
star forming clouds with masses down to $5\,\Msun$ and the slope of the
embedded star cluster mass function is observed to be close to 2 
\citep{LadaLada2003}.
However, the value of this minimum mass
might change with redshift, shifting to higher values due to more
violent star forming conditions.  \citet{Weidner2013a} also find evidence
that the power-law index $\beta$ of the mass function of the embedded
star cluster decreases with increasing SFR. However, up to now there is
no observational constraint or physical model that help to determine a 
preferred variation in elliptical galaxies.   
In order to understand the way in which a TH-IGIMF operates helping to 
achieve the correct [$\alpha/{\rm Fe}$] abundance ratios for
galaxies of different masses, we analyse the evolution of
the mean slope of the TH-IGIMF ($\langle \alpha_{\rm TH} \rangle$,
see eq. \ref{eq:THIGIMF-FIT}) 
weighted with the stellar mass formed
in each event, which is characterized by a given SFR,
for galaxies within different
mass ranges; galaxies in each mass range are selected according to 
their stellar mass at $z=0$. This is shown in Fig. \ref{fig:MeanSlope}
for models {\small SAGTH5B2}, {\small SAGTH5B21} and {\small SAGTH100B2}
(left-hand, middle and right-hand panels, respectively). 
The general evolutionary trend of  $\langle \alpha_{\rm TH} \rangle$
is very similar for all models, 
but the values of 
$\langle \alpha_{\rm TH} \rangle$ for a given $z=0$ stellar mass range 
are lower for
higher values of $M_{\rm ecl}^{\rm min}$ and for lower values of $\beta$.
Overall, regardless of the value of these free parameters, we can see that
$\langle \alpha_{\rm TH} \rangle$ becomes 
progressively lower for more massive galaxies, that is, 
the star formation events taking place in more massive galaxies
are characterized by flatter IMFs, which reflects that more
massive galaxies have higher SFRs. 
These levels of star formation rate and their
dependence with the stellar mass
lead to adequate values of the specific SFR (sSFR),
defined as SFR per unit
stellar mass. 
This is shown in the
the inset panel in the plot on the left,
where we can see that
the sSFR at $z=0$
decreases with increasing stellar mass,
in agreement with the general trend found by
\citet{Feulner05} for galaxies at higher redshifts.

The lowest values of the slope for each mass range
are reached at earlier epochs for lower mass galaxies,
shifting from $1\lesssim z \lesssim 2$ to 
$3\lesssim z \lesssim 4$, for galaxies with $z=0$ stellar masses
of $\sim 5.6 \times 10^{11}\,\Msun$ and  $\sim 1.8\times 10^{10}\,\Msun$,
respectively.
The shape of the dependence of $\langle \alpha_{\rm TH} \rangle$
with redshift for each mass range indicates that lower mass
progenitors have lowers levels of SF and hence their star formation
events are characterized by steeper IMFs. Thus, the progenitors
of the most massive galaxies have a value of
$\langle \alpha_{\rm TH} \rangle$ at $z \approx 7$ quite similar
to the corresponding value of galaxies within the mass range 
$3.1 \times 10^{10}\,\Msun < M_{\star} <1 \times 10^{11}\,\Msun$
at $z \approx 3$.
It is interesting that the model {\small SAGTH5B2}, which gives a 
[$\alpha/{\rm Fe}$]-stellar mass relation
in better agreement with observations than the other two models
({\small SAGTH5B21} and {\small SAGTH100B2}), 
has star formation events characterized by IMFs steeper than the Salpeter IMF
for low mass galaxies and top heavier for high mass galaxies, since the
general discussion found in the literature considers that an IMF top-heavier
than Salpeter IMF is needed to reproduce 
the observed [$\alpha/{\rm Fe}$]-stellar mass relation
\citep{Nagashima+2005b, Calura2009, Arrigoni2010}. 
\citet{Calura2009} propose that a top-heavy IMF should be 
considered when the SFR of the galaxies exceeds a value of  
100 $\Msun\,{\rm yr}^{-1}$. 
In that case, they obtain a positive slope of the [$\alpha/{\rm Fe}$]-$\sigma$
relation, although the dependence becomes flat for
galaxies with high velocity dispersion;
when no starbursts triggered 
by interactions are taken into account, this dependence also becomes flat 
for low velocity dispersion galaxies (see their figure 20).
The dichotomy in the dependence of the slope of the IMF with the SFR of the 
galaxy assumed in their model seems to prevent the development of a well
behaved [$\alpha/{\rm Fe}$]-$\sigma$ relation.
The comparison of our results with those of \citet{Calura2009} 
highlights the importance of a progressive and smooth  
variation of the slope in the IMF with the SFR
supported by an empirically tested model of galaxy-wide IMF like the TH-IGIMF. 
From our investigation, we find that the key 
to achieve the correct dependence of [$\alpha/{\rm Fe}$]
abundance ratios with stellar mass is  
the combination of two aspects. On one hand,
the difference
of the IMF slope $\langle \alpha_{\rm TH} \rangle$ between
galaxies with different masses, such that  $\langle \alpha_{\rm TH} \rangle$
becomes progressively smaller for more massive galaxies.
On the other, the values adopted by $\langle \alpha_{\rm TH} \rangle$, 
which not only affect the slope of the [$\alpha/{\rm Fe}$]-stellar mass
relation but also its normalization.

Models with a TH-IGIMF characterized by higher values of
$M_{\rm ecl}^{\rm min}$ ({\small SAGTH100B2}) and a higher value of 
$\beta$ ({\small SAGTH5B21}) give place to a
[$\alpha/{\rm Fe}$]-stellar mass relation with a positive slope
since they also present
relative differences in the values of  $\langle \alpha_{\rm TH} \rangle$
for galaxies within
different mass ranges, although these differences are smaller  
for the model with higher $\beta$. In the latter case,
the values of $\langle \alpha_{\rm TH} \rangle$ 
are larger than that of Salpeter IMF regardless of the galaxy mass,
while in model {\small SAGTH100B2} the TH-IGIMF is always 
top-heavier than Salpeter IMF.  
In both cases, the  [$\alpha/{\rm Fe}$]-stellar mass relation is flatter
than in {\small SAGTH5B2}.
The fact that model {\small SAGTH5B21} is characterized by
TH-IGIMF steeper than Salpeter IMF also for massive galaxies 
makes these galaxies to be
less enriched by $\alpha$ elements that they should be, leading to a flatter
[$\alpha/{\rm Fe}$]-stellar mass relation.
On the contrary, 
in model {\small SAGTH100B2}, virtually all star formation events 
contain a large number of massive 
stars above $8\,\Msun$, thus displaying an enhanced production of SNe CC
with the consequent over-abundance of $\alpha$-elements since
galaxies of different
mass have TH-IGIMF top-heavier than Salpeter IMF. 
These results point
out in the direction that galaxies with masses smaller
than $\sim 3\times 10^{10}\,\Msun $ should have IMF 
steeper than Salpeter IMF,
while galaxies with higher masses should take top-heavier IMF than Salpeter 
IMF in order to achieve the right chemical abundances. 

Our results allow to restrict the possible values of 
free parameters involved in the TH-IGIMF theory, that is, 
the minimum embedded star cluster mass and
the slope of the embedded mass function, supporting a value as 
small as $M_{\rm ecl}^{\rm min}=5\,\Msun$ for the former and 
of $\beta=2$ for the latter.\\

\subsection{SNe delay times and extended bursts}
\label{sec:starbTimeScales}

In order to understand the particular pattern developed by the 
[$\alpha/{\rm Fe}$] abundance ratios
of elliptical galaxies, 
both the duration of the starbursts in galaxies
within different mass ranges and the  
lifetime of progenitors of different types of SNe
must be taken into account.\\

\subsubsection{Extended bursts and impact 
of different assumptions on the initial starburst duration}

In section \ref{sec:ExtendedBurstsInSAG}, we introduced the implementation
of extended starbursts in \sag. We now analize the results
and their implications in the chemical evolution of elliptical
galaxies. We also investigate the role of the extended bursts
in the development of the  [$\alpha/{\rm Fe}$]-stellar mass relation.

\begin{figure}
\centering
\includegraphics[scale=0.49]{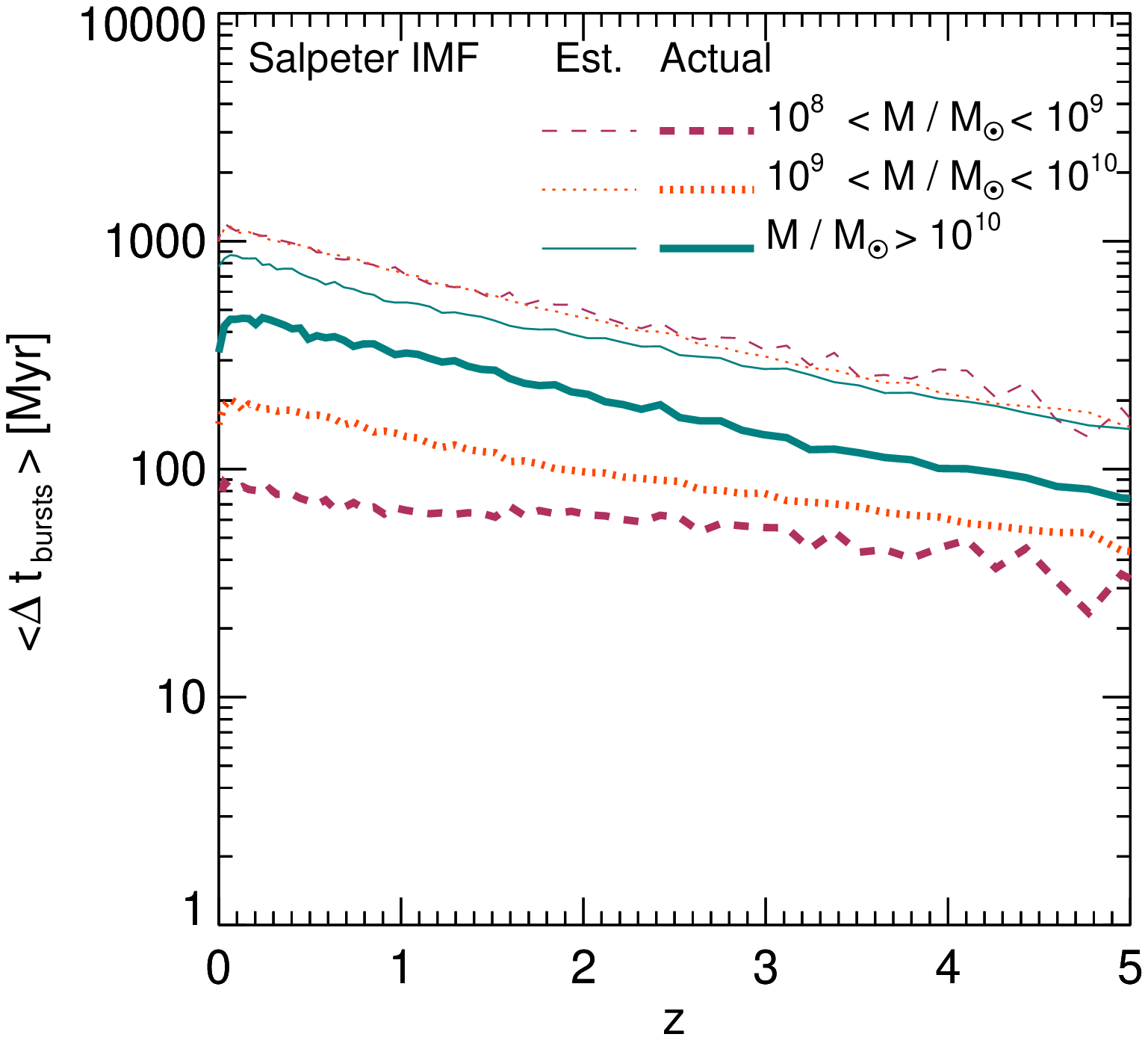}
\includegraphics[scale=0.49]{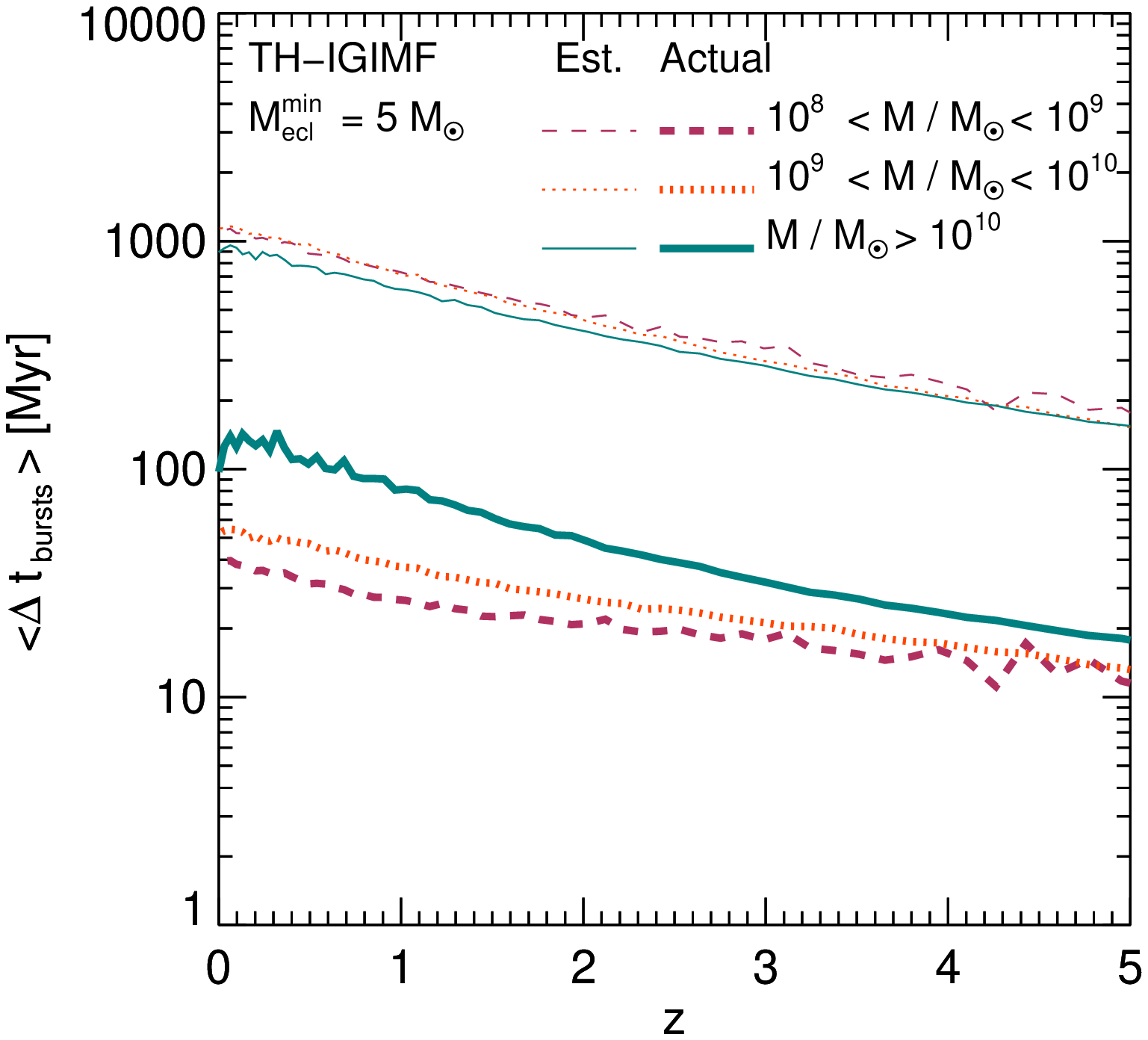}
\caption{{\it Top panel:} Mean duration of starbursts as a function of redshift for
galaxies generated by model {\small SAGS} in different mass ranges,
as indicated in the key;
galaxies are grouped according to their stellar mass at $z=0$.
Mean estimated burst time-scales and actual duration of starbursts 
are represented
by thin and thick lines, respectively.
{\it Bottom panel:} Same as the top panel, but for model {\small SAGTH5B2}. 
}
\label{fig:BurstsDurationSAGS}
\end{figure}

Fig. \ref{fig:BurstsDurationSAGS} presents the mean 
duration of starbursts taking place in galaxies within
different mass ranges
as a function of redshift; 
galaxies are grouped according to their stellar mass at $z=0$.
The redshift indicates the time in which
the starburst is triggered. We are considering here only starbursts that have
ended as a result of the exhaustion of
the bulge cold gas.
There are several aspects to note in this plot.
Firstly, the mean estimated time-scale of the starbursts (thin lines)
increases with decreasing redshift, 
with a very similar trend for galaxies of different stellar mass. 
This arises because of the natural growth of galaxies that
are characterized by larger dynamical times at more recent epochs.
Secondly,
the actual duration of the starbursts (thick lines)
are shorter than the estimated burst time-scales for galaxies 
in all mass ranges.
This is a result of the competition of several processes.
The bulge cold gas reservoir is common to all the starbursts
taking place simultaneously in a given galaxy. These starbursts may have been 
triggered at different times, contributing initially to increase
the bulge cold gas reservoir.
This reservoir is reduced by SNe feedback from 
stars in the bulge already formed in previous finished or on-going
starbursts.
The bulge cold gas is also reduced by the gradual growth of central BHs.
These processes 
make starbursts reach 
their end before than originally estimated; this difference is more pronounced
for less massive galaxies for which 
the cumulative effect of SNe feedback has stronger impact.

Considering starbursts triggered at $z\approx 0$ in model {\small SAGS} (top panel),
we find that their actual mean duration are of the
order of $9\times 10^{7}\, {\rm yr}$, 
$2\times 10^{8}\, {\rm yr}$, and
$5\times 10^{8}\, {\rm yr}$ 
for stellar masses within the ranges $10^8 - 10^9 \Msun$,
$10^9 - 10^{10} \Msun$, and  
larger than $10^{10} \Msun$, respectively.
These values can be compared with the exponential time-scale 
of starbursts at $z=0$ given by the semi-analytic model
used by \citet{Granato2000} (see their fig. 6),
which is assumed to scale with the dynamical time-scale of the spheroid 
formed in a major merger; this process
is the only channel of bulge formation that they consider.
Their values of starburst time-scales range from $5\times 10^{6}\, {\rm yr}$
to $1\times 10^{8}\, {\rm yr}$ for galaxies within the mass range
$10^8 - 3 \times 10^{11} \Msun$.
The mean estimated duration of starbursts in model {\small SAGS}
are of the same order of magnitude for all galaxies, being
larger than the exponential time-scale of starbursts
obtained by \citet{Granato2000}.

The fact that, regardless of the
redshift in which starbursts are triggered, the 
mean duration 
of the starbursts increase with stellar mass, 
contributes to explain the slightly positive slope in model {\small SAGS}. 
However, we emphasize here that the time-scale of galaxy
formation, that is assumed to drive the relation of [$\alpha/{\rm Fe}$] vs. 
mass in ellipticals by numerous works, is not linked to the duration 
of individual 
starbursts. The former is an integration over cosmic time of multiple star 
formation events.

This time-scale argument that explains the behaviour of the
[$\alpha/{\rm Fe}$]-stellar mass relation in the model is
valid as long as we consider a universal 
IMF. A variable IMF which slope depends on the SFR gives different 
relative numbers of SNe Ia and SNe CC progenitors for each
star formation event. This aspect plays an important role
in the build-up of the [$\alpha/{\rm Fe}$]-stellar mass
relation, as
we showed in Subsection \ref{sec:impactVarIMF}. The dependence with redshift
of starbursts in the case of the model {\small SAGTHB2} 
(bottom panel of Fig. \ref{fig:BurstsDurationSAGS})
is quite similar
to the model with a Salpeter IMF, 
but the duration of the bursts are approximately a factor of two shorter.

The shorter duration of the starbursts in {\small SAGTHB2} is explained 
by taking into account that the mean 
slopes of the IMF, 
$\langle \alpha_{\rm TH} \rangle$, become progressively 
larger than the slope of
the Salpeter IMF as we consider less massive galaxies
(see left panel of Fig. \ref{fig:MeanSlope}).
This fact leads to a low rate of SNe CC for low mass galaxies 
that have to be compensated
by large values of the efficiency of SNe feedback for the disc
and bulge ($\epsilon$ and $\epsilon_{\rm bulge}$, respectively),
which are free parameters of the model, in order to reproduce the
faint end of the luminosity functions.
These values are listed in 
Table \ref{table:table2}; 
we can see that they are 
larger than the corresponding values
for the calibrated model {\small SAGS} (Table \ref{table:table1}),
specially $\epsilon_{\rm bulge}$.

Hence, galaxies in model {\small SAGTH5B2} suffer, on average, a much stronger 
effect from 
SNe feedback arising from stars in the bulge than galaxies
in model {\small SAGS}.
This stronger effect contributes to deplete 
more rapidly the bulge cold
gas reservoir for starbursts, reducing considerably their actual duration.

Contrary to what is obtained from model {\small SAGS},
the mean actual duration of starbursts given by model {\small SAGTH5B2} 
and their dependence with the stellar mass
are in good agreement with the 
exponential time-scale
of starbursts at $z=0$ obtained  
by \citet{Granato2000}, whose results are described above in this section.
This good agreement is achieved 
from the consideration of a TH-IGIMF in our modelling.
The values obtained from our model for the most massive galaxies
are also consistent with
results from numerical simulations given by \citet{DiMatteo08},
who find that the activity of merger-driven starbursts
have a typical duration of a few $10^{8}\, {\rm yr}$.

After having analysed the features of the estimated and actual
mean duration of starbursts
resulting from the modelling adopted in this work,
we investigate the impact of a different assumption for the initial
estimation of the starburst time-scale on the mean actual 
starburst duration and the [$\alpha/{\rm Fe}$]-stellar mass relation.
The combination of this quantity
with the amount of bulge cold 
gas available
defines its SFR. Thus, models of galaxy formation
with a SFR dependent IMF, like the models with TH-IGIMF analysed here,
will be affected by the choice of this initial value.
SNe feedback
and BH growth 
determine the actual duration of the starburst, 
but they keep the SFR unaltered.

\begin{figure}
\centering
\includegraphics[scale=0.485]{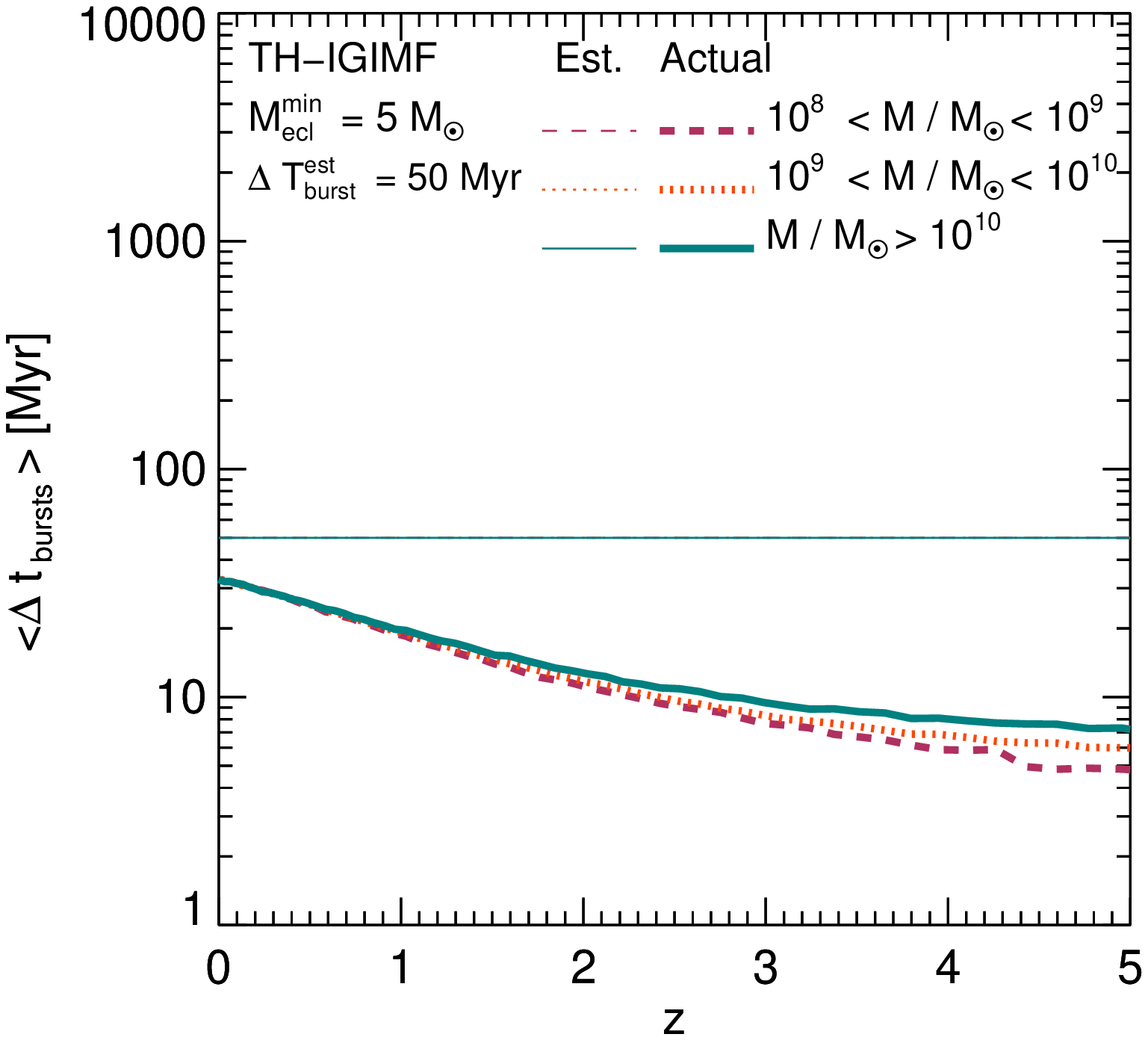}
\includegraphics[scale=0.435]{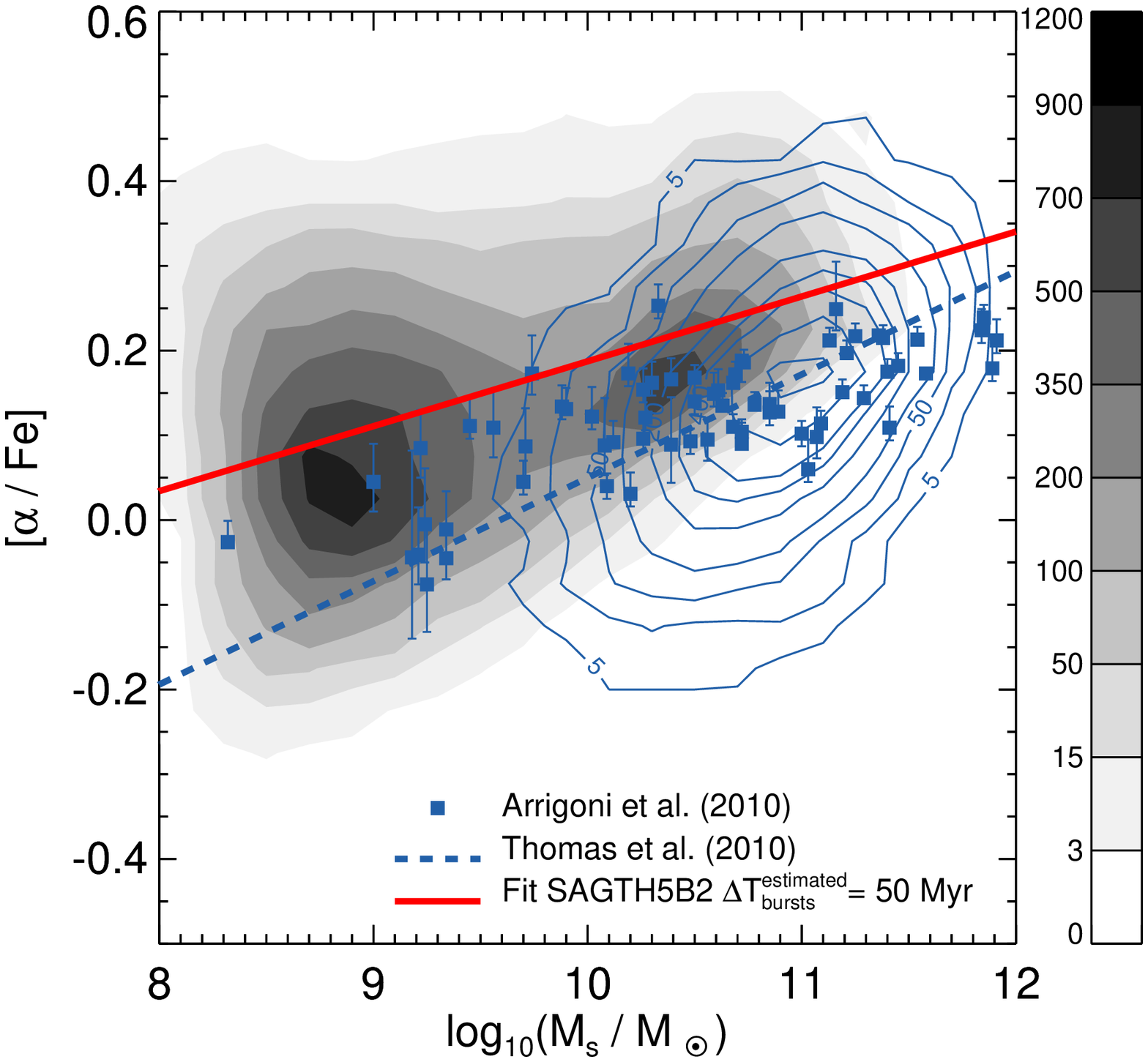}
\caption{{\it Top panel:}
Mean duration of finished starbursts as a function of the redshift 
in which they were triggered for
galaxies in different mass ranges
for model {\small SAGTH5B2}, as indicated in the key; galaxies are selected
according to their stellar mass at $z=0$. 
A fixed value of $50\,{\rm Myr}$ is assumed
for the initial
estimation of starburst time-scale of all galaxies.
{\it Bottom panel:} [$\alpha/{\rm Fe}$]-stellar mass relation for model
{\small SAGTH5B2} with a fixed initial value of estimated starburst 
timescale of $50\,{\rm Myr}$ for all galaxies.
}
\label{fig:BurstsDurationSAGTH5B2}
\end{figure}

We consider model {\small SAGTH5B2} and modify the 
estimation of the initial duration of starbursts, adopting  
the simplest case in which all galaxies
have the same initially estimated value for the duration of starbursts, 
$\Delta {\rm T}_{\rm burst}^{\rm estimated}=50\,{\rm Myr}$. 
The mean estimated and actual duration of starbursts as a funcion of redshift
for this modified model are shown in  
Fig. \ref{fig:BurstsDurationSAGTH5B2} for galaxies in different
mass ranges.  
The thin line denotes the estimated value of starburst duration
which is identical for all galaxies at all times.
The mean actual values are characterized by redshift evolution
which is pretty similar
for all galaxies, and  
only show a very mild dependence with stellar mass at high redshifts.
These values approach to the initially estimated ones
as redshift decreases indicating that the potential well of galaxies
become larger at lower redshifts with the consequent lower effect of
SNe feedback.
Since the initial estimated starburst time-scales are smaller
than those estimated considering the disc dynamical time-scale (see
lower panel of Fig. \ref{fig:BurstsDurationSAGS}),
the resulting actual starburst durations are also smaller.

Fig. \ref{fig:BurstsDurationSAGTH5B2} shows the impact of these shorter 
starbursts and their evolution on 
the [$\alpha/{\rm Fe}$]-stellar mass relation
obtained from the modified model {\small SAGTH5B2}.
s
The red solid line denote the linear fit to the
distribution of [$\alpha/{\rm Fe}$] abundance ratios with
stellar mass.
It has a slope of $a=0.076$, a bit smaller that one
corresponding to the model 
{\small SAGTH5B2}
that uses the dynamical time as an initial estimation of
the starburst duration ($a=0.088$).
Note that we are using the same set of parameters
as those corresponding to 
the model {\small SAGTH5B2}.

 All in all, different assumptions on the initial starburst duration do
not produce a significant change on the pattern of chemical abundances
of elliptical galaxies. These results provide additional support to our 
conclusions that the main responsible for the development of the proper
trend of [$\alpha/{\rm Fe}$] abundance ratios with stellar mass
is a SFR dependent TH-IGIMF, being $M_{{\rm ecl}}^{{\rm min}}=5\,\Msun$
and $\beta=2$ the preferred values for the free parameters of this model. \\

\begin{figure}
\centering
\includegraphics[scale=0.45]{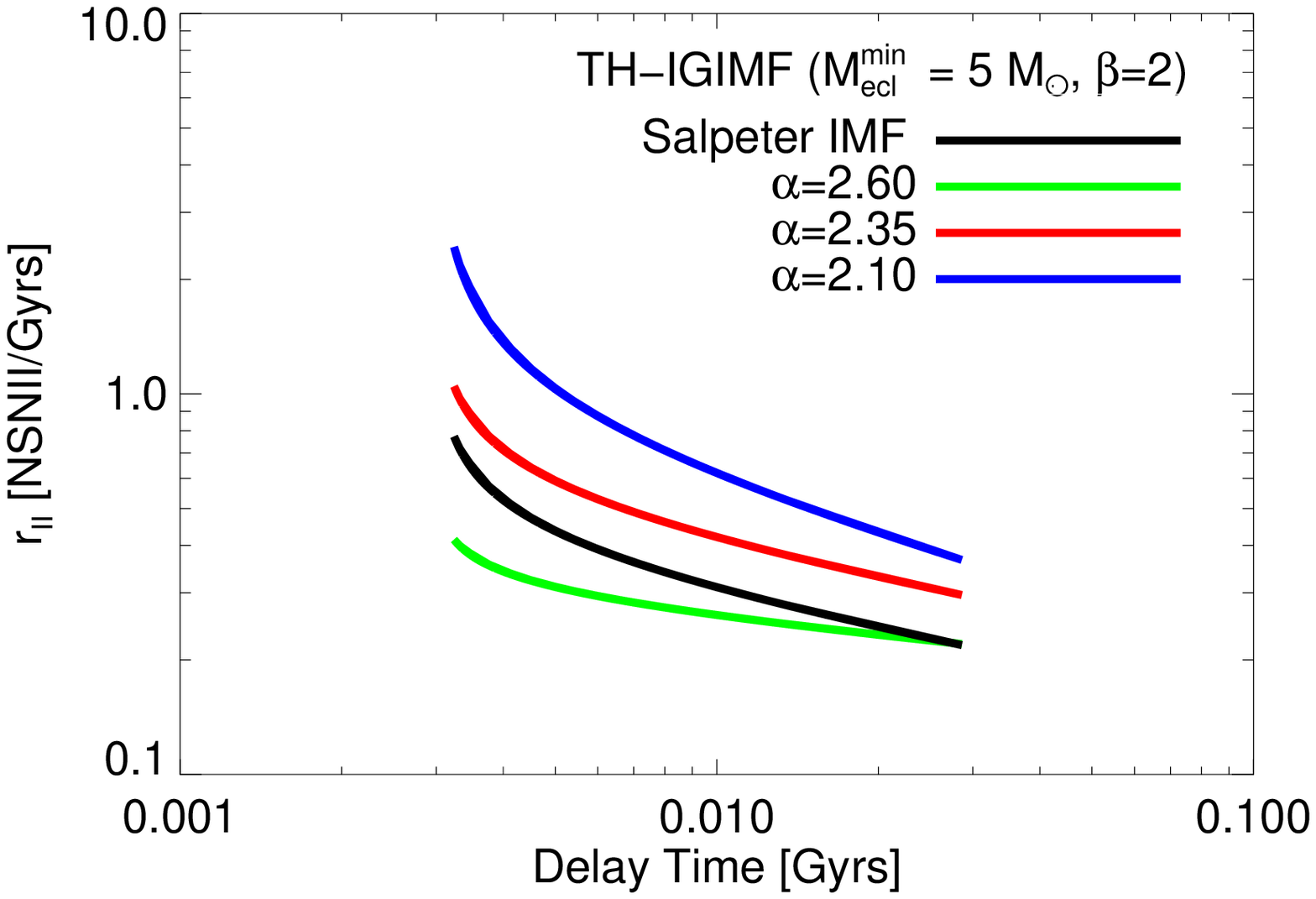}
\includegraphics[scale=0.45]{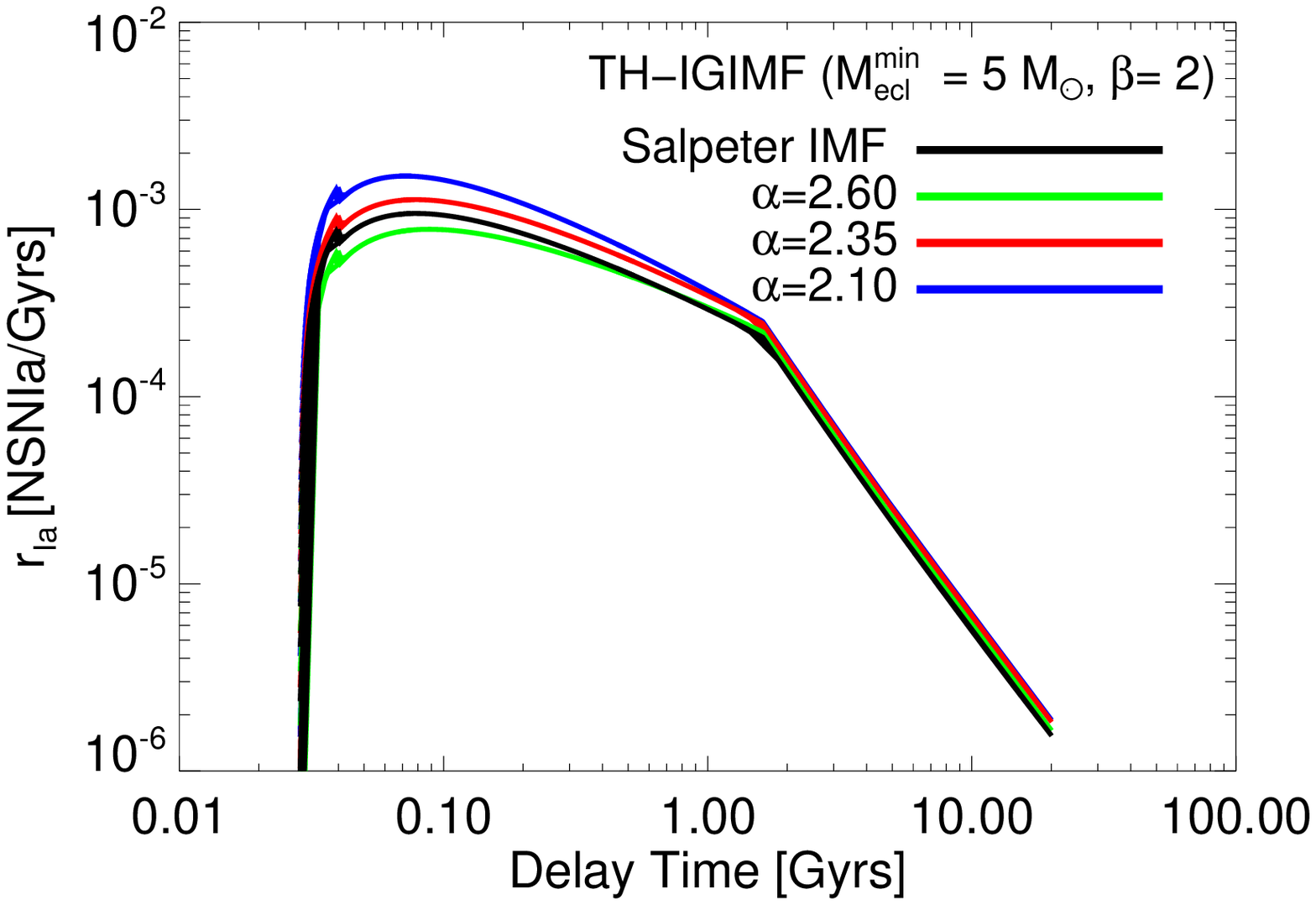}
\caption{Rates of SNe CC and SN Ia as a function of their delay time for
three different possible values of $\alpha_{\rm TH}$
for a TH-IGIMF with $M_{\rm ecl}^{\rm min}=5\,\Msun$ and $\beta=2$.
}

\label{fig:DTD_ECLMIN5Beta2_rIa}
\end{figure}

\subsubsection{SNe delay times}
\label{sec:SNeDelayTimes}

The top panel of 
Fig. \ref{fig:DTD_ECLMIN5Beta2_rIa} shows the rates of SNe CC 
as a function of their delay times, that is, 
the time-scale 
for explosion. Different coloured lines correspond
to the rates obtained from 
a TH-IGIMF with $M_{\rm ecl}^{\rm min}=5\,\Msun$ and $\beta=2$ and
different
possible values of $\alpha_{\rm TH}$. 
As expected,
flatter IMFs give place to larger SNe CC rates. 
These rates are compared with those
obtained from a Salpeter IMF (black line).
The SNe Ia rates corresponding to these four cases are presented
in the bottom panel of Fig. \ref{fig:DTD_ECLMIN5Beta2_rIa}.

As we mentioned
in Section \ref{sec:chem_model}, the SNe Ia rate is estimated according to the
formalism described in
\citet{Lia2002} that assumes that SNe Ia originate in binary systems
\citep{Greggio83}. 
The fraction of these binary systems,
$A_{\rm bin}$, is one of the free parameters of \sag.
For an assumed IMF, the value adopted by this parameter
determines the SNe Ia rate.
In the case of a universal IMF, like the Salpeter IMF adopted in model
{\small{SAGS}}, the value of $A_{\rm bin}$ is linked to that IMF,
constrained by the requirement of reproducing the
observed evolution of SNe Ia rates.
For a model with TH-IGIMF, the slope of the IMF
for stellar masses larger than $1.3\,M_{\odot}$ (see eq. \ref{eq:THIGIMF-FIT}) 
can vary within a wide range, according to the SFR of a given star formation
event. However, we consider a unique fraction of binary systems 
regardless of the slope of the IMF, $\alpha_{\rm TH}$. 
Thus, SNe Ia rates for each single stellar population differ for different
values of $\alpha_{\rm TH}$, as 
can be seen in the bottom panel of 
Fig. \ref{fig:DTD_ECLMIN5Beta2_rIa}, where the rates are given 
as a function of the SNe Ia delay times, which
are set by
the lifetime of the secondary star of the binary system.
Lower (higher) values of $\alpha_{\rm TH}$ (flatter (steeper) IMF) 
determined by
star formation events with higher (lower) SFR
give place to lower (higher) SNe Ia 
rates. The value adopted by $A_{\rm bin}$ in the calibration process
is the result of keeping the balance between the resulting SNe Ia rates
from all the different possible
values of  $\alpha_{\rm TH}$ that can occur in all SF events in galaxies
such that the average behaviour of the evolution of the SNe Ia rates
for the whole galaxy population resembles the observed one.

\begin{figure}
\centering
\includegraphics[scale=0.42]{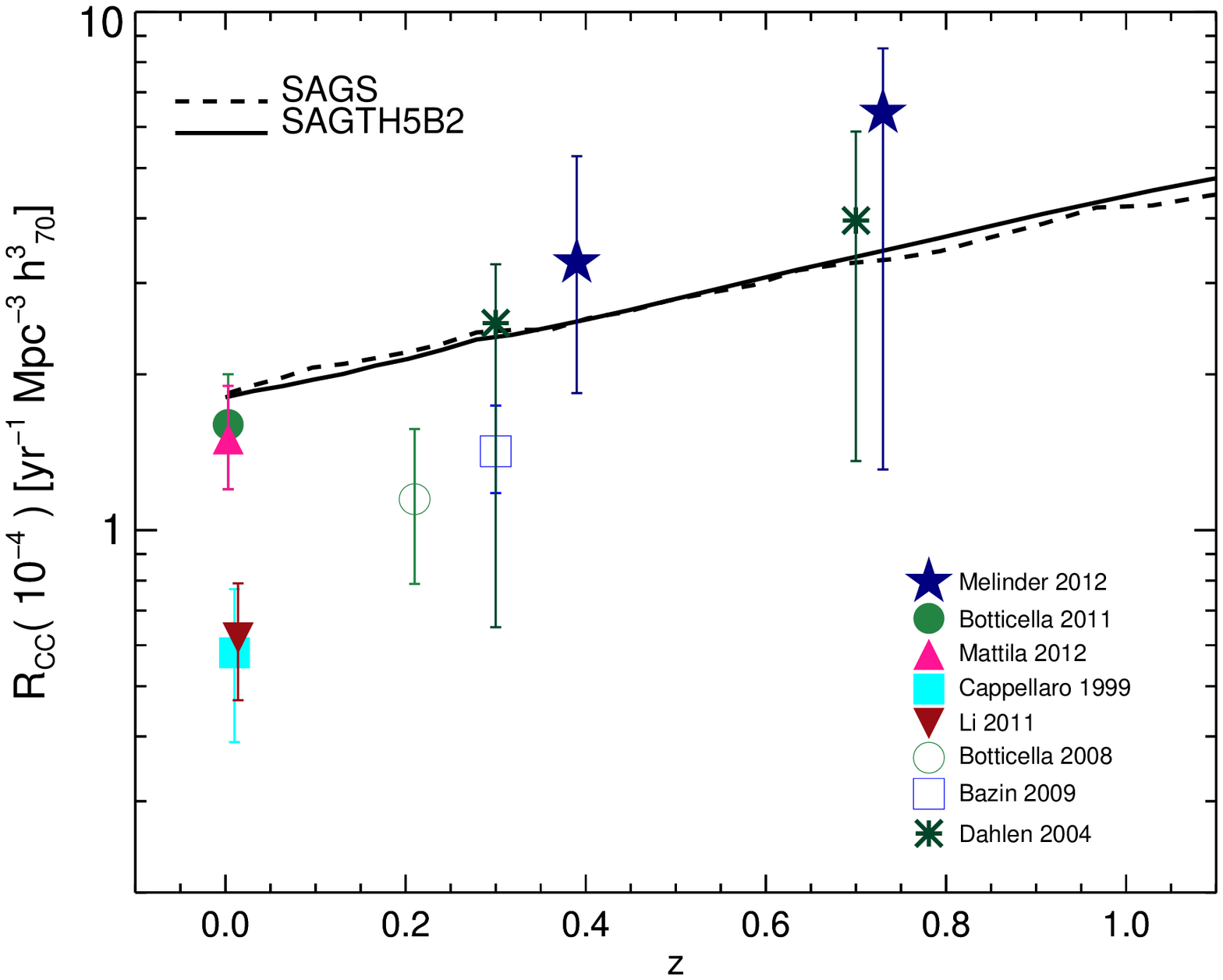}
\includegraphics[scale=0.42]{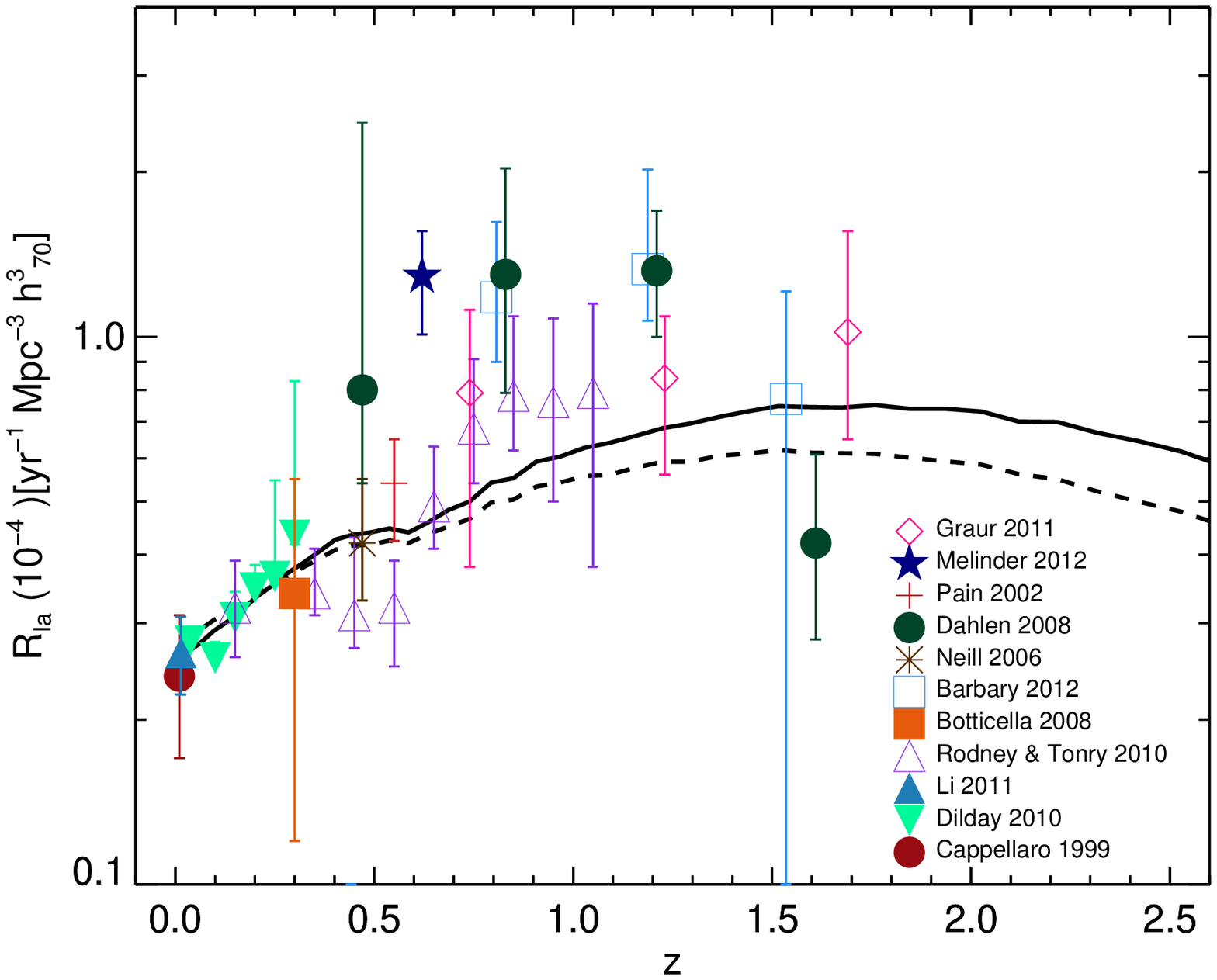}
\caption{Evolution of SNe CC rates (top panel) and SNe Ia rates (bottom
panel) given by models    
{\small SAGS} (dashed line) and {\small SAGTH5B2} (solid line), 
compared with the compilation of observational determinations 
by \citet{Melinder2012}.
}
\label{fig:SNRatesCal}
\end{figure}

Figure \ref{fig:SNRatesCal} shows the computed SNe rates 
as a function of redshift for both 
{\small SAGS} (dashed line) and {\small SAGTH5B2} (solid line)
models. 
The observational data corresponds 
to the compilation of \citet{Melinder2012}. 
The evolution in the SNe CC rates are depicted in the top panel. Almost no 
difference arises between the results of the calibrated models {\small SAGS} 
and {\small SAGTH5B2}. 
The observed rates show a dichotomy at low redshift; while the works 
of \citet{Botticella2012} and \citet{Mattila2012} estimate SNe CC rates 
of almost 
$2 \,(10^{-4})\, {\rm yr}^{-1}\, {\rm Mpc}^{-3}\, h^{3}_{70}$, 
the works of
\citet{Cappellaro1999} and 
\citet{Li2011} give SNe CC rates of the order of  
$0.6\,(10^{-4})\, {\rm yr}^{-1}\, {\rm Mpc}^{-3}\, h^{3}_{70}$. 
The model outputs are consistent with the higher values of 
SNe CC rates at $z=0$.

The SNe Ia rates shown in Fig. \ref{fig:DTD_ECLMIN5Beta2_rIa} for different
values of $\alpha_{\rm TH}$ are compared with the one emerging from Salpeter 
IMF. We can see that for a value of $\alpha_{\rm TH}$ equal to the
slope of Salpeter IMF, the TH-IGIMF gives larger values of SNe Ia rates
since the differences between both IMFs for stellar masses smaller that
$1.3\,\Msun$ lead to different normalization values.
Once both {\small SAGS} and {\small SAGTH5B2} have been calibrated, 
the SNe Ia rates
are similar in both models, as can be seen from the comparison of
dashes and solid lines in Fig. \ref{fig:SNRatesCal}.

The CC SNe in the model 
have delay times between $2$ and $30 {\rm Myr}$ (top panel of Fig. \ref{fig:DTD_ECLMIN5Beta2_rIa}), 
sufficiently short for the 
forming stars in a burst to lock most of the $\alpha$-elements produced during 
the explotions, since the actual duration of bursts have
longer timescales, of the order of tens of ${\rm Myr}$ 
(see Fig. \ref{fig:BurstsDurationSAGS}). 
On the other hand, the 
delay times of SNe Ia are in the range comprised between 
tens of ${\rm Myr}$ and some ${\rm Gyr}$, with the bulk of SNe Ia
having delay times between $20 {\rm Myr} - 2 {\rm Gyr}$ with a 
peak at
approximately $100 {\rm Myr}$.

This time-scale is of the order of the maximum value
of the starbursts duration, therefore many starbursts will not form
stars highly enriched with the main SNe Ia product, that is, iron.
However, since the starbursts do not consume the cold gas bulge instantaneouly,
it is frequent that several bursts coexist in a same galaxy and
a late triggered one might form stars rich in iron from gas polluted
by a previous generation of stars formed in the bulge.
Hence, the final [$\alpha/{\rm{Fe}}$] abundance ratio of an elliptical
galaxy is the result of the combination of the duration of starbursts,
SNe delay times and time-scale of galaxy formation. The latter
is analysed in detail for both {\small SAGS} and {\small SAGTH5B2} models
in Section \ref{sec:timescales}. 

\subsection[]{Further tests: mass-to-light ratios and luminosity functions}
\label{sec:masstolight} 

\citet{Matteucci94} suggested that the variable IMF hyphothesis to explain
the [$\alpha/{\rm{Fe}}$]-mass relation in elliptical galaxies should be 
tested on integrated luminosities and mass-to-light ratios (M/L). 

\begin{figure}
\centering
\includegraphics[scale=0.35, angle=270]{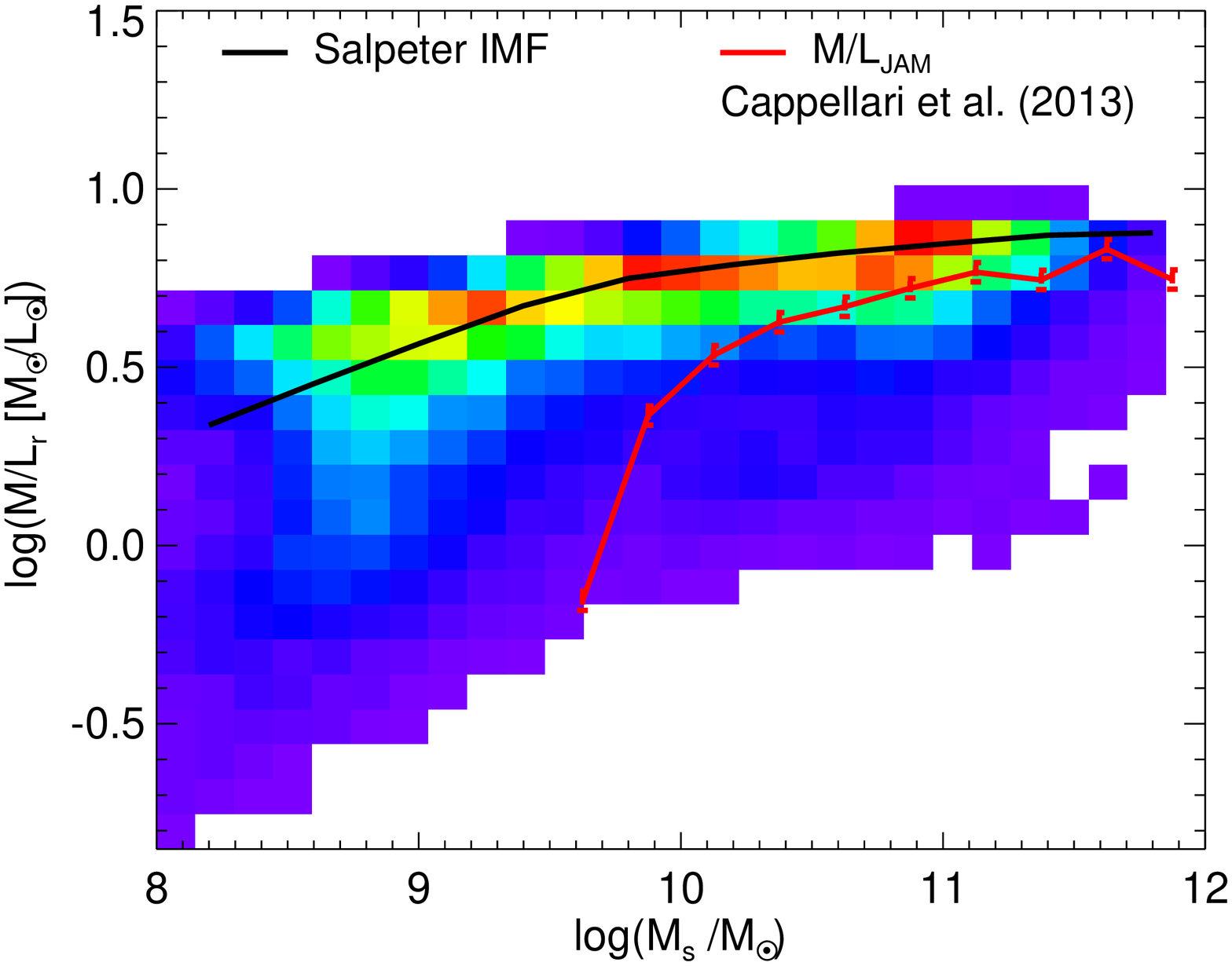}
\includegraphics[scale=0.35, angle=270]{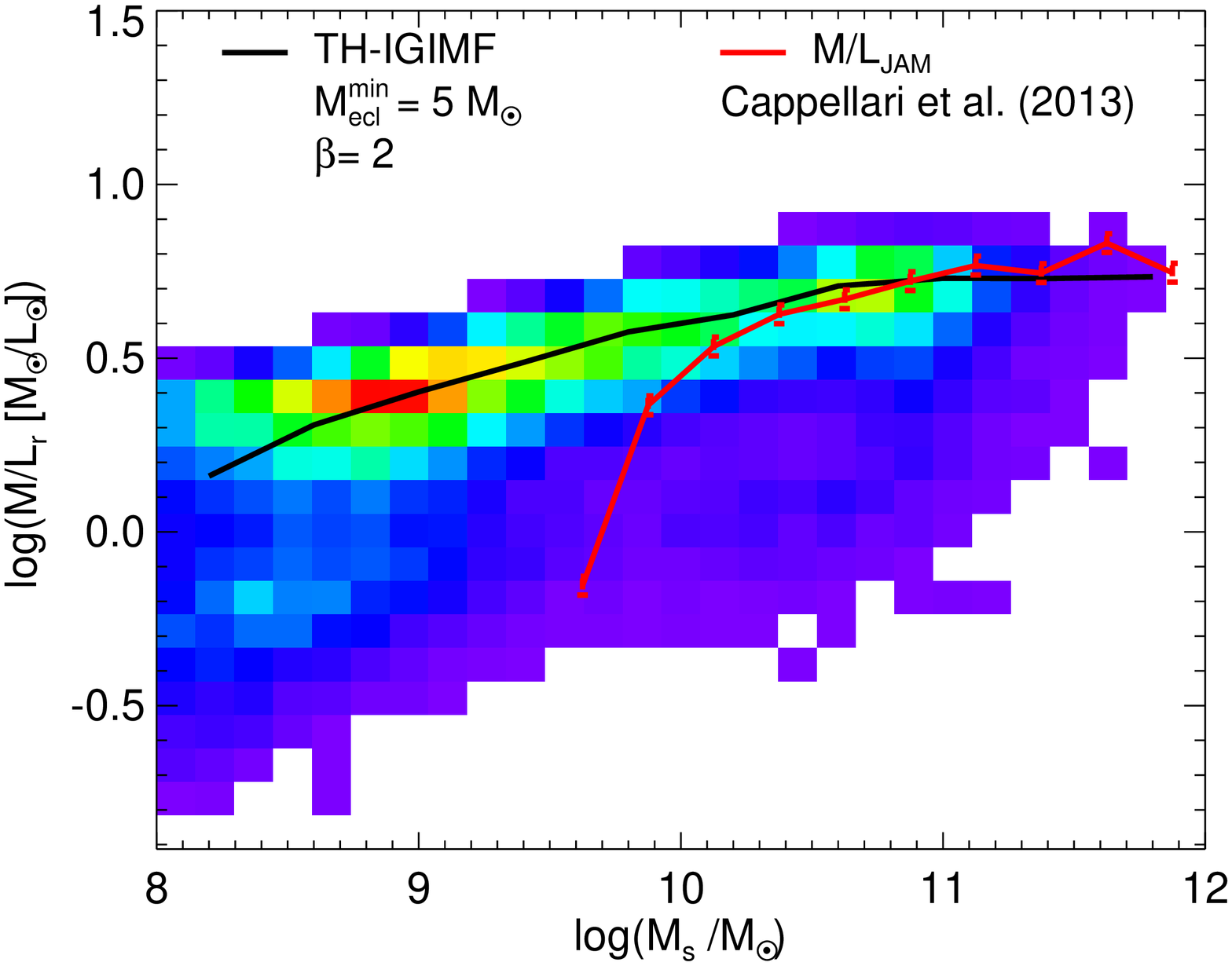}
\caption{Mass-to-light Ratios as a function of the stellar mass of elliptical
galaxies. Results from models  {\small SAGS} and {\small SAGTH5B2}
are shown in the top and bottom panels, respectively,
compared with observational determinations of \citep{Cappellari2013b}.
}
\label{fig:MLRatios}
\end{figure}

In a series of papers, the ATLAS3D collaboration study the properties of 
 a sample of 260 early-type galaxies \citep[see][]{Cappellari2011}.
 In paper XV \citep{Cappellari2013b} and XX \citep{Cappellari2013c} they 
concentrate on the mass plane 
(M, $\sigma_e$, $R_e$), its projections, and the distribution of galactic 
properties among them. They derive total (baryonic plus dark matter) dynamical 
masses and mass-to-light ratios ($M_{\rm JAM}$, $M/L_{\rm JAM}$ ) 
within a sphere of one effective radius of the early type galaxies in their 
sample. They point 
out that this quantity resembles and is directly comparable to the total stellar
 mass used in previous studies, given that the calculated fraction of DM inside
 that radius is ussually low, $~13\%$ on average. They choose the Sloan 
{\em r}-band 
to measure luminosities. It is worth noting that these quantities are 
unaffected
by assumptions on the IMF in galaxies because they are derived using dynamical
considerations and their calculations do not involve 
stellar population synthesis
models. In Fig. \ref{fig:MLRatios} we compare the  mass-to-light ratio
as a function of the stellar mass of elliptical galaxies resulting from  
our model 
with Salpeter IMF (top panel)
and {\small SAGTH5B2} (bottom
panel),
with the data 
extracted from Table 1 of \citet{Cappellari2013b}. For this purpose, we 
consider the stellar mass of model galaxies divided by the total luminosity in 
the {\em r}-band. The error at 1$\sigma$ of the 
estimations of M/L ratios from \citet{Cappellari2013b} is 0.027 dex. 
We can see that in the case of our best choice of the TH-IGIMF parameters, that is, $\beta=2$ and $M_{\rm ecl}^{\rm min}=5\,\Msun$, the
zero point of the relation is better reproduced 
than in model {\small SAGS},
with a good agreement with the observed M/L ratios
in a broad range of stellar masses. 
The departure from observations occurs for masses smaller than
$2\times 10^{10}\,\Msun$, for which the slopes of the IMF are steeper than
the one of Salpeter IMF (see Fig. \ref{fig:MeanSlope}).
The shape of the M/L ratios is similar for both {\small SAGS}
and {\small SAGTH5B2} models, with lower values for the latter
case and a higher dispersion towards lower values,
as a result of the calibration process that regulates
in a different way the star formation and feedback.
That is, it seems that the IMF helps to recover the correct values
of the M/L ratios for massive galaxies, but it 
cannot solve the problem of 
the excess at
lower masses with respect to observations,
which is present for any IMF.
This result supports the hypothesis of a TH-IGIMF in elliptical galaxies.

\begin{figure}
\centering
\includegraphics[scale=0.45]{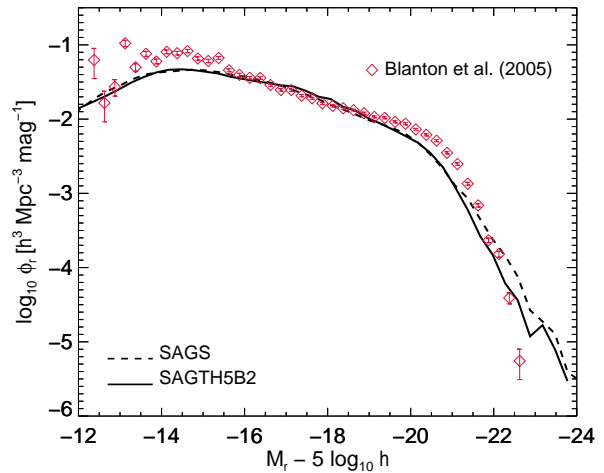}
\caption{ Luminosity function in the {\em r}-band 
of the population of galaxies generated by models {\small SAGS} (dashed line)
and {\small SAGTH5B2} (solid line)
compared with the observed determinations by \citet{Blanton2005} 
(open symbols with error bars).
}
\label{fig:LFCal}
\end{figure}

We perform a further test to the population of galaxies. We compare the 
results from models {\small SAGS} and {\small SAGTH5B2} with 
the observed {\em r}-band LF of \citet{Blanton2005}, used as a constraint to
calibrate the model. The {\em r}-band LFs 
resulting from both models are shown in Fig. \ref{fig:LFCal}. 
As can be seen, there are not major 
differences between them and the agreement with the observed LF
\citep{Blanton2005}
is quite good. A slight excess of luminous 
galaxies is present in both models, being more notorius in the 
model {\small SAGS}. The supression of SF in the most massive galaxies
cannot be achieved with the current modelling of AGN feedback.
Instead, a better agreement with obsevations could be obtained 
if starbursts during galaxy mergers are triggered only when the remnant galaxy 
becomes unstable, as proposed by \citet{padilla_flip_2014}. The luminosity 
function together with the mass-to-light ratios represent a good test to 
the implementation of a variable IMF, given that different stellar populations
are characterized by distinct signatures of light that must be constrained 
together with the [$\alpha/{\rm Fe}$]-mass relation in order to obtain 
consistent results in the properties of galaxies when modifying an important
distribution function like the IMF of stars. 

\section[]{[$\alpha/{\rm Fe}$]-mass relation: influence of formation time-scales}
\label{sec:timescales} 

In this work, we have demonstrated that
a TH-IGIMF is needed to reproduce
the trend of [$\alpha/{\rm Fe}$] abundance ratios
with stellar mass shown by elliptical galaxies.
However, one aspect also considered to explain the built-up of this
relation is the time-scale of galaxy formation. This argument claims that  
more massive galaxies are formed in a shorter time-scale than less massive 
ones \citep[e.g.][]{Thomas2005, Recchi+2009}. In this sense, 
this relation is usually considered as another consequence of the 
downsizing.
Observational evidence indicates that elliptical galaxies assemble most of 
their mass at high redshift. A common technique used nowadays is the 
stellar archaeology \citep{Thomas2005, Neistein2006}. In this line of work, 
spectral indices are measured and used to compute parameters of stellar 
populations of observed galaxies with population synthesis models. One common
result of applying this method is that stellar populations in massive galaxies
are older than in their less massive counterparts, disregarding the galaxy 
morphology \citep{Panter2007}. Moreover, elliptical galaxies present the 
oldest stellar populations, given that they show little or none star 
formation activity, although T10 find that approximately $10$ per cent
of their sample of ellipticals are actively forming stars and define a peak 
at young ages in the age distribution. 
\citet{Neistein2006} find that this aspect of the downsizing behaviour of 
galaxies can be explained naturally in the 
frame of hierarchical clustering in terms of the progenitors of the main 
haloes, defining a minimum mass of progenitors that are capable of forming 
stars. However, other aspects also considered as downsizing, like the shift 
with time in the characteristic mass of star forming galaxies 
\citep{Cowie-Barger2008}, or 
the chemo-archaeological downsizing \citep{Fontanot2009} 
can not be explained with this line of reasoning. 

Here, we are focusing on the chemo-archaeological downsizing.
The aim of this section is to analyse the formation time-scales of 
our sample of model galaxies in order 
to determine whether
they are partially responsible for leading to a positive 
slope in the [$\alpha / {\rm Fe}]$-stellar mass relation or not.
 
For each galaxy 
selected according to its stellar mass at $z=0$, 
we define the formation time-scale, $\Delta T_{50}$,
as the period of time elapsed since the redshift at which the first star
formation event occurs and the fomation redshift, $z_{\rm form(50)}$.
The latter is defined as the 
redshift at which the galaxy acquires $50$ per cent of its total 
stellar mass formed {\it in-situ},  
 where the term in-situ refers to the stellar mass formed in a galaxy by 
quiescent SF or in starbursts, without taking into 
account the stars of accreted satellites. 
This definition ensures that 
the analysed galaxies and their stellar populations are formed from 
enriched gas belonging to their main progenitors, 
making easier the interpretation of the influence of 
star formation time-scales
and ejection of SNe products on the 
[$\alpha / {\rm Fe}]$-stellar mass relation.
Note that the total stellar mass formed in-situ considered for
the present analysis is tracked in a separate variable within the code, and 
does not take into account the effect of mass loss due 
to the recycling process. 

\begin{figure}
\centering
\includegraphics[width=0.4\textwidth, angle=270]{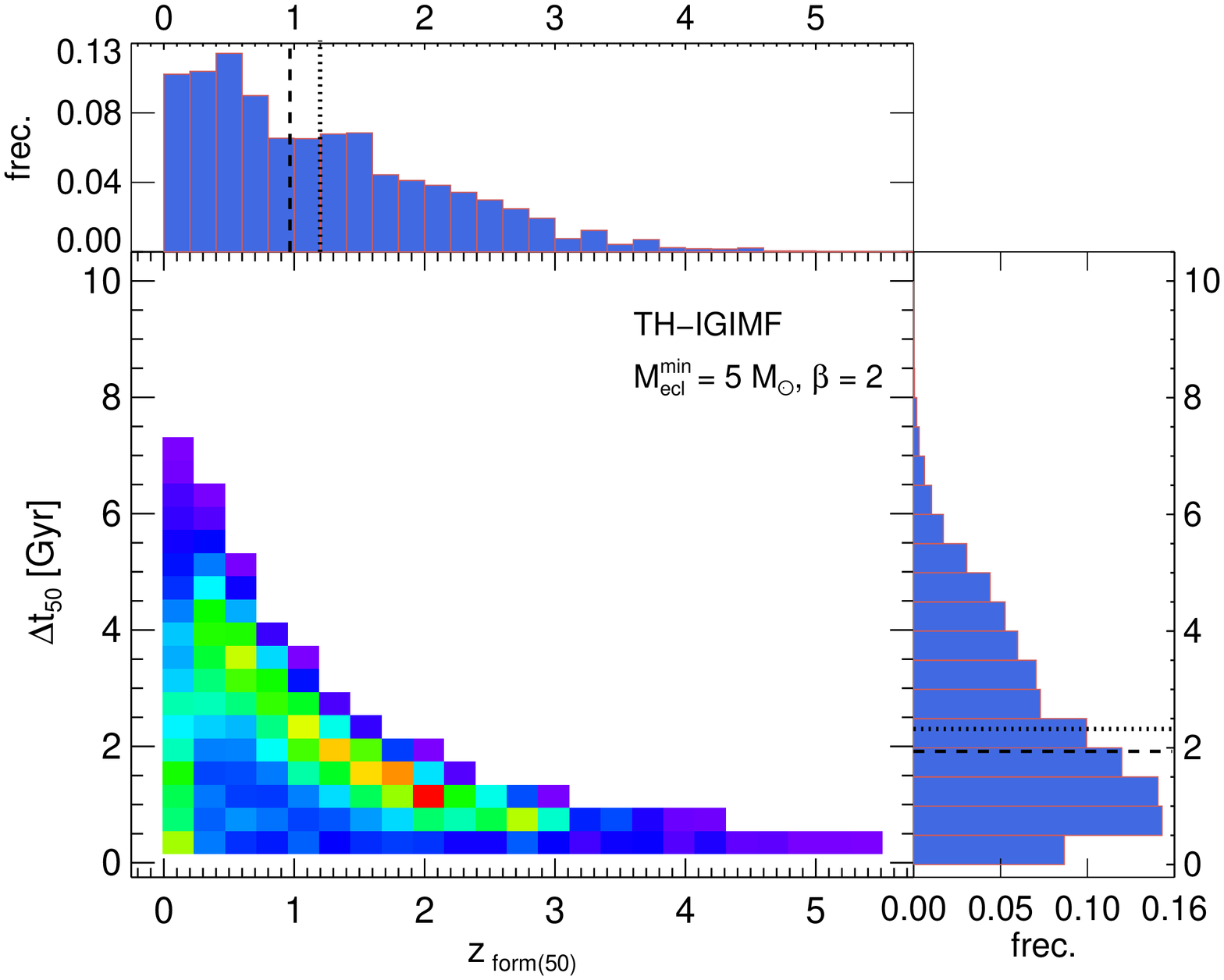}
\includegraphics[width=0.4\textwidth, angle=270]{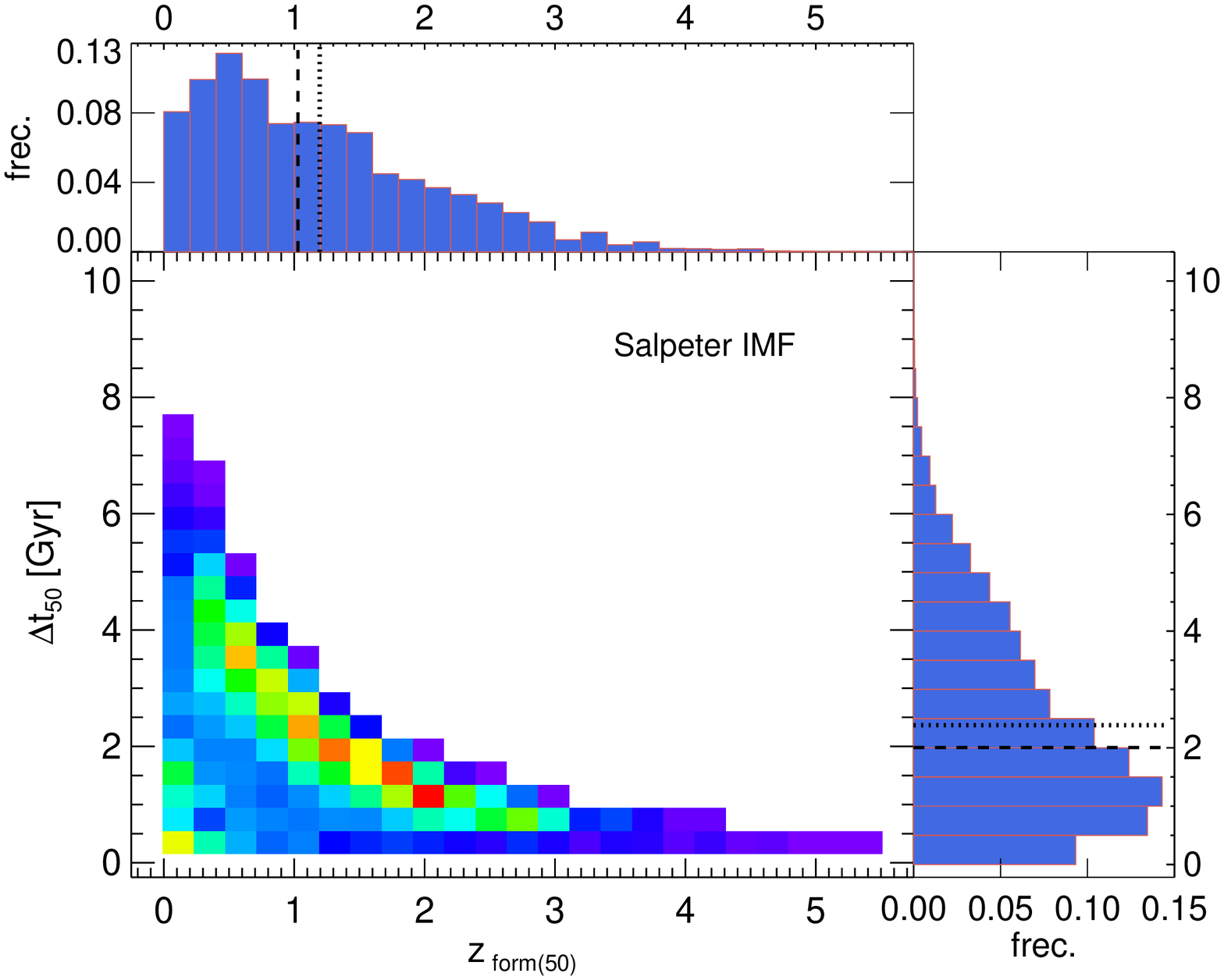}
\caption{Formation time-scales, $\Delta T_{50}$, 
of elliptical galaxies from the models
versus their formation redshifts, $z_{\rm form(50)}$ 
(see plain text for the definition)
for galaxies from model {\small SAGTH5B2} (top panel) and  {\small SAGS}
(bottom panel). The histograms 
on the right of central panels are the frequency distributions of 
the formation time-scales of the galaxy population, and those on the top 
correspond to the  
formation redshifts.  
Dotted and dashed lines represent the mean and median values of the 
distributions, respectively.
}
\label{fig:DeltaTz50}
\end{figure}

Fig. \ref{fig:DeltaTz50} shows 
the formation time-scale, $\Delta T_{50}$, as a function
of the formation redshift, $z_{\rm form(50)}$, for our whole sample of 
early-type galaxies obtained from models with two different IMFs.
We consider the models {\small SAGTH5B2} (top panel) and {\small SAGS} 
(bottom panel) 
which involve a TH-IGIMF with 
parameters 
$M_{\rm ecl}^{\rm min}=5\,\Msun$ and $\beta=2$
and the Salpeter IMF, respectively.
The former is our selected best case of TH-IGIMF, as discussed in the
previous section. 
We can see that, regardless of the IMF considered,
there is a clear trend in which $\Delta T_{50}$ decreases as $z_{\rm form(50)}$
increases, as depicted by the colour-density plot. This indicates that
galaxies that form at higher redshift complete the 
formation process in shorter time-scales,
consistent with \citet*{Lagos09}. 
More specifically,
galaxies that have formed at $z_{\rm form(50)}\approx 2$ or before
took less than $2\,{\rm Gyr}$ to achieve $50$ per cent of their stellar 
mass formed in-situ. 

At lower redshifts,
the dispersion in the formation time-scales becomes larger.  
The histograms on the top and on the right of the central panel,
in both the top and bottom plots,
are the frequency distributions 
of the formation redshift and the formation time-scale, respectively. We 
see that the bulk of early type galaxies form between 
$z_{\rm form(50)}\approx 1$ and $0$ and in time-scales 
within a range $1\,{\rm Gyr} \lesssim \Delta T_{50} \lesssim 4\,{\rm Gyr}$.
If we consider the global population of galaxies formed at different
redshifts, we find
that most of the galaxies
have formation time-scales less than $2\,{\rm Gyr}$, 
as it is evident from the histograms on the right.
The mean and median values of the distributions of $\Delta T_{50}$
(dotted and dashed lines in the figure, respectively)
are $2.32\,{\rm Gyr}$ and $1.94\,{\rm Gyr}$ for model
{\small SAGTH5B2}, and $2.51\,{\rm Gyr}$ and $2.02\,{\rm Gyr}$ for
model {\small SAGS}, respectively.
For the distributions of $z_{\rm form(50)}$, we have
mean and median values of $1.2$ and $0.97$ 
for {\small SAGTH5B2}, and $1.18$  and $0.97$
for {\small SAGS}, respectively.

\begin{figure}
\centering
\includegraphics[width=0.5\textwidth]{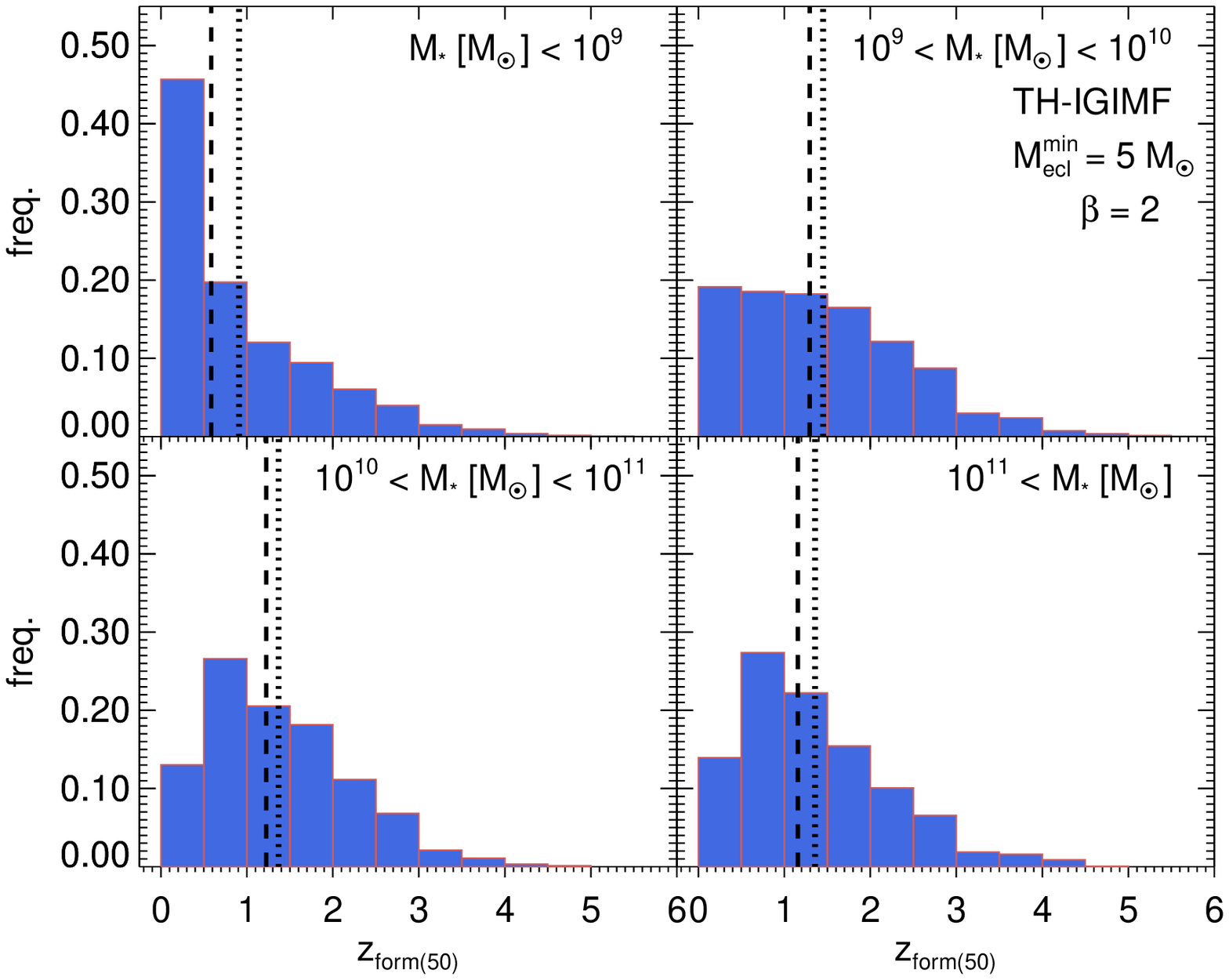}
\includegraphics[width=0.5\textwidth]{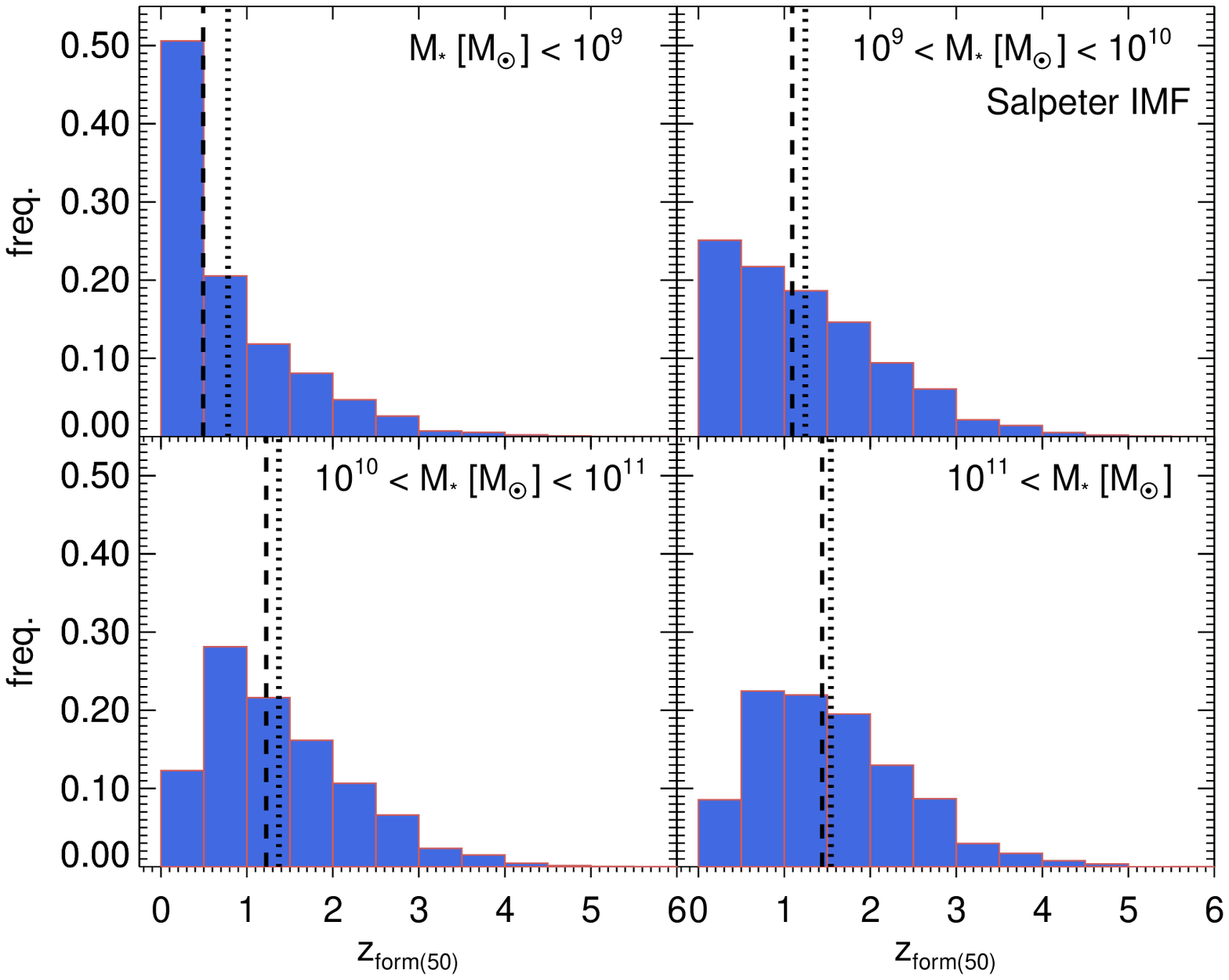}
\caption{
Distribution of formation redshifts, $z_{\rm form(50)}$, for
galaxies of different stellar mass, as indicated in the key
for models
{\small SAGTH5B2} (top panel) and {\small SAGS} (bottom panel).
Dotted and dashed lines represent the mean and median values of the 
distributions, respectively.
}
\label{fig:freqzform50MassBin}
\end{figure}

The distributions of
$z_{\rm form(50)}$ for
galaxies of different mass are shown in Fig. \ref{fig:freqzform50MassBin} 
for models
{\small SAGTH5B2} and {\small SAGS}
(top and bottom panels, respectively).
In agreement with a previous study by \citet{Lagos09},
both models present similar shapes of their $z_{\rm form(50)}$ distributions 
for different mass
ranges. 
We see that the least massive galaxies ($M_{\star} < 10^{9}\,\Msun$)
achieve the $50$ per cent of their stellar
mass formed in-situ at lower redshifts than more massive galaxies,
with the bulk of them forming between $z\approx 0$ and $z\approx 0.5$. 
The downsizing trend is less pronounced for galaxies more
massive than $\approx 10^{9}\,\Msun$.
The mean and median values of the 
distributions (denoted by dotted and
dashed lines, respectively) are pretty similar for galaxies with masses
within the ranges $10^{9} - 10^{10} \,\Msun$ and 
$10^{10} - 10^{11} \,\Msun$, and $10^{11} \,\Msun$,
with values comprised within the interval 
$1 \lesssim z_{\rm form(50)} \lesssim 1.5$.
However, 
the peak of the corresponding distributions
shifts from the interval $0 \lesssim z_{\rm form(50)} \lesssim 0.5$ to 
$0.5 \lesssim z_{\rm form(50)} \lesssim 1$, making evident a
mild downsizing trend. The shift of the peak of the distributions
is more evident in the model {\small SAGS} (lower panel). 
This is in qualitatively agreement with \citet{Lagos09} and 
\citet{DeLucia2006}.
In the latter work, it is shown that massive galaxies form
their stars earlier than less massive
systems but that they assemble their stars later.
They define 
the formation redshift as the redshift
when $50$ per cent of the stars that make up the galaxy
at redshift $z = 0$ are already formed, and 
the assembly time as the redshift when $50$ per cent
of the final stellar mass is already contained in a
single object. 
In our definition of formation time, we are not considering the
contribution of the stellar mass formed in accreted satellite galaxies.
We have tested the impact of this accreted mass on the histograms
shown in Fig. \ref{fig:freqzform50MassBin}, and we do not find any  
significant change, neither for {\small SAGTH5B2} nor
for {\small SAGS}. This is expected taking into account that the mass 
contributed by 
satellites is a small fraction of the total final mass
of the galaxy, being of the order of $\approx 10$ per cent
for the vast majority of galaxies, in agreement with the analysis done by 
\citet{Jimenez2011}.
 
\begin{figure}
\centering
\includegraphics[width=0.5\textwidth]{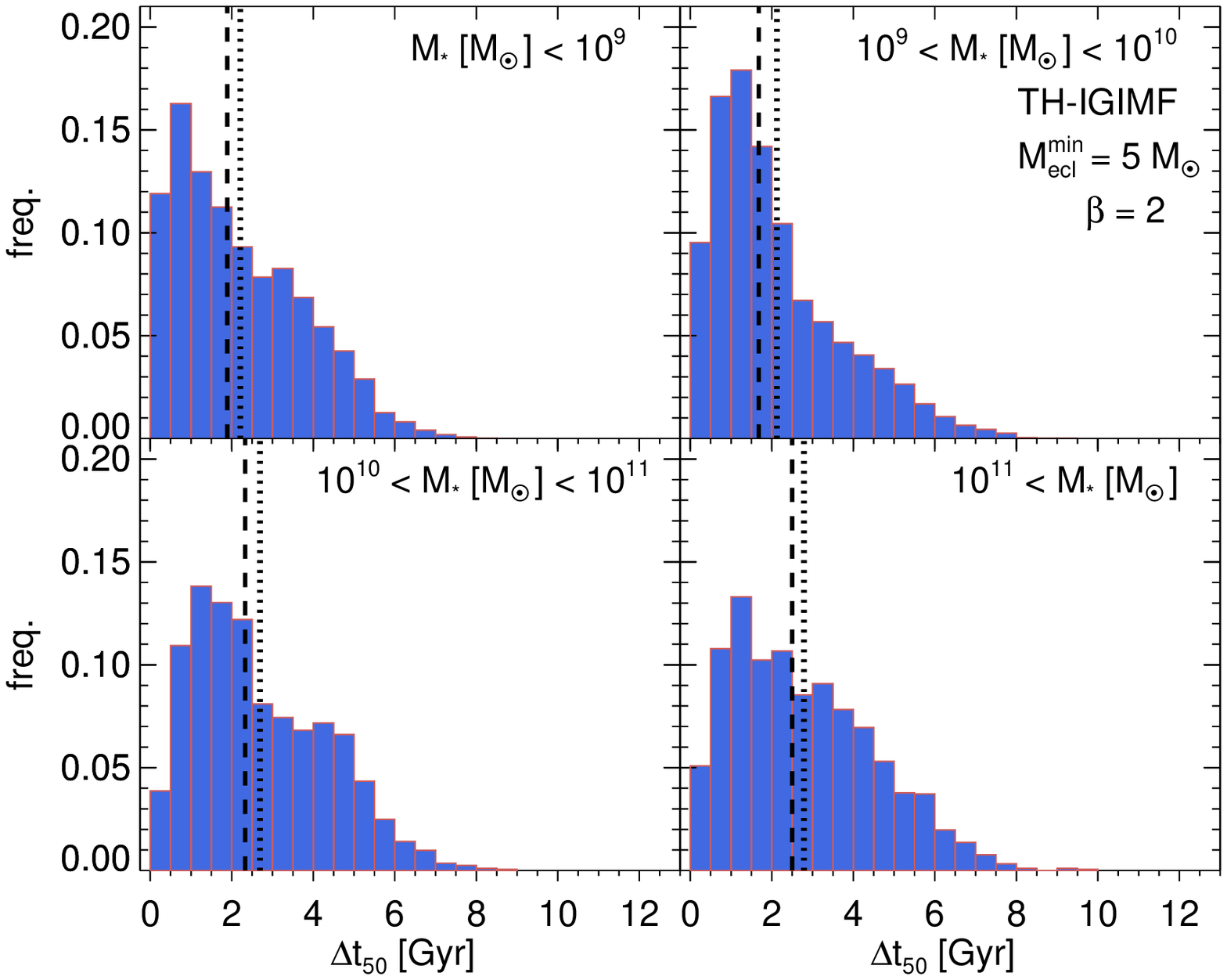}
\includegraphics[width=0.5\textwidth]{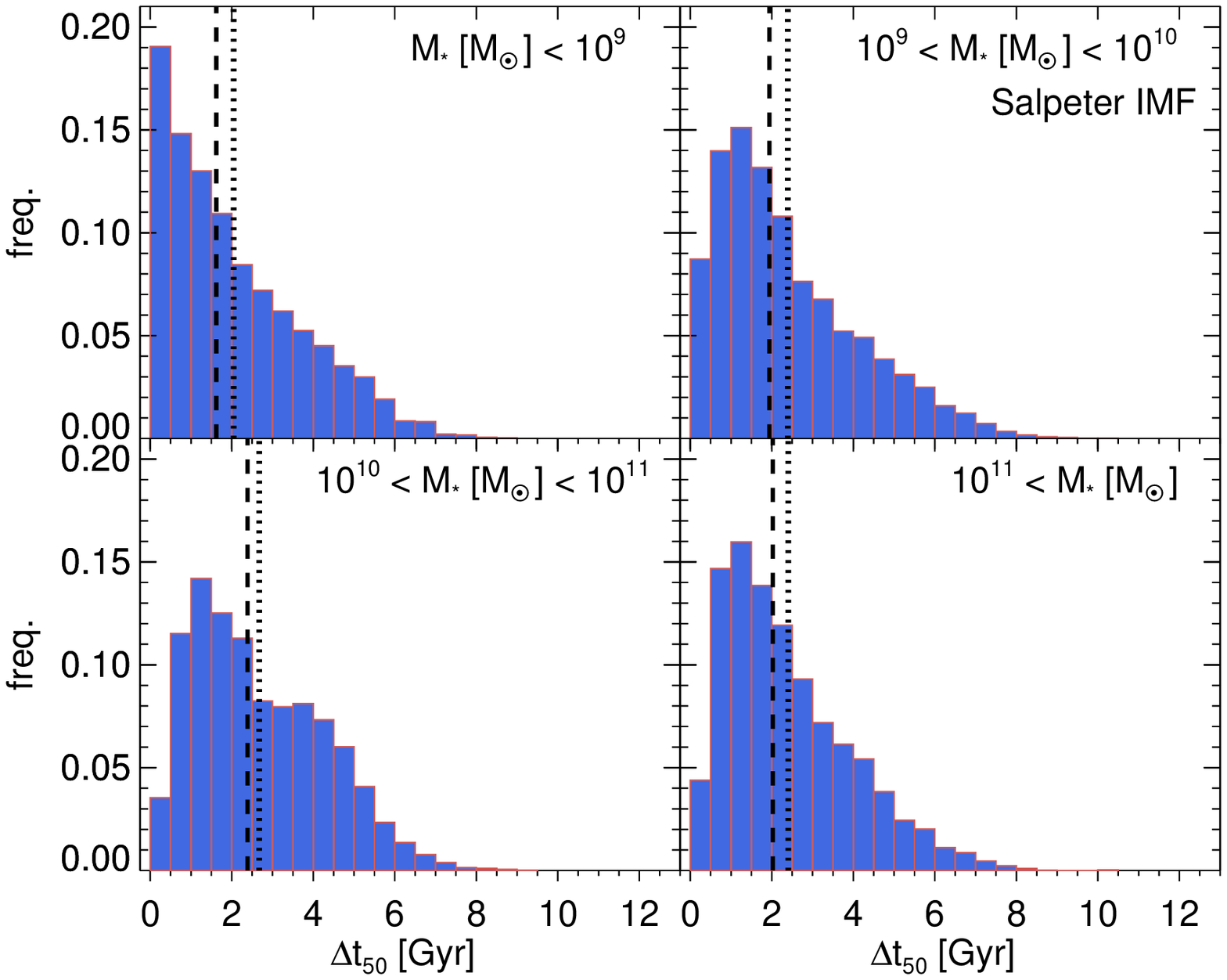}
\caption{
Distribution of formation time-scale, $\Delta T_{50}$, for
galaxies of different stellar mass, as indicated in the key
for models
{\small SAGTH5B2} (top panel) and {\small SAGS} (bottom panel).
Dotted and dashed lines represent the mean and median of the 
distributions, respectively.
}
\label{fig:freqDeltaT50MassBin}
\end{figure}

We now focus on the distributions of 
$\Delta {\rm T}_{50}$ 
considering galaxies in different mass ranges.
This is shown in Fig. \ref{fig:freqDeltaT50MassBin}
for both {\small SAGTH5B2} and {\small SAGS}
models (top and bottom panels, respectively).
As in the distributions for the whole galaxy population
(Fig. \ref{fig:DeltaTz50}),
the mean and median values (dotted and dashed lines, respectively)
are very close to each other. The main aspect to emphasize is that 
they are very similar to the values corresponding to the
complete sample of galaxies ($\approx 2 - 2.5 \,{\rm Gyr}$)
and do not present a great dependence with stellar mass;
only most massive galaxies ($M_{\star}>10^{11}$) 
in model {\small SAGTH5B2}
have slightly larger
mean and median values of $\Delta {\rm T}_{50}$.
This situation is pretty similar for both models, indicating that the
time-scales of galaxy formation are not strongly affected by the choice 
of the IMF. To put in context these galactic formation time-scales, we make a 
comparison with the time-scales needed to reproduce the right [$\alpha/{\rm Fe}$]-mass relation 
in the model of \citet{PipinoMatteucci2004}.
 Their models reproduce right slopes of [$\alpha/{\rm Fe}$] as a
function of galactic mass  assuming early infall of gas
 with time scales of the order of 0.4 ${\rm Gyr}$ and galactic
winds that last no longer than 0.7 ${\rm Gyr}$, 
within the paradigm of monolithic collapse, which explains the 
differences with our time-scales of galaxy formation.

The mild dependence of the formation time-scales with stellar mass that
emerges from our models, with a Salpeter IMF or a TH-IGIMF,
shows a trend opposite to the one found by 
\citet{DeLucia2006}. Based on the analysis of the star formation histories
of galaxies obtained from a semi-analytic model, they show 
that massive elliptical galaxies have shorter star formation time-scales
than less massive ones,
in qualitative
agreement with the downsizing scenario
\citep{Cowie1996} and 
with the stellar mass time-scale dependence required by 
\citet{Thomas2005} in order to explain the
observed [$\alpha/{\rm Fe}$]-stellar mass relation. 
The different trends obtained from our models and those 
of \citet{DeLucia2006} 
are not explained by the fact that 
we are not considering the star formation taking place in accreted satellites
in our definition of formation time-scale; as in the distributions
of formation redshift, we evaluate the influence of the stellar mass
in accreted satellites finding no significant change.
Note, however, that
the extensions of the time interval 
in which the star formation takes place present small differences
between galaxies in different mass ranges, both in our sample of model galaxies
and in those of \citet{DeLucia2006}. 

This mild
downsizing trend that emerges from our model, as shown by the
$z_{\rm form(50)}$ distributions and the almost null differences
of $\Delta T_{50}$ between galaxies of different masses
minimize the relevance of
the most accepted explanation to interpret
the larger
[$\alpha/{\rm Fe}$] abundance ratios in more massive galaxies,
which is generally attributed to their relatively short
formation time-scales as inferred from the analysis of observational data
\citep[e.g.][]{Thomas2005}. 
If this difference in the formation timescales
were present in real galaxies, it would contribute to the development of a 
positive slope of the relation as shown in other galaxy formation models
\citep[e.g.][]{PipinoMatteucci2004}.
However, this explanation of the chemo-archaeological downsizing
could be discarded if we take into account the
lack of strong evidence of downsizing
in terms of differential
evolution of the stellar mass assembly or for differential
evolution of the SFR density for galaxies of different mass
as emerge from the analysis of several observational data set
\citep{Fontanot2009, Sobral2014}.
Our models support
the possibility that massive galaxies reach higher [$\alpha/{\rm Fe}$]
abundance ratios because they are characterized by top-heavier IMFs.
Our study reinforces this idea because of
the similarity of the $z_{\rm form(50)}$ and $\Delta {\rm T}_{50}$
distributions
and of the relation between them for models with a
Salpeter IMF and a TH-IGIMF, indicating that the time-scales
of galaxy formation do not play a decisive role in determining
the slope of the
[$\alpha/{\rm Fe}$]-stellar mass relations followed
by galaxies generated by these two models.
The most relevant result
of our work is that a TH-IGIMF is necessary to reproduce the observed trend.
We have to bear in mind that,
although our model present a mild downsizing
in terms of the characteristic redshift of galaxy formation
and time-scales of star formation, the latter is the quantity that
matters in the interpretation of
the [$\alpha/{\rm Fe}$]-stellar mass
relation. However,
the main observational evidence of downsizing in terms of these time-scales
is inferred from the
relation itself.

\begin{figure}
\centering
\includegraphics[width=0.4\textwidth,angle=270]{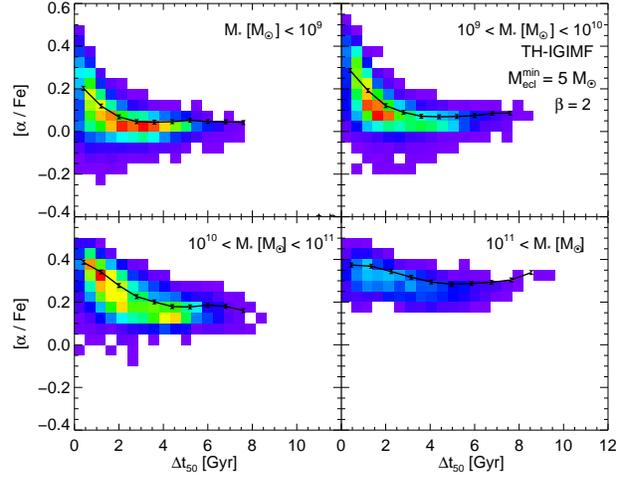}
\caption{
Distributions of [$\alpha/{\rm Fe}$] abundance ratios
as a function of the formation time-scale, $\Delta {\rm T}_{50}$,
for galaxies within different mass ranges, as indicated in the key,
for model {\small SAGTH5B2},  
characterized by $M_{\rm ecl}^{\rm min}=5\,\Msun$.
The solid line in each panel connects the mean values of 
[$\alpha/{\rm Fe}$] abundance ratios for different bins of 
$\Delta {\rm T}_{50}$.
}
\label{fig:AlphaDeltaT50MassBin}
\end{figure}

From this important result, we consider galaxies
from model {\small SAGTH5B2} and analyse the 
link between the
distribution of [$\alpha/{\rm Fe}$] abundance ratios 
and the formation time-scale $\Delta {\rm T}_{50}$
in order to understand the origin of the main features 
of the [$\alpha/{\rm Fe}$]-stellar mass relations
of galaxies with a TH-IGIMF characterized by $M_{\rm ecl}^{\rm min}=5\,\Msun$ and
a slope the mass function of embedded clusters $\beta = 2$.
This is shown in Fig. \ref{fig:AlphaDeltaT50MassBin}, where the different
panels correspond to galaxies within different mass ranges.
In all cases, we can see that larger values of [$\alpha/{\rm Fe}$] 
abundance ratios are achieved for shorter formation time-scales.
The important aspect to emphasize is that the time-scales argument
allows to explain the dispersion of [$\alpha/{\rm Fe}$] values 
for galaxies of a given mass, but is not the reason of the
development of an [$\alpha/{\rm Fe}$]-stellar mass relation
with a positive slope. This dependence between the [$\alpha/{\rm Fe}$] 
abundance ratios and the stellar mass is only possible by considering
a SFR dependent IMF like the TH-IGIMF tested here.

Linking the result shown in Fig. \ref{fig:AlphaDeltaT50MassBin},
which indicates that galaxies with shorter formation time-scales
have larger values of [$\alpha/{\rm Fe}$]
abundance ratios,
with that appreciated from Fig. \ref{fig:DeltaTz50}, 
where it is clear that older galaxies (those with 
higher $z_{\rm form(50)}$) are formed in shorter time-scales,
we conclude that older galaxies achieve larger values of [$\alpha/{\rm Fe}$]
abundance ratios.
This is consistent with the observational data analysed by 
\citet{SanchezBlazquez2006}, who find 
that the models that better reproduce
the observed slopes are those in
which the $\alpha$-elements vary
more than the Fe-peak
elements along the
sequence of velocity dispersion;
moreover, older galaxies at a given velocity dispersion
show higher [$\alpha/{\rm Fe}$] abundance ratios.
These results also support previous findings by \citet{Trager2000},
who show that  
younger ellipticals have
higher [Fe/H] abundances than older ones.

\section[]{Discussion}
\label{discussion}

The trend of $[\alpha /{\rm Fe}]$ with stellar mass in early type galaxies 
has been studied in the past, both observationally and theoretically. 
Regardless of the observational methods, there is agreement 
that the relative abundances of light elements
(produced almost 
exclusively by SNe CC) with respect to iron (produced mainly by SNe Ia) 
increase 
monotonically with galaxy central velocity dispersion (hence, galaxy mass).
On the theoretical side, a considerable deal of work has been devoted to study
this aspect of chemical evolution within the framework of the cosmological 
model $\Lambda$-CDM, mainly because it has been pointed out 
as a problem of the model,
showing tension between the expected hierarchical bottom-up growth of 
structure and the downsizing behaviour of galaxies. 
Two main hypotheses have been proposed to explain the observed trend,  
the time-scale of galaxy formation and 
the variable IMF.
In the time-scale argument, the stellar 
populations in galaxies with higher stellar masses are assumed to form 
in shorter time-scales, avoiding the enrichment of their ISM by SNe Ia that 
are caracterized by long explosion delay-times. In the case of the variable IMF 
hypothesis, the variable high-mass to low-mass stars ratio in 
the initial distribution of stars in different galaxies is assumed to play 
a crucial role in the building-up of the observed abundance pattern 
in early-type galaxies, given that this ratio determines the SNe CC to
SNe Ia ratio. 

In the last years, there have been several studies of massive early
type galaxies suggesting that their stellar populations might 
have been formed
with IMFs different from those found in the solar neighbourhood. 
There are
indications that the light of stars in massive early type galaxies is consistent
with populations formed with a bottom-heavy IMF, as given by 
the analysis of their spectral features  
\citep{vanDokkum2010, vanDokkum2012, Ferreras2013, LaBarbera2013, Conroy2013b}; 
in other works, the mass-to-light ratio of massive
early type galaxies is constrained by
using dynamical tools  
\citep{Cappellari2012, Dutton2012, Tortora2013} or
 strong gravitational lensing
\citep{Auger2010, Treu2010}.
However, these methods present a degeneracy in the sense 
that a higher ratio of less massive stars implies either a bottom heavy IMF 
or a top heavy IMF, since massive stars end their lives in short time-scales 
and contribute 
to the total emitted light only during star formation events and shortly after. 
\citet{Weidner2013c} study the problem with a toy model that proposes a top heavy
IMF at high redshift during early starbursts conditions and a bottom heavy IMF 
in the subsequent evolution. In their model, the high star formation 
regime drives the fragmentation of the ISM favouring the formation of low-mass
stars. One important aspect they remark is that the bottom heavy IMF proposed
by several works, like \citet{Ferreras2013}, can underestimate severely the 
amount of metals present in massive elliptical galaxies, given that most metals
are ejected to the ISM by stars with masses above $1\,\Msun$. 

In general, when the argument of a variable IMF is considered
to explain the $[\alpha /{\rm Fe}]$-stellar mass relation, 
the proposed IMF is treated as a free parameter \citep{Arrigoni2010},  
or is varied with exploratory aims 
\citep{Thomas1999, Nagashima2005, Calura2009},
following no particular theory and leaving unexplored a vast region of the 
corresponding parameter space. In this work, we test the well defined theory
regarding the integrated initial mass function of stars in galaxies with top
heavy IMFs in star clusters during starbursts (see Subsection \ref{sec:varIMF} 
for a brief description
and references for the original papers). 
We have shown
that the $[\alpha/{\rm Fe}]$ abundance ratio is well reproduced 
by assuming a SFR dependent TH-IGIMF; 
our results support a value of $5\,\Msun$
for the minimum embedded star cluster mass
and of $\beta=2$ for the slope of the embedded cluster mass function, 
which are free parameters in this theory.
 \citet[][ R09 hereafter]{Recchi+2009} have already 
considered the IGIMF theory to evaluate the impact in the build-up of the 
[$\alpha /{\rm Fe}$] - $\sigma$ relation and explore different values for the 
slope of the embedded cluster mass function ($\beta$ $1, 2$ and $2.35$). 
They make use of a previous version of the IGIMF, where the effect of 
high levels of star formation in galaxies was not considered.
When these high levels of SF are attained,
the IMF whithin clusters can become top heavy due to 
effects of crowding. 
In such conditions,
the SFR dependence changes 
with the consequent reduction in the values of the slopes of the IGIMF in the TH-IGIMF version.
There are important differences between the 
models used in R09 and
those used in this paper. We make use of a semi-analytical model
with merger trees that considers the evolution of elliptical galaxies in the
framework of a $\Lambda$-CDM Universe. The analytical model used in R09 
considers
the evolution of elliptical galaxies from the point of view of the monolithic
collapse.  Another fundamental aspect is the treatment
of the stellar and AGN feedback.  Our model takes into account the 
feedback produced by stars consistently with the choice of the 
TH-IGIMF slope. This means that lower values of the
slopes of the 
TH-IGIMF lead to higher numbers of SNe CC that reheat cold gas,
regulating the 
formation of new generations of stars and placing strong constraints on the
free parameters of the TH-IGIMF theory. For example, for 
$M_{{\rm ecl}}^{{\rm min}}=100\,\Msun$ or even for the slope of the embedded star
cluster $\beta=2.1 $, 
we find it difficult to calibrate
the model
given the high levels of SNe feedback. Feedback of AGNs in galaxies,
which is not considered in the models of R09, 
is also an important aspect because it 
prevents massive galaxies to 
continue forming stars at low redshifts. This contributes to determine the mild downsizing 
trend seen in our model although a more strong effect would be expected.

An associated problem to the $\alpha$-element abundances
in early-type galaxies is related to the calcium (Ca) abundances. 
It is commonly 
expected that all the light elements produced by $\alpha$-capture track
each other 
as a result of their common origin. 
However, the Ca abundance with respect to iron
does not follow the magnesium (Mg) or titanium (Ti) abundance 
relative to iron as a function of central
velocity dispersion (or galaxy mass) in early type galaxies. While the 
spectral indices
sensitive to Mg or Ti show an increase of their abundances with increasing mass,
some indices sensitive to Ca show a decrease of the abundance of this element
with mass \citep{Saglia2002, Cenarro2003, Cenarro2004}. Recently
\citet{Worthey2013} confirm this puzzling trend with new observations and
evaluate different hypothesis to explain it. They find that a steepening
of the IMF at low-mass stars is ruled out by constraints on galaxy colours.
Interestingly, an IMF favouring the formation of massive stars is in agreement 
with their results, provided that the used stellar 
yield of Ca depends on the progenitor mass of SNe CC. Yields from 
\citet{Woosley1995} show a decrease of Ca with respect to iron with 
increasing mass of the progenitor star.
However, \citet{Nomoto2006} and \citet{Kobayashi2006} show
no dependence of the Ca yield with progenitor mass. If this trend of the SNe CC
stellar yields of Ca is real, a top heavy IMF, as the one tested in this 
paper, would be needed to explain the calcium correlation with mass of early 
type galaxies, since the formation of massive stars would be favoured during
the formation of more massive galaxies.

Recent work by \citet{Conroy2013} show more interesting trends between 
other individual elements. They study the abundance patterns of barium (Ba) 
and strontium (Sr)
and find that the relative abundance ratio [Ba/Mg] show a decrease with respect
to [Mg/Fe] and, although not shown in their work, they claim a strong 
correlation between [Mg/Ba] and velocity dispersion in a sample of early-type 
galaxies drawn from the SDSS. The importance
of this finding lies in the fact that Ba is an
element that is produced by the s-process in low and intermediate mass stars
during the phase of asymptotic giant branch (AGB) in their evolution. 
Thus, the abundance ratio [Mg/Ba]
can be used as a probe of short formation time-scales as well as an indicator
of high mass to low mass stars ratio, just like the [Mg/Fe] ratio. But this is
more than an alternative measure because it permits to set aside the 
influence of SNe Ia. Consequently, if real, this new abundance pattern  
would allow to rule out some explanations for the $[\alpha/{\rm Fe}]$-stellar mass 
relation in 
early-type galaxies involving different delay-times distributions for SNIa
\citep[e.g.][]{Yates2013}. 
In a future paper, we will analyse the behaviour
of these individual trends in our semi-analytic model in order to shed 
light on these issues.

We point out here a drawback in our chemical implementation, also
present in other semi-analytic models \citep[e.g.][]{Yates2013}.
We consider, for simplicity, that yields from winds of low- and 
intermediate-mass stars, estimated    
as the integral of the metal ejection taking place during their 
entire evolutions, 
are deposited in the ISM at the end of their lives. Low and intermediate
mass stars (LIMS) have long lives, and most of them still exist at the time 
of the 
final simulation output. Therefore, we consider that the gradual 
ejection of metals from LIMS could
affect our results. 
We will also tackle this 
problem in a future work ,
 analysing also the dependence of the stellar
yields with the metallicity of the stars.

\section[]{Conclusions}
\label{conclusions}

We use a
model
of galaxy formation that combines a
dissipationless cosmological
{\em N}-body simulation
with the semi-analytic model of galaxy
formation and evolution \sag~\citep{Cora06,lcp08,Tecce2010} 
to study the physical aspects behind the build-up of
the $[\alpha/{\rm Fe}]$-stellar mass relation of elliptical galaxies,
known as chemo-archaeological downsizing \citep{Fontanot2009}.
A version of the semi-analytic model \sag~that involves
the Salpeter IMF (model {\small SAGS}) is not able
to reproduce the right slope of the $[\alpha/{\rm Fe}]$-stellar mass
relation once the
modelling of several aspects of \sag~have been improved (like the 
starburst time-scales).

Earlier attempts to model the abundance of $[\alpha/{\rm Fe}]$ 
have failed to reproduce the observed relation with stellar mass, 
unless when invoking ad hoc hypotheses, 
like a change in the slope of the global IMF \citep{Arrigoni2010}, an 
arbitrary SFR-dependent IMF \citep{Calura2009}, starbursts in progenitor 
galaxies triggered by close encounters \citep{Calura2011} or
empirically adopted SNIa delay times \citep{Yates2013}. 
Motivated by our finding that
a universal IMF does not naturally allow
to reproduce the right slope in the observed trend between the
[$\alpha/{\rm Fe}$] abundance ratio and the stellar mass of elliptical galaxies,
we explore the impact of a variable
IMF on their
[$\alpha/{\rm Fe}$]-stellar mass
relation.
We have achieved $[\alpha/{\rm Fe}]$
abundance ratios for different stellar mass 
consistent with those found in observations, by
 implementing in our
semi-analytic model a SFR dependent IMF that results from
a physically motivated theory (TH-IGIMF, WKP11). 
The consistency of this implementation and the fact that many other
observational galaxy properties and relations are simultaneously reproduced,
make our conclusions more robust. 
We find that the use of a 
TH-IGIMF to represent
the galaxy-wide IMF is necessary for a proper treatement of the
chemical enrichment of galaxies. 
We summarize the main results of our work:
\begin{itemize}

\item{
A galaxy formation model with a TH-IGIMF can recover the
positive slope of the observed [$\alpha/{\rm Fe}$]-stellar mass
relation (Fig. \ref{fig:AlphaFeTHIGIMF}).
This is the case for 
the two test values considered for the minimum embedded star cluster mass
($M_{\rm ecl}^{\rm min} = 5$ and $100,\Msun$), and the two values considered 
for the slope of the embedded star cluster mass function $\beta = 2$ and $2.1$
which are free parameters involved in the TH-IGIMF theory.
However,
the agreement with the observational trend is particularly good for
$M_{\rm ecl}^{\rm min} = 5\,\Msun$ and $\beta = 2$.
Thus, our results 
favour a small value for 
the minimum embeded star cluster mass and the standard value for the 
slope of the embedded star cluster mass function.
}

\item{
Regardless of the value of $M_{\rm ecl}^{\rm min}$ and $\beta$, the mean slope
of a TH-IGIMF
($\langle \alpha_{\rm TH} \rangle$) becomes
progressively lower for more massive galaxies (Fig. \ref{fig:MeanSlope}). 
Since this slope is SFR-dependent, this behaviour is a manifestation of the
fact that  
more massive galaxies have higher SFR.
The relative difference
of the IMF slope for galaxies with different masses
is key to reproduce the observed 
[$\alpha/{\rm Fe}$]-stellar mass relation.
This is demonstrated by 
our best model, characterized by 
$M_{\rm ecl}^{\rm min}=5\,\Msun$ and $\beta = 2$; 
in this case, the TH-IGIMF is steeper than the Salpeter IMF
for half of the  mass ranges, showing that
a top heavy IMF in all galaxies is not a necessary condition in the 
build-up of the [$\alpha/{\rm Fe}$]-stellar mass relation as generally believed.
Our results indicate 
that galaxies with masses smaller
than $\sim 3\times 10^{10}\,\Msun $ should have IMF
steeper than Salpeter IMF,
while galaxies with higher masses should take top-heavier IMF than Salpeter
IMF in order to achieve the right chemical abundances.
}

\item{
A mild downsizing trend is evident for galaxies generated by models
with different IMF. 
The distributions of the formation redshift ($z_{\rm form(50)}$) 
and formation time-scale
($\Delta T_{50}$) of elliptical galaxies generated from models 
{\small SAGTH5B2} and 
{\small SAGS}, characterized by a TH-IGIMF with 
$M_{\rm ecl}^{\rm min}=5\,\Msun$ and $\beta = 2$ and the Salpeter IMF, 
respectively, 
have similar features and dependence with stellar mass 
(Figs. \ref{fig:freqzform50MassBin} and \ref{fig:freqDeltaT50MassBin})
supporting the fact 
that the choice of the IMF has negligible effect on the
time-scales of galaxy formation.
In both cases, the
least massive galaxies ($M_{\star} < 10^{9}\,\Msun$)
achieve the $50$ per cent of their stellar
mass formed in-situ at lower redshifts than more massive ones,
with the bulk of them forming between $z\approx 0$ and $z\approx 0.5$;
the peak of the distributions for galaxies more massive than 
$\approx 10^{9}\,\Msun$
shifts from the interval $0 \lesssim z_{\rm form(50)} \lesssim 0.5$ to
$0.5 \lesssim z_{\rm form(50)} \lesssim 1$. The distributions of
$\Delta T_{50}$ are very similar for galaxies of different mass,
displaying values corresponding to the
complete sample of galaxies ($\approx 2 - 2.5 \,{\rm Gyr}$). 
This mild downsizing trend present in models either with 
a universal or variable IMF indicates that the time-scale for galaxy 
formation does not play a decisive role in determining the slope of the 
[alpha/Fe]-stellar mass relation, which does show important changes 
in our results when changing to the TH-IGIMF. It should be born in mind, 
though, that the mild downsizing is a feature of semi-analytic models. 
If the time scale for star formation is allowed to vary more freely, 
then it also helps to produce a dependence of [alpha/Fe] on stellar mass. 
However, in this case, other properties of the galaxy population in a 
cosmological context might not behave properly.  
Using a semi-analytic model, we are able to fit the [alpha/Fe]-stellar 
mass relation at the same time as the z=0 LF produces concordant 
results with observations. This requirement poses extra limitations in SAMs, 
which we show can still provide a reasonable [alpha/Fe]-stellar mass relation. 
Our conclusion does not rule out the influence of downsinzing on the 
development of this relation; it is just that this effect is not strongly 
manifested in galaxy formation models.
}

\item{
Older galaxies (those with
higher $z_{\rm form(50)}$) are formed in shorter time-scales;
galaxies that have formed at $z_{\rm form(50)}\approx 2$ or before
have $\Delta {\rm T}_{50} \lesssim 2\,{\rm Gyr}$ (Fig. \ref{fig:DeltaTz50}).
On the other hand, galaxies with shorter formation time-scales
have larger values of [$\alpha/{\rm Fe}$]
abundance ratios (Fig. \ref{fig:AlphaDeltaT50MassBin}).
Hence, older galaxies achieve larger values of [$\alpha/{\rm Fe}$]
abundance ratios in agreement with observational data
\citep{SanchezBlazquez2006,Trager2000}. This correlation is
present for galaxies of different mass, thus 
explaining the dispersion of the [$\alpha/{\rm Fe}$]-stellar mass relation
at a given stellar mass. 
}

\item{
Mass-to-light ratios of massive elliptical galaxies have similar
shapes  
in both models {\small SAGTH5B2} and 
{\small SAGS}, showing an excess for low mass galaxies 
($\lesssim 2\times 10^{10}\,\Msun$) with respect to
observations (Fig. \ref{fig:MLRatios}).
However,  
model SAGTH5B2 allows to achieve the correct values of M/L ratios for
massive galaxies, 
leading to a better agreement with observational data. The bright end of 
the {\em r}-band LF also behaves better in model SAGTH5B2
(Fig. \ref{fig:LFCal}). 
}

\end{itemize}

From the interpretation of observational data
\citep[e.g.][]{Thomas2005}, a
generally accepted explanation for the build-up of 
the [$\alpha/{\rm Fe}$]-stellar mass relation has been 
the relatively short
formation time-scales of massive galaxies.  
Based on this time-scale argument, we would obtain
larger values of [$\alpha/{\rm Fe}$] for these galaxies since
they have lower
possibilities to form stars from cold gas contaminated by products of SNe Ia.
The detailed analysis performed in this work had the purpose of
determining if the galaxy formation time-scales are
responsible for the positive
slope in the [$\alpha / {\rm Fe}]$-stellar mass relation.
From this investigation, we can conclude that 
this time-scale argument 
only allows to explain the dispersion of
[$\alpha / {\rm Fe}]$ values for galaxies of a given mass, 
but cannot account for the observed 
[$\alpha / {\rm Fe}]$-stellar mass relation.
Our results, based on a hierarchical galaxy formation model, 
show that the presence of a SFR-dependent TH-IGIMF is a valid explanation 
for the relation between the [$\alpha / {\rm Fe}]$ abundance ratios 
and stellar mass.
A variable IMF which slope depends on the SFR gives different
relative numbers of SNe Ia and SNe CC progenitors for each
star formation event, which is a key aspect
in the build-up of the [$\alpha/{\rm Fe}$]-stellar mass
relation.
Massive galaxies reach higher [$\alpha/{\rm Fe}$]
abundance ratios because they are characterized by top-heavier IMFs
as a result of their higher SFR.

\section*{Acknowledgments}
We thank the referee for his/her helpful comments that allowed to improve this
paper.
This work was partially supported by the Consejo Nacional de Investigaciones Cient\'{\i}ficas y
T\'ecnicas (CONICET, Argentina), 
Universidad Nacional de La Plata (UNLP, Argentina), 
Instituto de Astrof\'{\i}sica de
La Plata (IALP, Argentina) and 
Secretar\'{\i}a de Ciencia y Tecnolog\'{\i}a de la Universidad
Nacional de C\'ordoba (SeCyT-UNC, Argentina).
AO gratefully acknowledges support from FONDECYT project 3120181.
IDG acknowledges support from  Comisi\'on Nacional de Investigaci\'on Cient\'{\i}fica y Tecnol\'ogica (CONICYT, Chile) and Pontificia Universidad Cat\'olica de Chile (PUC) projects PFB-06 and ACT-86 for an academic stay at PUC.
SAC acknowledges grants from CONICET (PIP-220), Argentina,
Agencia Nacional
de Promoci\'on Cient\'{\i}fica y Tecnol\'ogica (PICT-2008-0627), Argentina,
and Fondecyt, Chile.
AMMA acknowledges support from CONICYT Doctoral Fellowship program.
TET acknowledges funding from GEMINI-CONICYT
Fund Project 32090021, Comit\'e Mixto ESO-Chile and Basal-CATA (PFB-06/2007).
GB acknowledges support from the National Autonomous 
University of M\'exico, through grant IB102212-RR182212.
This project made use of the Geryon cluster at the Centro de Astro-Ingenier\'ia UC for most of the
calculations presented.

\newpage
\bibliographystyle{mn2e}
\bibliography{gargiulo}

\bsp

\label{lastpage}



\appendix

\section{\\Free parameters and analytic recipies in the model} 
\label{App:AppendixA}

In the present appendix, we give a short description of the role
of each of the free parameters involved in the model {\small SAG} and the
recipies used to follow the formation and evolution of galaxies.

The efficiency of star formation, $\alpha_{\rm SF}$, is a
free paraemter involved in
the quiescent mode of star formation and follows the expression proposed by 
\citet{Croton2006}, that is,
\begin{equation}
{\frac{dM_\star}{dt}} = \alpha {\frac{M_{\rm cold} - M_{\rm cold,crit}}{t_{\rm dyn}}},
\end{equation}
with
\begin{equation}
M_{\rm cold,crit} = 3.8\times10^9 \left(\frac{V_{\rm vir}}{200\kms}\right) \left(\frac{3\,R_{\rm disc}}
{10\,{\rm kpc}}\right) \Msun,
\end{equation}
\noindent where $t_{\rm dyn}=V_{\rm vir}/3R_{\rm disc}$ 
is the dynamical time of the galaxy, $V_{\rm vir}$
the circular velocity at the virial radius and $R_{\rm disc}$ 
the disc scale length \citep{Tecce2010}. 
If the mass of cold gas exceed $M_{\rm cold,crit}$, star formation occurs. 

The SNe feedback efficiency, $\epsilon$, 
is a free parameter that controls the amount of cold gas reheated by the 
energy generated by SNe CC. The amount of reheated mass produced by 
a star forming event 
that generates a stellar population of mass $\Delta M_\star$ is assumed to be
\begin{equation}
\Delta M_{\rm reheated} = \frac{4}{3} \epsilon {\frac{\eta E}{V_{\rm vir}^2}} \Delta M_{\star}
\label{eq:feedbackSN}
\end{equation}
\noindent where $E=10^{51}\,{\rm erg}\, {\rm s}^{-1}$ is the energy released 
by a SNe, and $\eta$ is the number of 
SNe which depends on the IMF adopted, estimated as 
\begin{equation} 
\eta = \frac{\int^8_0 \phi(m)\, m \,\,{\rm d}m}{\int_0^{\infty} \phi(m) \,{\rm d}m} ,
\end{equation}
\noindent where $\phi(m)$ is the adopted IMF and $m$ 
the stellar mass. This quantity 
is constant for the Salpeter IMF (models {\small SAGS-c1} and {\small SAGS-c2}), 
and varies accordingly to the adopted slope of
TH-IGIMF in each event of star formation for models {\small SAGTH5B2}, 
{\small SAGTH5B21} and {\small SAGTH100B2}. 
The SNe feedback from stars
in the disc and in the bulge is treated separately.  
The efficiency $\epsilon$ of eq.
\ref{eq:feedbackSN} corresponds to stars borned in the disc, and
a new free parameter is introduced to regulate the SN feedback from
stars in the bulge, to which we refer to as $\epsilon_{\rm bulge}$. 
In both cases, the amount of reheated mass is given by eq. \ref{eq:feedbackSN}.

There are two free parameters related to the growth of the BHs in the model. 
In first place,
a BH is fed by the inflow of cold gas provided by perturbations to the 
gaseous disc 
resulting from galaxy mergers or disc instabilities. 
When a galaxy merger occurs, 
the central BHs are assumed to merge instantly. 
In all the versions of \sag~calibrated for this
work, a fraction of the cold gas from the merger remnant, 
$f_{\rm BH}$, is supplied for the BH growth 
as
\begin{equation}
\Delta M_{\rm BH} = f_{\rm BH} {\frac{M_{\rm sat}}{M_{\rm cen}}} {\frac{M_{\rm cold,sat} + M_{\rm cold,cen}}
                {(1 + 280 \kms / V_{\rm vir})^2}}
\end{equation}
where $M_{\rm cen}$ and $M_{\rm sat}$ are the masses of the merging 
central and satellite 
galaxies, and $M_{\rm cold, cen}$ and $M_{\rm cold,sat}$ 
are their corresponding masses of cold gas.
When a starburst is triggered through a disk instability occurring 
in a galaxy, only
the cold gas of the unstable galaxy is involved. 
The BH accretes cold gas gradually as long as there is
bulge cold gas available (see Subsection \ref{sec:starbTimeScales}). 
In the latter cases, the BH growth and star formation in starburst
mode are coupled in such a way that their actions, together with SNe feedback
from the stars formed in the bulge, determine the duration of the starburst.
In second place, BHs 
grow from hot gas during the gas cooling processes
taking place in central galaxies (type 0) as 
\begin{equation}
{\frac{dM_{\rm BH}}{dt}} = \kappa_{\rm AGN} {\frac{M_{\rm BH}}{10^8 \Msun}} {\frac{f_{\rm hot}}{0.1}}\left({\frac{V_{\rm vir}}{200 \kms}}\right)^2
\end{equation}
\noindent 
where $f_{\rm hot}=M_{\rm hot}/M_{\rm vir}$, 
being $M_{\rm hot}$ and $M_{\rm vir}$ the hot gas and virial 
masses, respectively; the free parameter
$\kappa_{\rm AGN}$ is the efficiency of accretion. 

The last two free parameters of the calibration process are the fraction of 
binaries, $A_{\rm bin}$, that originates SNe Ia, already introduced in 
Subsection \ref{sec:chem_model},
and the factor involved in the 
distance scale of perturbation to trigger disc instabilities, $f_{\rm pert}$
(see Subsection \ref{sec:ellip_form}).

\end{document}